 \documentclass[authoryear,review]{elsarticle}
\usepackage{lineno}
\pdfoutput=1
\usepackage{color}
\usepackage{hyperref}
\hypersetup{
	linkcolor  = red,
	citecolor  = blue,
	urlcolor   = cyan,
	colorlinks = true,
}

\usepackage{graphicx}
\usepackage{rotating}
\usepackage{tikz}

\usepackage{amssymb} 
\usepackage{amsfonts} 
\usepackage{amsmath}
\usepackage{enumerate}
\usepackage{fancyhdr}
\usepackage{sidecap}
\usepackage[textfont=small,justification=justified,labelfont=bf,up]{caption}
\usepackage{subfig}
\usepackage{wrapfig}
\usepackage{float}
\usepackage{quoting}
\quotingsetup{font=small}
\usepackage{amscd}
\usepackage{array}
\usepackage{multirow}
\usepackage{siunitx}

\newcommand{\R}{\mathbb{R}}
\DeclareMathOperator*{\argmax}{arg\,max}
\newenvironment{system}%
{\left\lbrace\begin{array}{@{}l@{}}}%
	{\end{array}\right.}

\begin{document}
\begin{frontmatter}
	
	\title{Application of optimal data-based binning method to spatial analysis of ecological datasets}
	
	%% Group authors per affiliation:
	\author[1]{Anna Tovo\corref{mycorrespondingauthor}}
	\cortext[mycorrespondingauthor]{Corresponding author}
	\ead{annatovo@math.unipd.it}
	\author[1]{Marco Formentin}
	\author[1]{Marco Favretti}
	\author[2]{Amos Maritan}
	\address[1]{Department of Mathematics, University of Padova, Padova, Italy}
	\address[2]{Department of Physics and Astronomy, University of Padova, INFN and CNISM, Padova, Italy}

	\begin{abstract}
		Investigation of highly structured data sets to unveil statistical regularities is of major importance in complex system research. The first step is to choose the scale at which to observe the process, the most informative scale being the one that includes the important features while disregarding noisy details in the data.   In the investigation of spatial patterns, the optimal scale defines the optimal bin size of the histogram in which to visualize the empirical density of the pattern. 
		 In this paper we investigate a method proposed recently by K.~H.~Knuth to find the optimal bin size of an histogram as a tool for statistical analysis of spatial point processes. We test it through numerical simulations on various spatial processes which are of interest in ecology. 
		We show that Knuth optimal bin size rule reducing noisy fluctuations performs better than standard kernel methods to infer the intensity of the underlying process. Moreover it can be used to highlight relevant spatial characteristics of the underlying distribution such as space anisotropy and clusterization. We apply these findings to analyse cluster-like structures in plants' arrangement of Barro Colorado Island rainforest.
	\end{abstract}
	
	
\end{frontmatter}

%----------------------------------------------------------------------------------------
%	ARTICLE CONTENTS
%----------------------------------------------------------------------------------------
	
%\linenumbers

\section*{Introduction}
Nowadays, a huge quantity of data structured on different time and space scales are easily available.
Analysis of these massive databases reveals that despite their diversity and complexity, natural phenomena are characterized by the emergence of regularities that are largely independent of biological and physiological details. One is the tendency, observed both in ecological communities and in human activities, to form spatial or temporal clusters (\cite{He,Condit,Plotkin,Adorisio}).
However, classifying a spatial point pattern as clustered rather than regular can be a challenging task because 
establishing the main features of its spatial density function strongly depends on the \emph{scale} through which we look at it.\\
More generally, it is well known that the form of a data-based density function may depend on the algorithm (binning rule) used for the binning of the data (\cite{Etienne}). In our view the main flaw of many binning rules (\cite{Sturges, Yule,Doane,Freedman,Stone,Scott}) is that they assume some knowledge on the data distribution. For example Sturges rule (\cite{Sturges}) assumes that the data are normally distributed. This is a key point if you have in view applications to ecological datasets. In many cases it is not reasonable to assume such knowledge and the process generating the dataset must be considered unknown. Therefore any criteria based on some prior knowledge of the true density should not be applied as it often introduces a degree of arbitrariness that may produce biased conclusions.\\
In this paper we are concerned with the statistical analysis of spatial patterns describing the location of plants in a tropical forest study area. 
We intend to use a method proposed by K.~H.~Knuth (\cite{Knuth}) to find the optimal bin size of a two-dimensional histogram.  Knuth non parametric method selects the optimal scale from the data without any assumption on the underlying process that generated the data. We show that the Knuth method can be used to highlight relevant spatial characteristics on the underlying distribution such as space anisotropy and clusterization. We tested it against most currently used (Epanechnikov) kernel method for two-dimensional datasets and one-dimensional (Stone binning rule) and it results to be more efficient in detecting Complete Spatial Random processes and avoiding sample fluctuations. Therefore our analysis validates it as a reliable method for determining the intensity function of a pattern. Additionally, it is not subject to the virtual aggregation phenomenon (see below Sec. \ref{vafen}). It correctly detects homogeneity or the presence of a gradient in the density function and the relative difference of the rectangular bin sides is a measure of the anisotropy of the pattern. It also allows to infer quantitative (cluster size) information on both first and second-order statistics. Thus, it is not only a rule to choose the bin size in which to organize the data. Indeed our analysis  proves that Knuth bin size is a good indicator of how finely structured is the dataset and that it can be used as a trusted tool for the preliminary statistical analysis of a spatial dataset. We show which are the relevant information contained in the size and the shape of the optimal bin and how they are related to the spatial features of the process/dataset. \\
We test our findings on an ecological dataset consisting of the spatial coordinates of individuals belonging to 300 different species of plants located in a 50~ha rectangle of the Barro Colorado Island rainforest (BCI). In particular, we study cluster-like structures in plants' arrangement.\\
The present analysis should help inform future investigations of temporal or spatial features of different complex systems in ecology and human dynamics (\cite{Simini,Formentin, Sanli}).

%------------------------------------------------

\section{On optimal binning rules}
There exist diverse rules to determine the optimal number of bins of a histogram. Some of the most known (\cite{Scott1,Freedman,Stone,Scott}) rely on the minimization of the $L^2$ norm between the histogram and the true underlying density on which they assume some prior knowledge. Assuming such prior information is not reasonable for ecological datasets and the process generating the data must be considered unknown. Hence, any criteria determining the optimal bin size based on prior knowledge on the true density should not be applied. Moreover, some methods work well for unimodal densities while they are known to be suboptimal for multimodal ones. In particular, Freedman and Diaconis rule (\cite{Freedman}) is not valid for uniform or piece-wise constant density functions whereas Sturges rule (\cite{Sturges}) is not suitable when the data exhibit skewness or any other non-normality. To overcome these difficulties, K.~H.~Knuth (\cite{Knuth}) proposed a method based on \emph{Maximum a posteriori estimation} and \emph{Bayes' Theorem}. The main advantage is that no prior information about the density from which the data are sampled is assumed. Moreover, the Knuth method is more efficient in avoiding sample fluctuations and it best approximates the probability density in the $L^p$ norm sense, with $p=1,2$. See in Supporting Information the comparison between the Knuth method and the one developed by C.~J.~Stone (\cite{Stone}) and in Sec.~\ref{sec2} below the comparison with standard kernel methods. 

\section*{The Knuth method} \label{sec:Knuth_method}
\label{Ksec}
We briefly present the Knuth method referring to \cite{Knuth} for further details.
Let us suppose to be working with a set of $ N $ 2-dimensional data sampled from an unknown probability density function we wish to estimate. In applications to spatial analysis of ecological datasets each 2-dimensional datum may represent the location of a tree.\\
Let us then cover our data span $V$ with a rectangular grid of $ M_x\times M_y $ rectangular bins, where $ M_x $ is the number of bins along the $ x $-axis and $ M_y $ the number along the $ y $-one. We assume no measurement uncertainty about the data hence $ V $ is precisely known. We set $\underline{M}=(M_x,M_y)$ and $M=M_x\cdot M_y$. We denote with $a_x$ and $a_y$ the width of each bin along the $x$ and $y$ axis respectively and with $ a= a_x\cdot a_y$ the bin area. We call $ h_k $, where $ k=1, ..., M $ is the bin label, the correspondent histogram column's height. After the normalisation of the volume of the histogram, $h_k$ represents the constant value of the probability density function over the region of the bin. The volume of each histogram column $\pi_k=h_ka$ is the probability mass of each bin,
i.e. the probability of finding a datum in the range dictated by the $ k^{th} $ bin. 
Therefore the piece-wise constant density function of the histogram with $ M $ bins is:
\begin{linenomath}
	\begin{equation}
	\label{h_hist}
	h(x,y)=\sum_{k=1}^{M}h_k\Pi_k(x,y),
	\end{equation}
\end{linenomath}
where $ \Pi_k(x,y) $ is the boxcar function, defined as
\begin{linenomath}
	\begin{equation}
	\label{boxcar}
	\Pi_k(x,y)=
	\begin{system}
	1\  \mbox{if  $(x,y)$  falls  within the $k^{th}$  bin} \\
	0\ \mbox{otherwise}  \\
	\end{system}
	\end{equation}
\end{linenomath}
Notice that due to the normalisation, only $ M-1 $ probabilities masses are independent.

\paragraph{The likelihood function}
When we arrange the data into a histogram, the probability that a datum falls within the $ k^{th} $ bin is given by the probability mass of that bin. The associated probability density $h_k$ is called likelihood function.\\
Denoting with $ \underline{\pi}={(\pi_1,...,\pi_{M-1})} $ the independent probabilities masses, with $ \underline{d}={(d_1,...,d_N)} $ the vector of 2-dimensional data points, with $ n_k $ the number of these latter contained in the $ k^{th} $ bin and assuming that sampled data are independent, their joint likelihood reduces to the product of $ N $ factors
\begin{linenomath}
	\begin{equation}
	\label{likelihood}
	p(\underline{d}|\underline{\pi},\underline{M})=\biggl(\dfrac{M}{V}\biggl)^N\pi_1^{n_1}\pi_2^{n_2}...\pi_{M-1}^{n_{M-1}}\bigg(1-\sum_{k=1}^{M-1}\pi_k\bigg)^{n_M}
	\end{equation}
\end{linenomath}

\paragraph{Prior and Posterior Probability}
The prior probability of the number of bins represents our knowledge of it a priori. As we do not know anything but the total range $ V $ of the data, it is reasonable to set
\begin{linenomath}
	\begin{equation}
	\label{priori}
	P(\underline{M})=
	\begin{system}
	C^{-1}\ $ if $\ 1\leq M\leq C \\
	\ \ 0\ \ \  \ $otherwise$
	\end{system}
	\end{equation}
\end{linenomath}
where $ C $ is the maximum number of bins we wish to consider. 
Similarly, we take as the prior probability of the bin masses $ \pi_1,...,\pi_{M-1} $ the uniform probability on the simplex defined by the corners of an $(M-1)$-dimensional hypercube with unit side lengths due to the normalisation condition:
\begin{linenomath}
	\begin{equation}
	\label{Jeffreys}
	p(\underline{\pi}|\underline{M})=\dfrac{\Gamma \biggl(\dfrac{M}{2}\biggl)}{\Gamma \biggl(\dfrac{1}{2}\biggl)^M}\biggl[\pi_1\pi_2...\pi_{M-1}\biggl(1-\sum_{k=1}^{M-1}\pi_k\biggl)\biggl]^{-1/2}
	\end{equation}
\end{linenomath}
where $ \Gamma (\cdot) $ is the Gamma function and $0\leq\pi_i\leq1,\ i=1,...,M-1$. Eq. $(\ref{Jeffreys})$ is called Jeffreys's non-informative prior (\cite{Box}) and it expresses complete ignorance about the form of the histogram.\\
To infer the posterior probability density for the number of bins $ M $ we use Bayes' Theorem, which leads to
\begin{linenomath}
	\begin{equation}
	\label{pMpid}
	p(\underline{\pi},\underline{M}|\underline{d})\propto p(\underline{\pi}|\underline{M})p(\underline{M})p(\underline{d}|\underline{\pi},\underline{M}).
	\end{equation}
\end{linenomath}
Inserting the expressions (\ref{likelihood}), (\ref{priori}) and (\ref{Jeffreys}) into the equation (\ref{pMpid}) above, we get the joint posterior probability for the bin parameters $ \underline{\pi} $ and pair $\underline{M}$: 
\begin{linenomath}
	\begin{equation}
	\label{posterior_bin}
	\begin{split}
	p(\underline{\pi},\underline{M}|\underline{d})&\propto \biggl(\dfrac{M}{V}\biggl)^N\Gamma \biggl(\dfrac{M}{2}\biggl)\Gamma \biggl(\dfrac{1}{2}\biggl)^{-M}\cdot\\
	&\cdot\pi_1^{n_1-\frac{1}{2}}\pi_2^{n_2-\frac{1}{2}}...\pi_{M-1}^{n_{M-1}-\frac{1}{2}}\biggl(1-\sum_{k=1}^{M-1}\pi_k\biggl)^{n_M-\frac{1}{2}}.
	\end{split}
	\end{equation}
\end{linenomath}
Notice that we disregarded $ p(\underline{M}) $ since it is constant.\\
By integrating equation \ref{posterior_bin} over all admissible values of $ \pi_1,...,\pi_{M-1} $ we get the posterior probability for the number of bins $ M $:
\begin{linenomath}
	\begin{equation}
	\label{posterior_M}
	p(\underline{M}|\underline{d})\propto\biggl(\dfrac{M}{V}\biggl)^N\dfrac{\Gamma (\frac{M}{2})}{\Gamma (\frac{1}{2})^M}\dfrac{\prod_{k=1}^{M}\Gamma(n_k+\frac{1}{2})}{\Gamma(N+\frac{M}{2})}.
	\end{equation}
\end{linenomath}
According to the method of \textit{Maximum a-posteriori Estimation}, the optimal choice for $ \underline{M} $, the one that provides the best agreement of the model with the observed data, is the one which maximizes the logarithm of the posterior probability.
Thus, the optimal number of bins is
\begin{linenomath}
	\begin{equation}
	\hat{\underline{M}}=\argmax_{\underline{M}} {\log p(\underline{M}|\underline{d})}.
	\end{equation}
\end{linenomath}
A great advantage of this mathematical formalism, is that, once we have computed the optimal binning grid $\underline{M}$, from \eqref{posterior_bin} and \eqref{posterior_M} (see \cite{Knuth} for details) we can analytically compute the mean and the variance of the bin probabilities which are given, respectively, by
\begin{linenomath}
	\begin{equation}
	\mu_k=\biggr(\frac{M}{V}\biggr)\cdot\biggr(\frac{n_k+1/2}{N+M/2}\biggr)
	\end{equation}
\end{linenomath}
and
\begin{linenomath}
	\begin{equation}
	\sigma^2_k=\biggr(\frac{M}{V}\biggr)^2\cdot\biggr(\frac{(n_k+1/2)(N-n_k+(M-1)/2)}{(N+M/2+1)(N+M/2)^2}\biggr)
	\end{equation}
\end{linenomath}
and which allow to construct the optimal histogram with the proper error bars.

\section{Knuth method for estimation of a Point Process's intensity}	
\label{sec2}
\begin{figure*}[p]
	\centering
	\includegraphics[width=36mm]{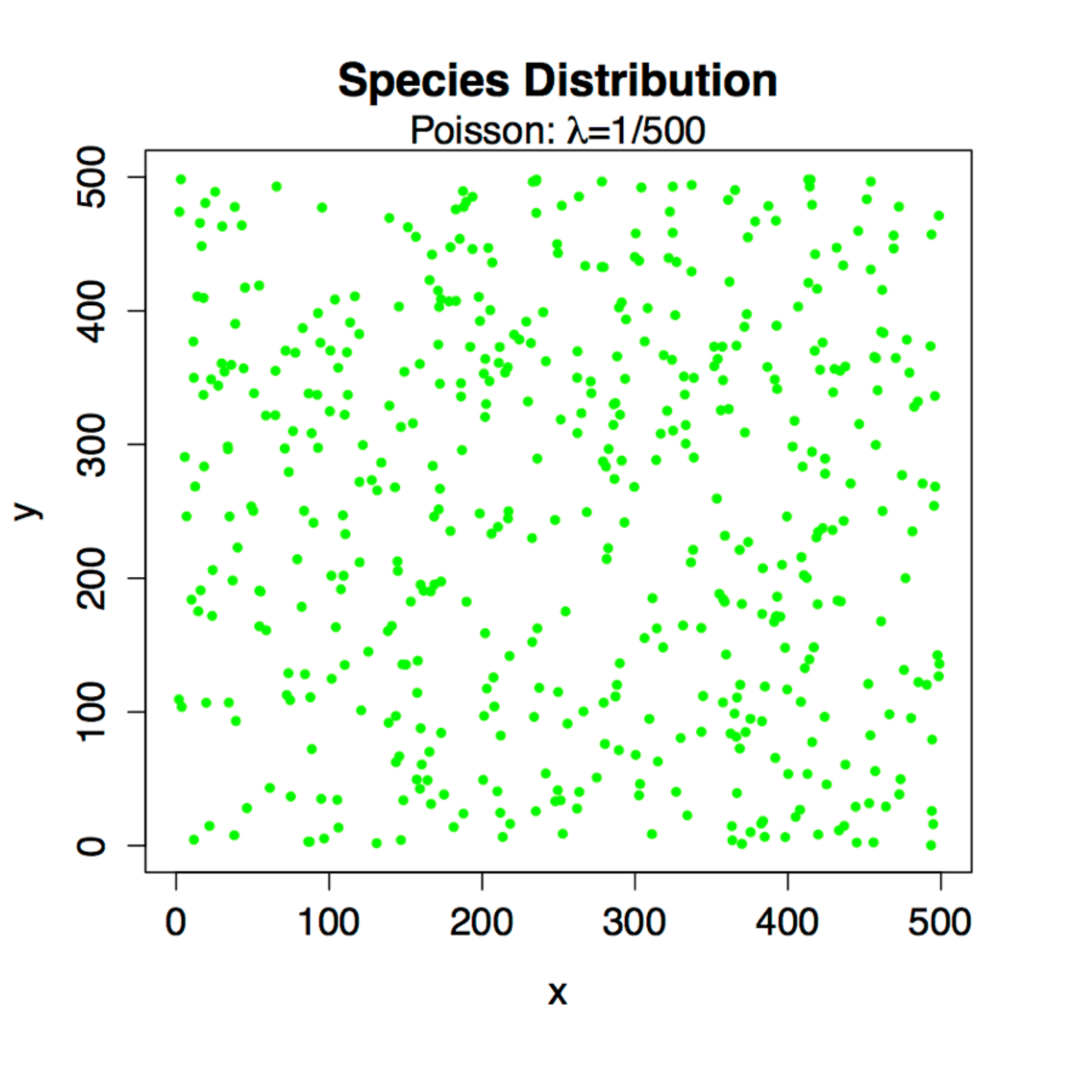}
	\subfloat[{}\label{fig:Poisson_density_est}]
	{\includegraphics[width=42mm]{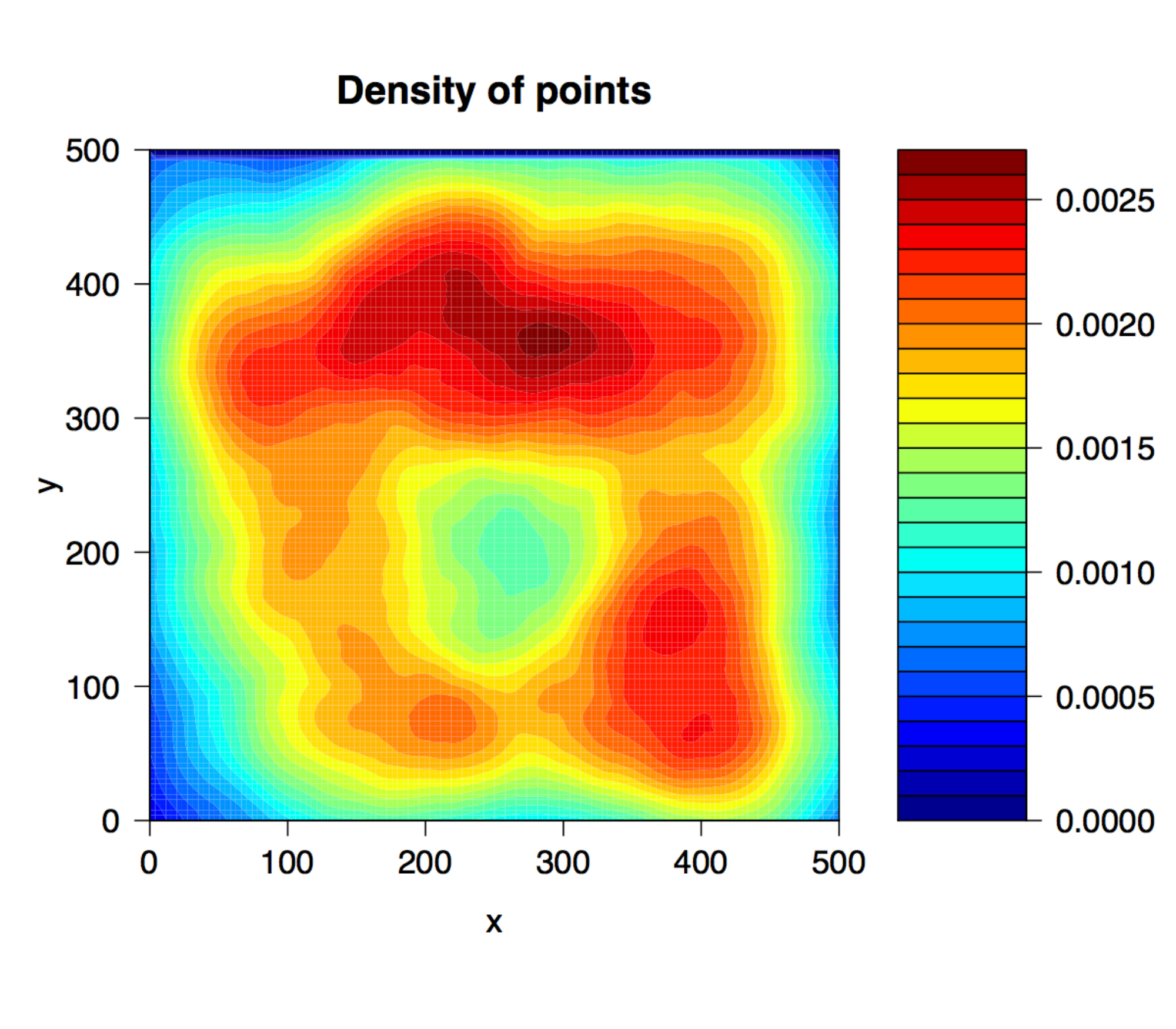}}
	\includegraphics[width=42mm]{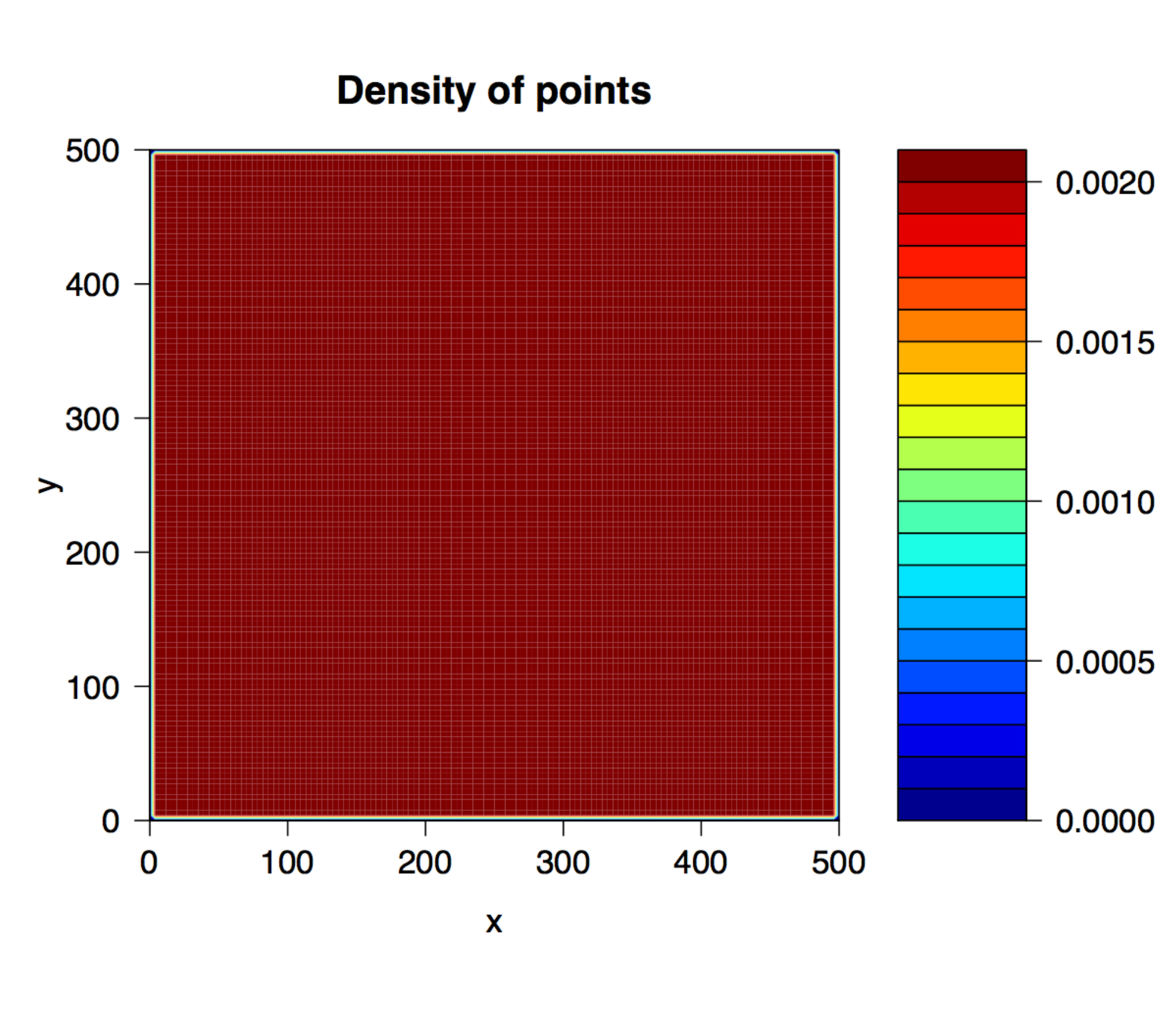}\\
	\vspace*{-0.3cm}
	\includegraphics[width=36mm]{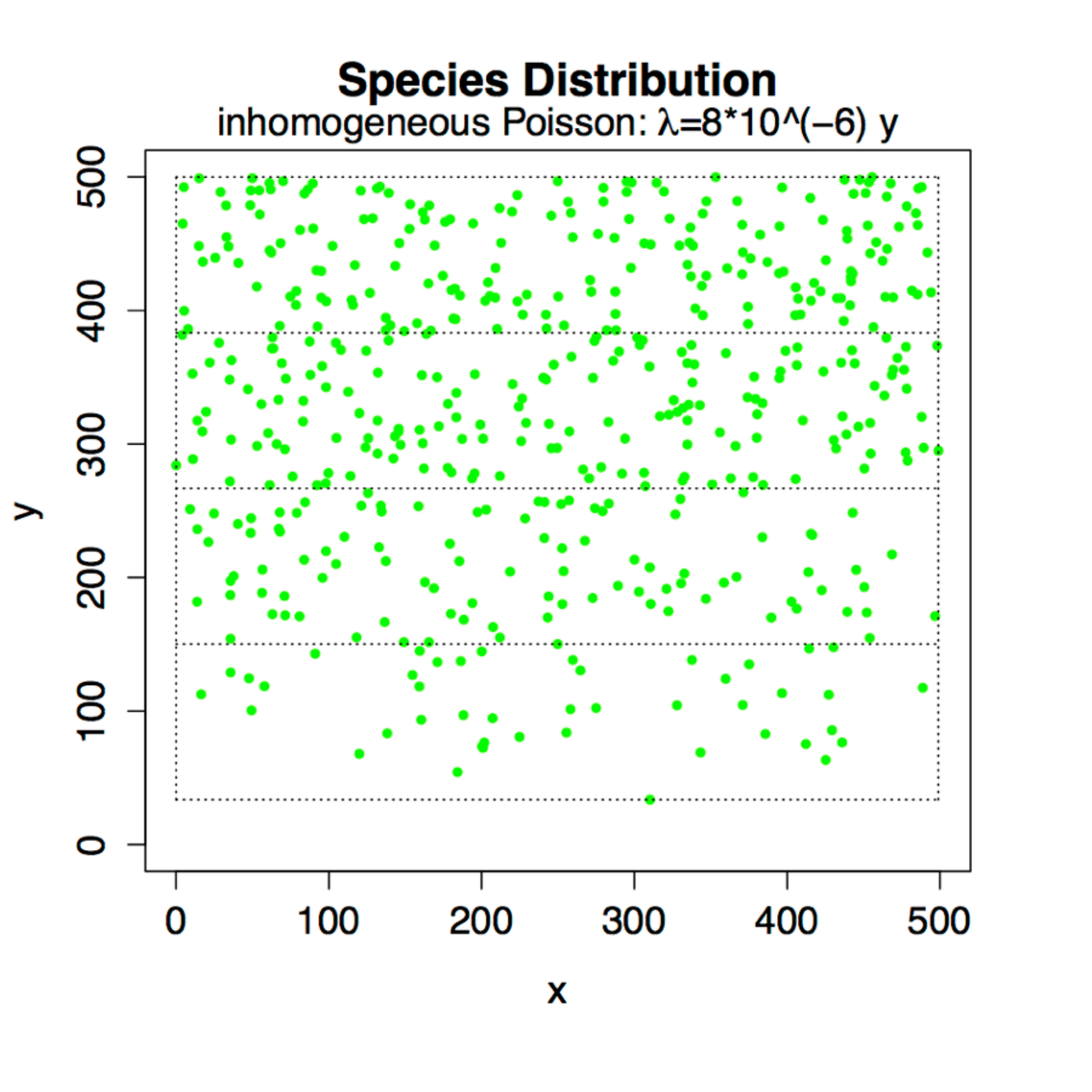}
	\subfloat[{}\label{fig:inPoisy_density_est}]
	{\includegraphics[width=42mm]{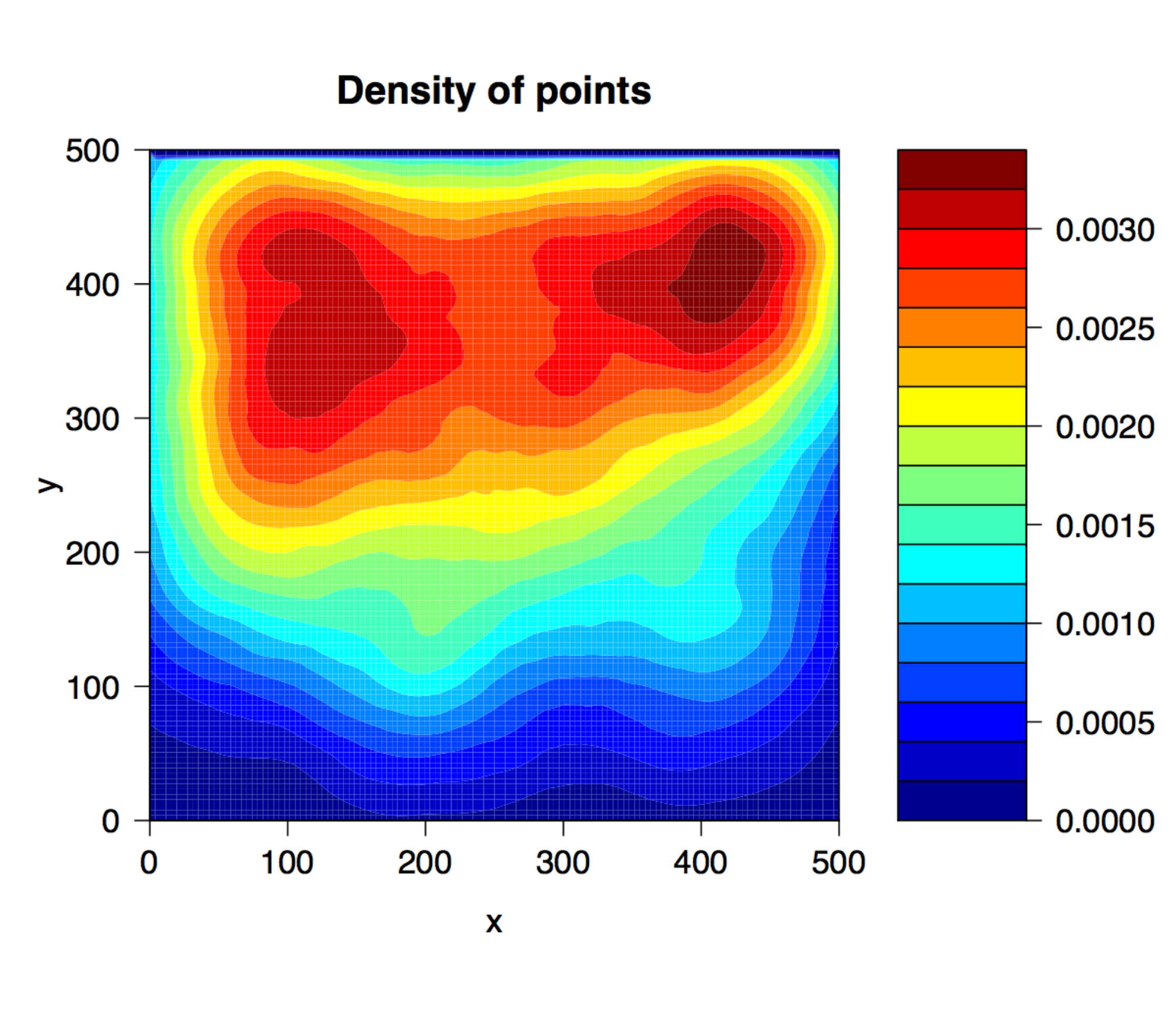}}
	\includegraphics[width=42mm]{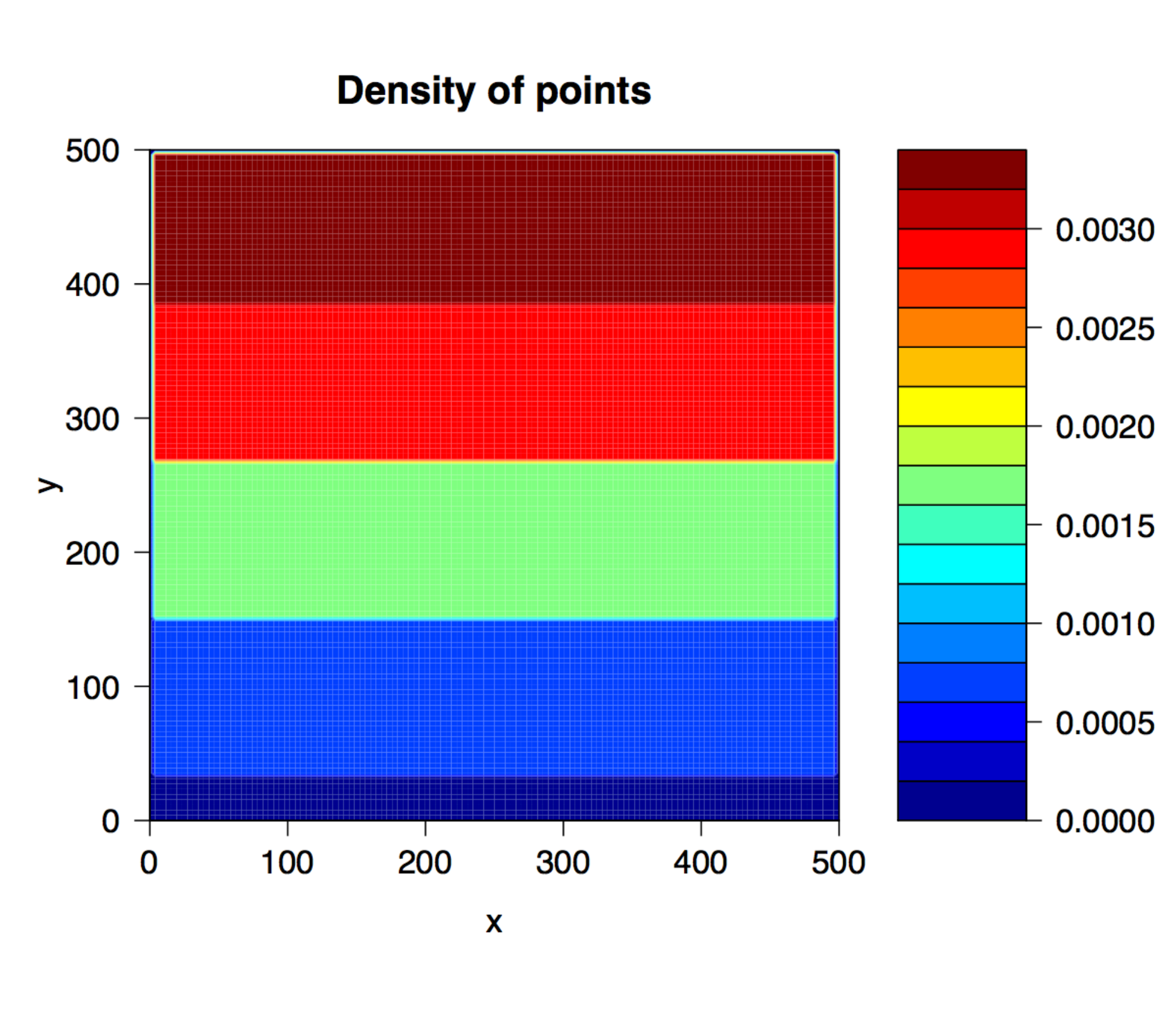}
	\vspace*{0.3cm}
	\caption{\small On the top: estimation of the intensity function of a Poisson point process. Epanechnikov kernel with bandwidth $R=4.5/\sqrt{\lambda_W}$ (see \cite{Chapmann}) results to be more sensitive to sampling fluctuations, while Knuth method arrange data in a unique bin, perfectly detecting the homogeneity of the density function. On the bottom: estimation of the intensity function of an inhomogeneous Poisson point process. Epanechnikov kernel with same bandwidth as above results again sensitive to sampling fluctuations, while Knuth method detects the homogeneity of the density function on the $x$-axis and the gradient along the $y$-axis.
	}\label{fig:kernel_Knuth}
\end{figure*}

Given a point process $\phi$ defined on the plane, its \textit{intensity function} $\lambda$ or \textit{density} is a first-order statistic measuring the mean number of points per unit area. Denoting with $\mathcal{N}_\phi(W)$ the number of points falling within a region $W\in\mathbb{R}^2$, we have
\begin{linenomath}
	\begin{equation}
	\mathcal{N}_\phi(W)=\int_W \lambda (x) dx.
	\end{equation}
\end{linenomath}
We are interested in estimating, from a sample of the process, its density $\lambda$. Notice that if the process is assumed to be homogeneous, then its density is constant within the observation window $W$ and it is therefore approximated by $\lambda_W=\mathcal{N}_{\phi}(W)/A_W$, where $A_W$ is the area of the window.\\
Using the Knuth method, we can approximate the intensity of the possibly non homogeneous processes by a piece-wise constant function: the density in a point $x$ is the height of the column over the bin containing $x$ in the optimal histogram. We compare Knuth answer with kernel methods which are widely used in literature (\cite{Illian,Shiffers,Chapmann}) (see Supporting Information for a comparison with Stone's non kernel method).\\
The idea of kernel methods is to estimate $\lambda(x),\ x\in\R^2$ by looking at the number of points falling within the small disk $\mathcal{B}(x,R)$ of radius $R$ (called bandwidth) centred in $x$ and by dividing it by the area of the disk: 
\begin{linenomath}
	\begin{equation}
	\hat{\lambda}(x)=\dfrac{1}{\pi R^2}\sum_{x_i\in\phi}k_R(|x-x_i|)
	\end{equation}
\end{linenomath}
where $x_i$ is a point of the process $\phi$ and $k_R(|x-x_i|)$ is the kernel function. A popular choice for $k_R$ is the \textit{Epanechnikov kernel}, defined as:
\begin{linenomath}
	\begin{equation}
	k_R^E(|x-x_i|)=
	\begin{system}
	2\Big(1-\frac{|x-x_i|^2}{R^2}\Big)\qquad $if$ \ |x-x_i|\leq R\\
	0\qquad\qquad\qquad\qquad $otherwise$ 
	\end{system}
	\end{equation} 
\end{linenomath}
This function weights the points $x_i\in\phi$ within the disk $\mathcal{B}(x, R)$ according to their distance from the centre $x$: the smaller $|x-x_i|$, the bigger the weight.\\
A weak point of kernel methods is that a general recipe for the choice of the bandwidth $R$, which gives the right smoothing of a rugged intensity function, does not exists (\cite{Illian} p.~115, \cite{Diggle} p.~116). Some authors suggest the rough estimate $R\sim  1/\sqrt{\lambda}$ and a subsequent finer tuning of $R$ using visual inspection (\cite{Chapmann}, p.~97). On the contrary, the Knuth non parametric method select the optimal scale from the data without any assumption on the underlying process that generated the data. Notice that also simpler methods for assessing the CRS hypothesis like the quadrats count can be considered as parametric methods since their effectiveness depends on the choice of the size of the quadrats. \\
Below we compare kernel and the Knuth methods on datasets generated from both homogeneous and inhomogeneous processes to test their reliability at 1) reproducing the intensity of the process and 2) capturing the eventual presence of a gradient in the density functions.
\paragraph{Detection of CSR and inhomogeneity}
We start applying the Knuth method and the Epanechnikov kernel estimation to a generated CSR pattern within a window $W$ of area $500\times500$ units. On the left of \figurename~\ref{fig:kernel_Knuth}\subref{fig:Poisson_density_est} we plot the CSR (Poisson) pattern generated with $\lambda=1/500$, while in the two graphs on the right we can see the intensity function estimated by the Epanechnikov kernel and by the Knuth method. This latter arranges the data in a unique bin, so that the intensity function it returns is constant within the plot. Therefore it perfectly detects the underlying homogeneous structure of the process. By contrast, the kernel method results to be sensitive to sampling fluctuations. Robustness of the Knuth method in detecting CSR processes is tested in Supporting Information. 
We then compare the two methods on an inhomogeneous Poisson point process where the true density function $\lambda$ increases with the $y$ coordinates (see \figurename~\ref{fig:kernel_Knuth}\subref{fig:inPoisy_density_est}). Although from the Epanechnikov kernel method it is evident that the number of points falling within the upper region of the window is bigger with respect to the lower region, it results to be still affected by sample fluctuations. By contrast, the Knuth method arranges the data into a $1\times4$ grid, perfectly detecting the homogeneity of the process along the $x$-axis and the density gradient along the $y$-axis. 

\section{Detection of clusterised, dispersed and inhomogeneous patterns }

\subsection{Knuth method description of cluster features}
\label{sec:Knuth_cluster}
\begin{figure*}[p]
	\centering
	\subfloat[{}\label{fig:fit_square}]
	{\includegraphics[scale=0.40]{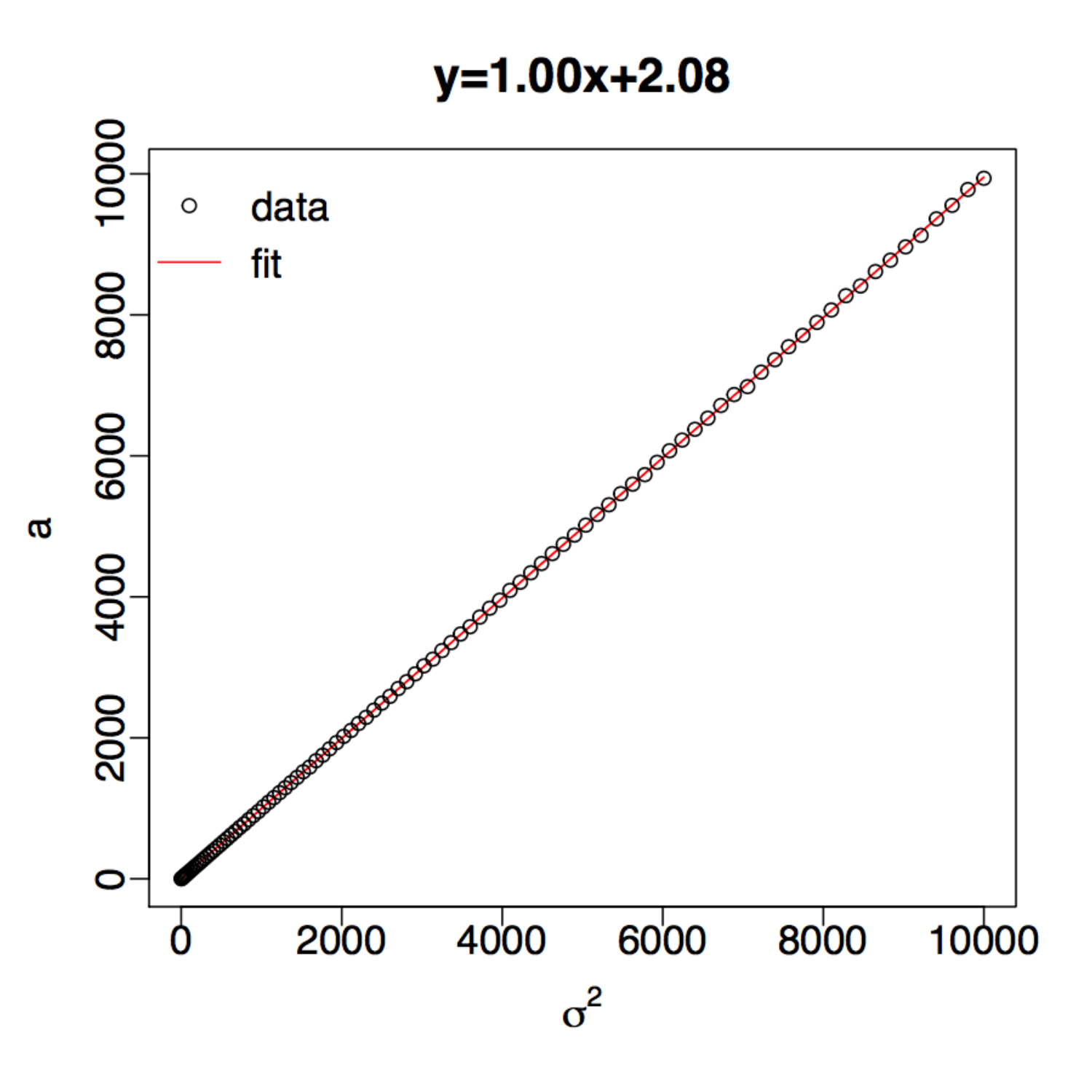}}
	\subfloat[{}\label{fig:fit_Matern}]
	{\includegraphics[scale=0.40]{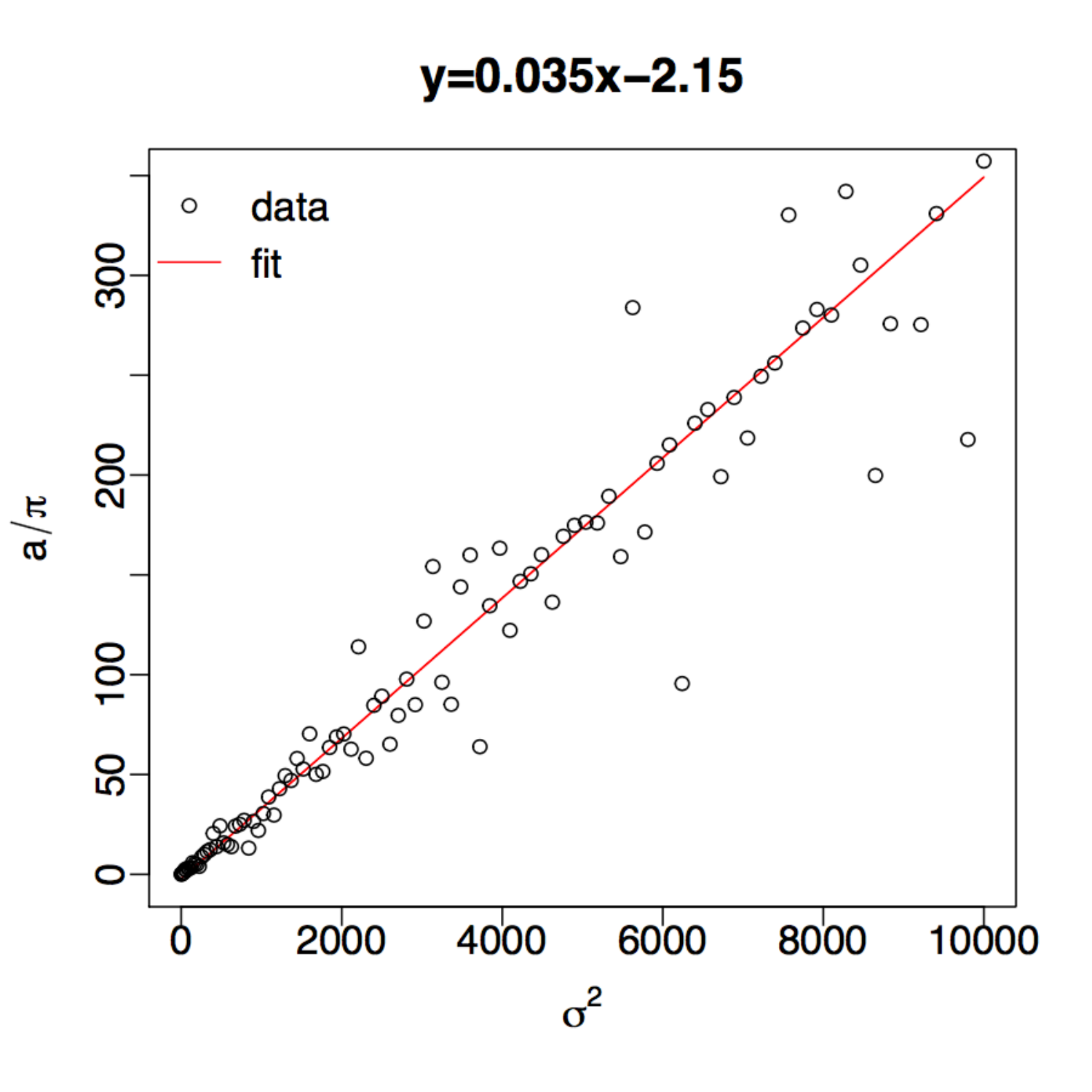}}\\
	\vspace*{-0.5cm}
	\subfloat[{}\label{fig:fit_Gaussian}]
	{\includegraphics[scale=0.40]{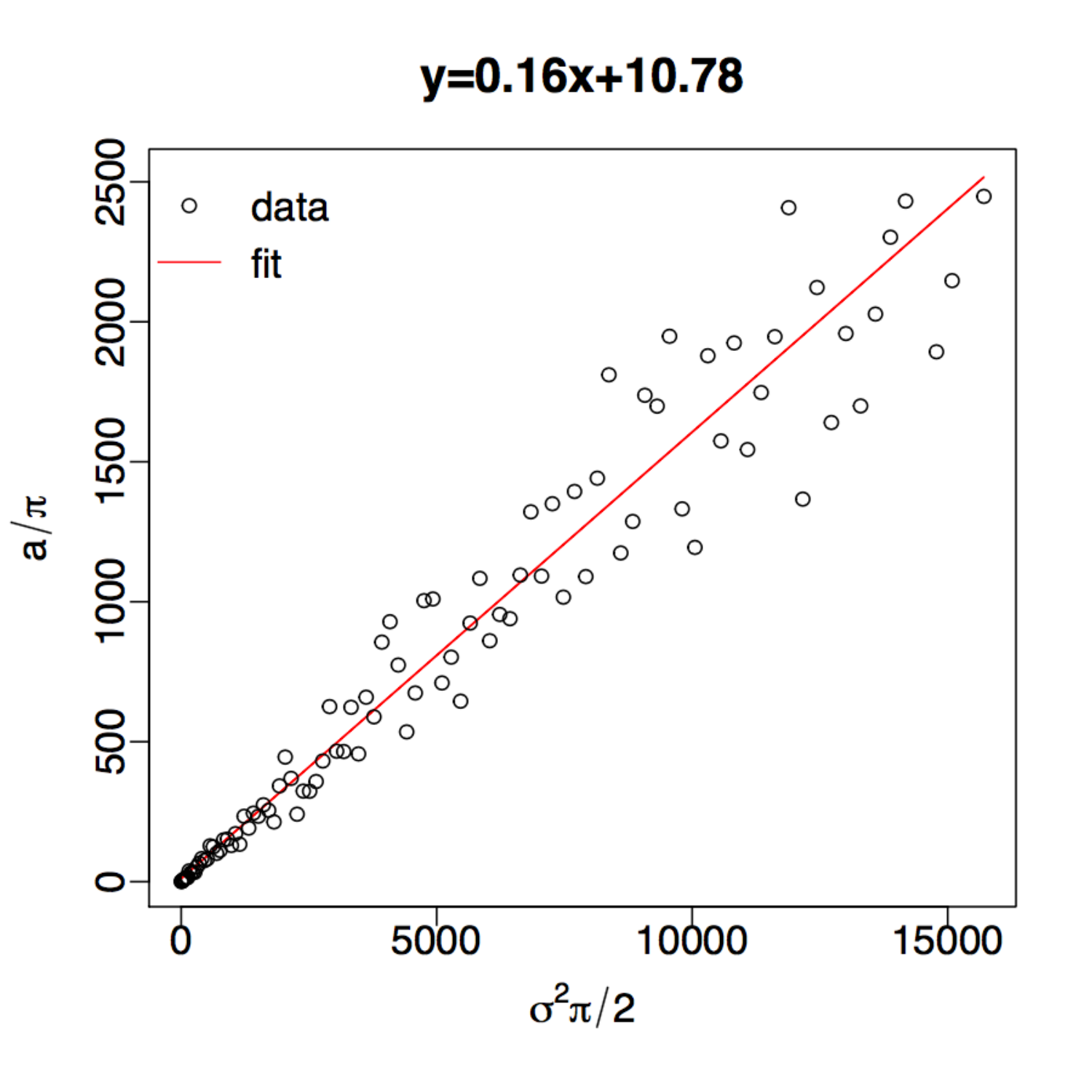}}
	\vspace*{0.3cm}
	\caption{Analysis with $ R $ software of the linear relation between Knuth optimal bin area $a$ and the characteristic size $\sigma$ of three type of clusters: (a) square and (b) circular with uniform density and (c) circular with Gaussian density. Each dataset consisted of 1000 points and for each case $\sigma$ varies from 1 to 100. In the first case we plot $a$ against the square of the cluster side $\sigma^2$, getting a determination coefficient of $R^2=1.00$. In the second case we plot $a/\pi$ against the squared radius of the cluster $\sigma^2$ and we get $R^2=0.90$ and in the last case, plotting $a/\pi$ against the square of the mean distance of a point from the cluster centre $\sigma^2\cdot\pi/2$, we get $R^2=0.94$. The correlation between the two variables was very good in all cases.}\label{fig:Knuth_sigma}
\end{figure*}

As a preliminary analysis we investigated how the Knuth method reproduces the features of three different type of clusters: square, circular with constant density and circular with Gaussian density.  We consider plots of area $A_0=1000\times500$ units  as is the BCI. 
For each type of cluster, we have generated 100 datasets setting the number of individual equal to $ N=1000 $ and arranging them in a \emph{unique} cluster positioned in the centre of the window. The characteristic size $\sigma$ of the cluster (respectively the side, the radius and the standard deviation) varied from 1 to 100. For each dataset, we have computed Knuth optimal binning area $ a $ and the correlation between the characteristic size of the cluster $\sigma$ ($x$-axis) and its optimal bin size representation $a$ ($y$-axis). 
Results are displayed in \figurename~\ref{fig:Knuth_sigma}.\\
In all cases, the determination coefficient shows a strong correlation between the cluster size and the optimal bin area $(R^2>0.9)$. In the first case (square clusters) the slope equals 1, meaning that the square cluster is arranged in a unique bin and that the density of points is correctly detected as constant. In the other two cases, the slope is far from 1, meaning that the cluster is described using a higher number of bins. In particular, in the second case, the Knuth method aims at reproducing the circular boundary of the cluster while in the last case, the optimal bin serves at reproducing both the circular boundary of the cluster and the Gaussian shape of the density (see Supporting Information for the correspondent histograms and density plots). We also tested the Knuth method on uniform and Mat\'ern clusters (see \figurename~\ref{fig:Knuth_sigma} and Supporting Information).

\subsection{Second-order statistics for estimation of cluster size and hard core radius} 

In spatial ecology, the CSR hypothesis mostly comes to fail due to several reasons: on one side, changes in environmental conditions, such as in physical features of the landscape or in chemical composition of the soil, may lead to inhomogeneous patterns. On the other side, different seed dispersal mechanisms may favour the formation of clumped structures as well as dispersed one. These pattern characteristics are revealed by second-order statistics, which take into account the correlations between pair of points due to possible interactions. Below we briefly recall the most common second order statistics and their use.\\
The probably most used second-order statistics for homogeneous pattern is the Ripley's \emph{K}-function $K(r)$ (\cite{Wiegand,Illian,Chapmann}), which is the expected number of points falling within a distance $r$ from a point of the process, divided by the intensity function $\lambda.$ Since under CSR hypothesis $K(r)=\pi\cdot r^2$ scales quadratically with the distance, it is usually substituted with the \emph{L}-function $L(r)=\sqrt{K(r)/\pi}-r$, which takes constant zero value for the CSR model. Both Ripley's \emph{K} and \emph{L} functions are regarded in literature as cumulative statistics (\cite{Chapmann}), which do not permit to properly infer pattern characteristics at any specific scale.\\
This limitation can be avoided by taking the pair correlation function $g(r)=K'(r)/2\pi r$ (\cite{Adorisio,Azaele}). It is defined as the ratio between the density of points falling within a small ring of radius $r$ centred at a point of the pattern and the constant density function $\lambda$ of the process assumed to be homogeneous.\\
Since values of $g(r)$ greater than 1 indicate clustering, whereas values less than 1 indicate dispersion, the point where the pair correlation function intercepts the line $y=1$ gives a rough estimate of the average diameter of clusters for clumped patterns or the average hard core radius for overdispersed ones.\\
As pointed out in (\cite{Shiffers,Chapmann}), both Ripley's \emph{K} and pair correlation function are \emph{window-dependent}, in the sense that their ability in detecting the scale of clustering or dispersion depends on the window within which we arrange the data. For example, if the plot contains empty spaces, these statistics put in evidence a clustered structure which is not due to the pattern's generating process. This phenomenon is called \emph{virtual aggregation} and it affects how these statistics perform in inhomogeneous pattern. In this latter case, the apparent clumping nature of the pattern is due to the fact that both $K$ and $g$ are normalised with the empirical intensity $\lambda$, which is wrongly assumed as constant. Shiffers' \emph{K2-index} (\cite{Shiffers}), using the derivative of the pair correlation function, is less sensitive to this phenomenon. 

A main concern of this paper is to test the ability of the Knuth method to deal with clustered, dispersed and inhomogeneous patterns and to cope with the virtual aggregation phenomena which may arise when considering non homogeneous patterns. We have thus selected a model of clustered process, the modified Thomas process and a dispersed one with an hard core repulsion radius. They are briefly described below.\\
The Poisson Cluster Process or, more often, its simplest variant the modified Thomas process (\emph{mTp}) (\cite{He,Condit,Plotkin,Morlon,Azaele2}) (see Supporting Information for the formal definition of the process) is one of the most used model in literature to describe the clumping mechanism of plants' species. 
This process, in addition to being mathematically tractable (\cite{Illian,Diggle}) is more efficient than others to capture important biological curves such as the \textit{Species-area curve} (see, e.~g. \cite{Plotkin}) or to model species occupancies at different spatial scales (\cite{Azaele2}). Instead, as shown in \cite{Morlon}, it inadequately reproduces the \textit{distance}$-$\textit{decay relationship}, indicating that some of the assumptions of \emph{mTp} do not hold in nature.\\
The hard core process is generated from a uniform distribution with the additional constraint that if a point comes to fall within a fixed hard core distance from a pre-existing one, it is rejected. This model is of importance in ecology to describe reproductive mechanism where the seeds are shot apart from parents to avoid species-specific predators (\textit{Janzen-Connell effect}, see \cite{Janzen,Connell,Adorisio})

\subsection{Interplay between second-order statistics and the Knuth method}
\label{vafen}
Despite their ability to detect significant departures from CSR processes, the pair correlation function and the \emph{K2}-index lack in reliability at determining the cluster size when dealing with inhomogeneous patterns being subject to the virtual aggregation phenomenon.\\
To show this, we computed the pair correlation function and the \emph{K2}-index of Shiffers for three generated patterns, respectively: a) a modified Thomas process, b) an overdispersed pattern with fixed hard core radius and c) an inhomogeneous Poisson process. 

\begin{sidewaysfigure}
	\centering
	\vspace*{-0.8cm}
	\includegraphics[width=43mm]{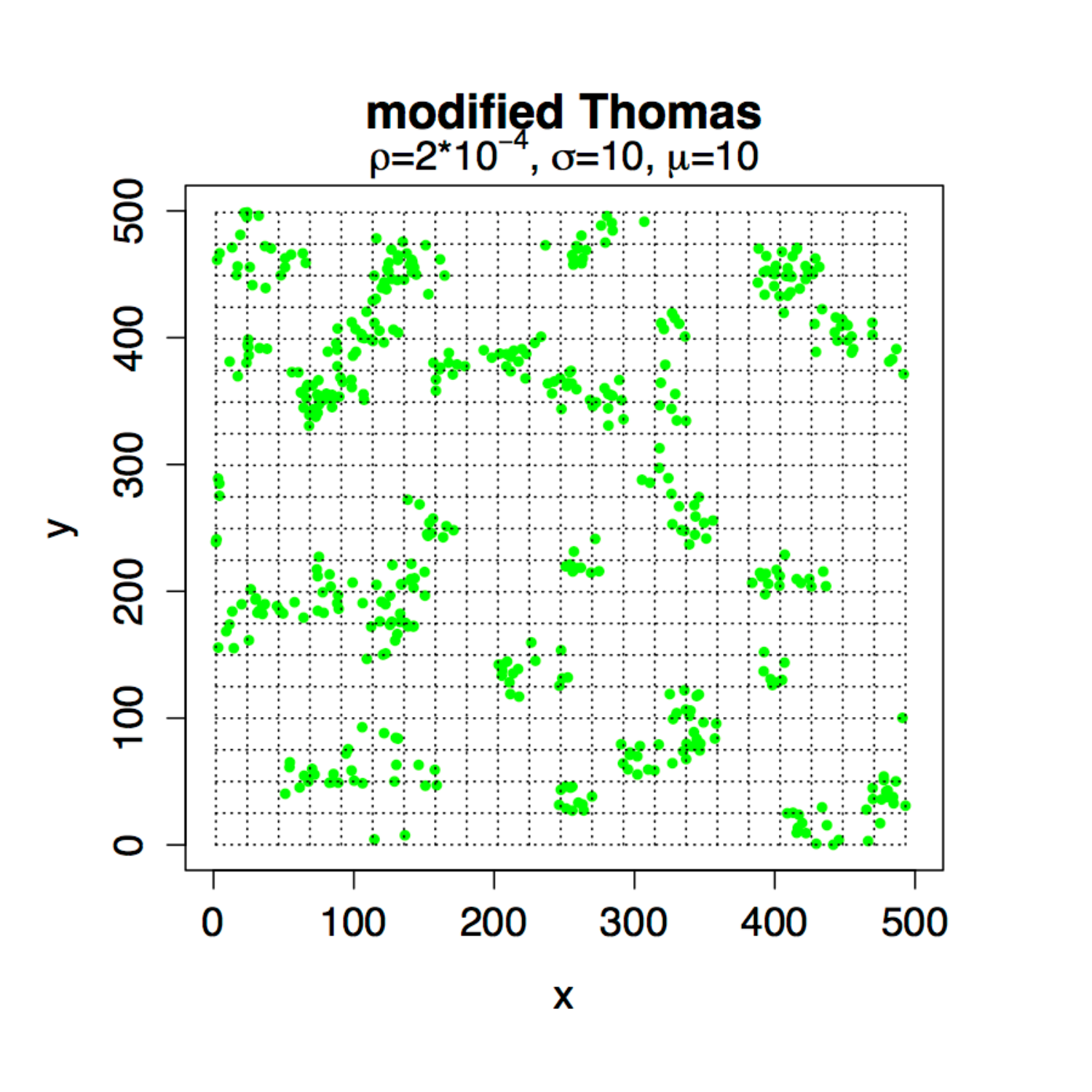}\hspace{0.5cm}
	\includegraphics[width=43mm]{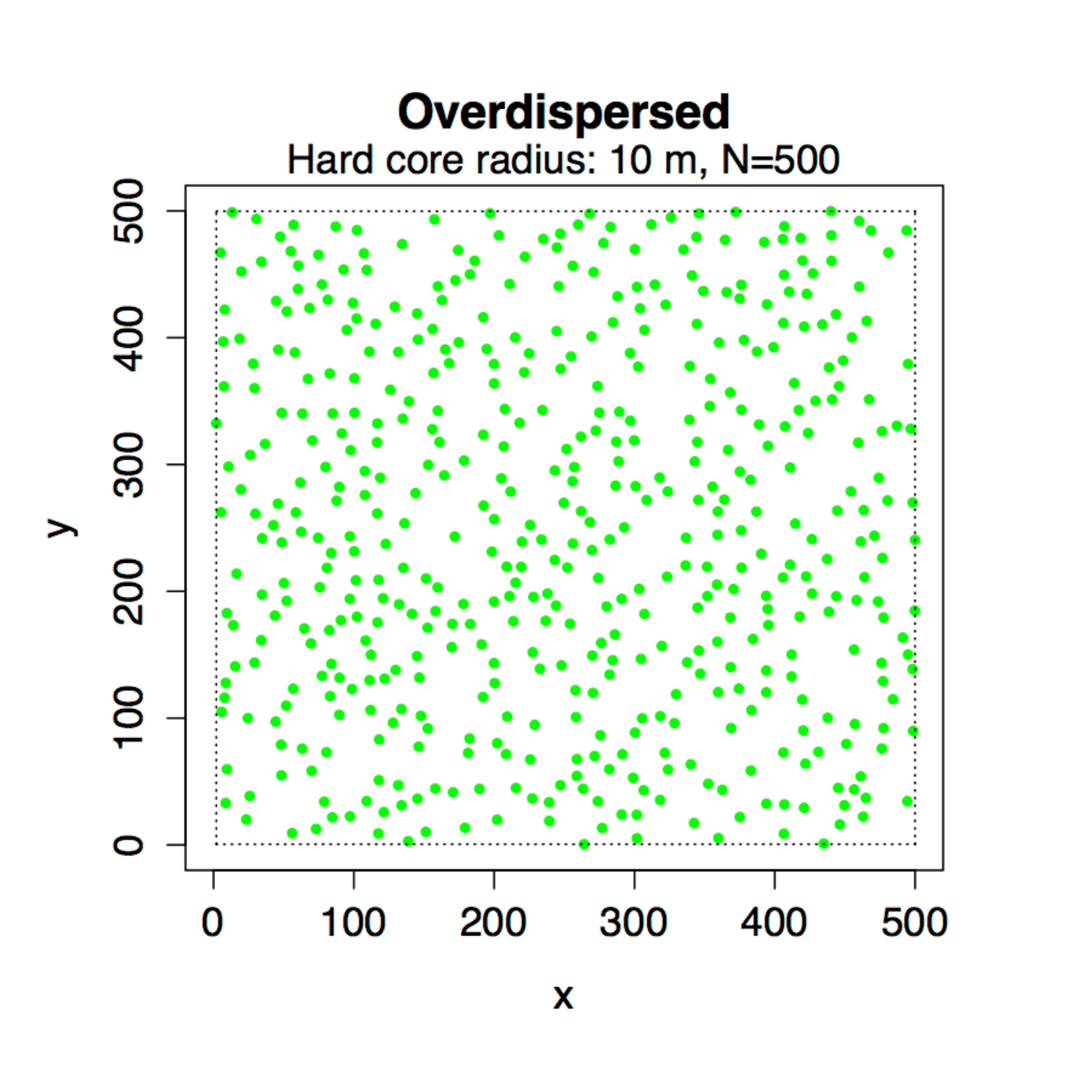}\hspace{0.5cm}
	\includegraphics[width=43mm]{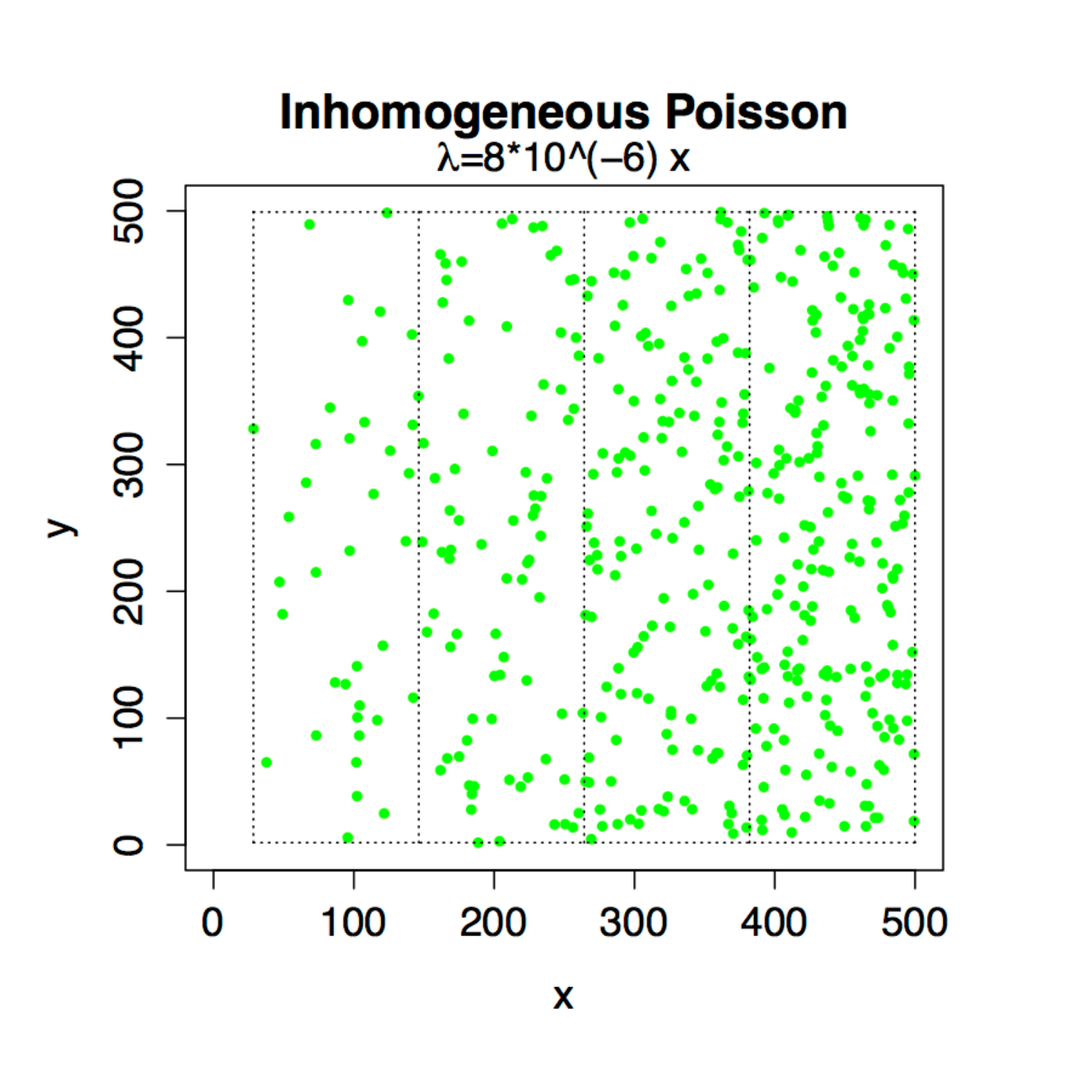}\\
	\vspace*{-0.3cm}
	\includegraphics[width=55mm]{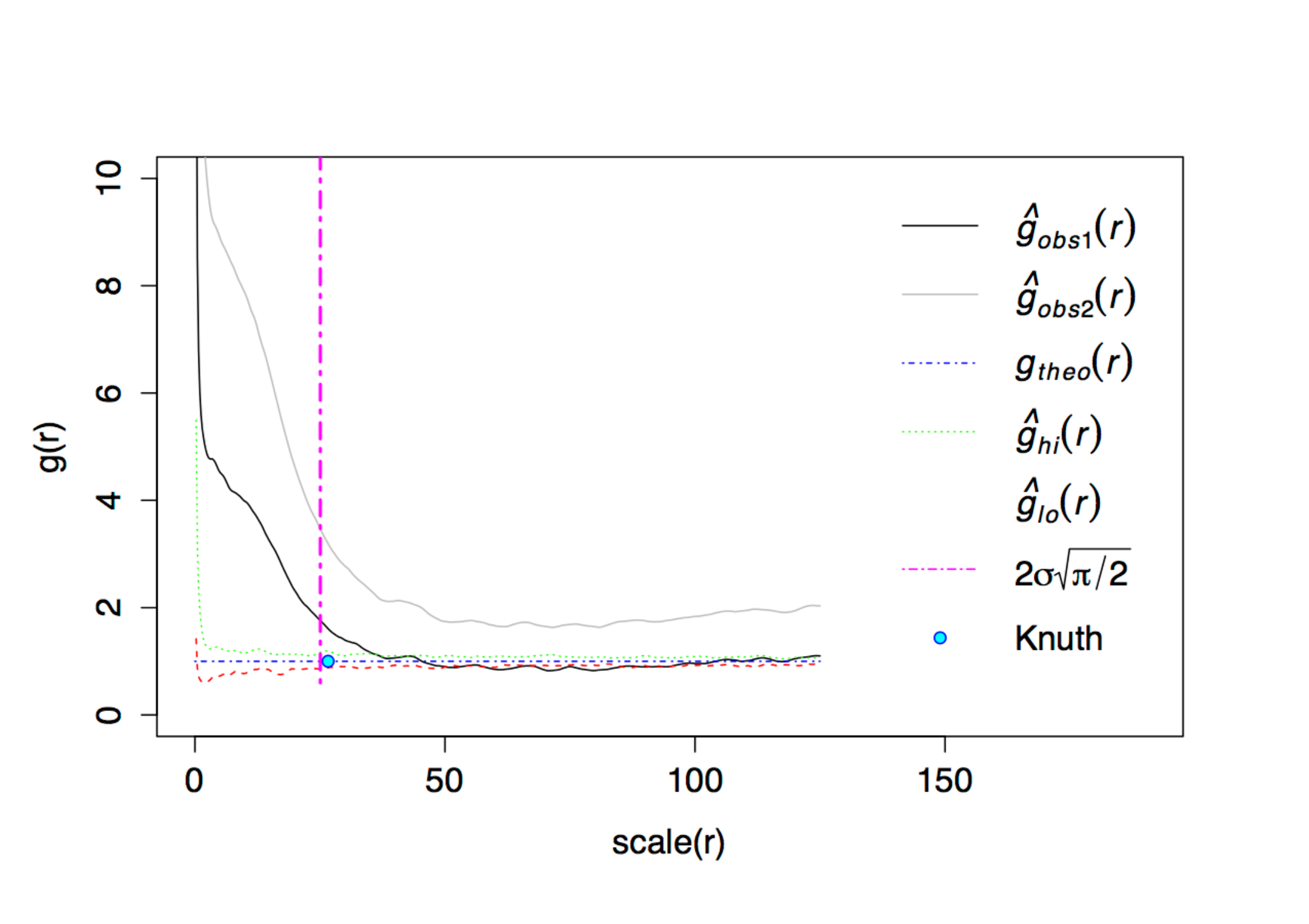}\hspace{-0.3cm}
	\includegraphics[width=55mm]{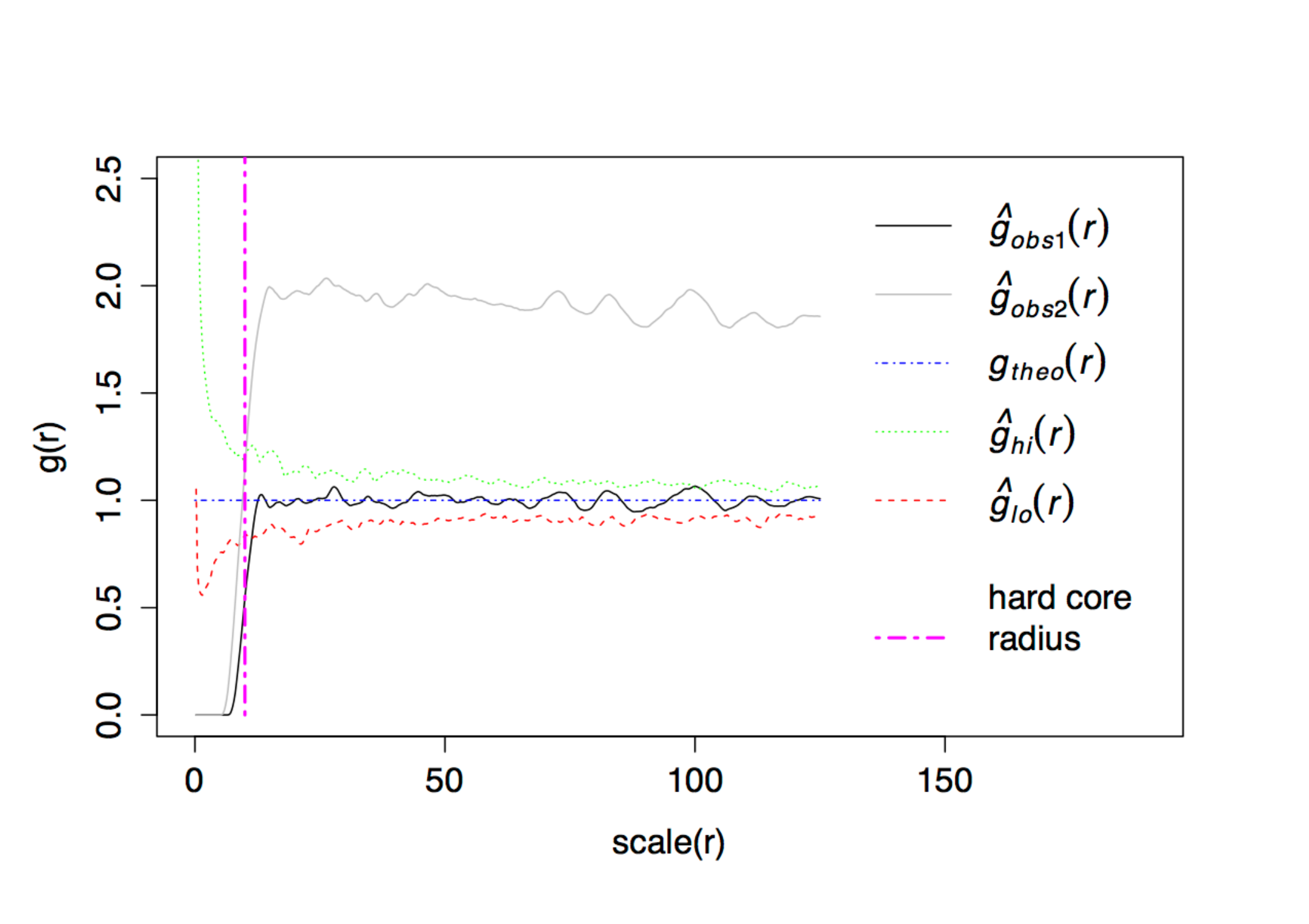}\hspace{-0.3cm}
	\includegraphics[width=55mm]{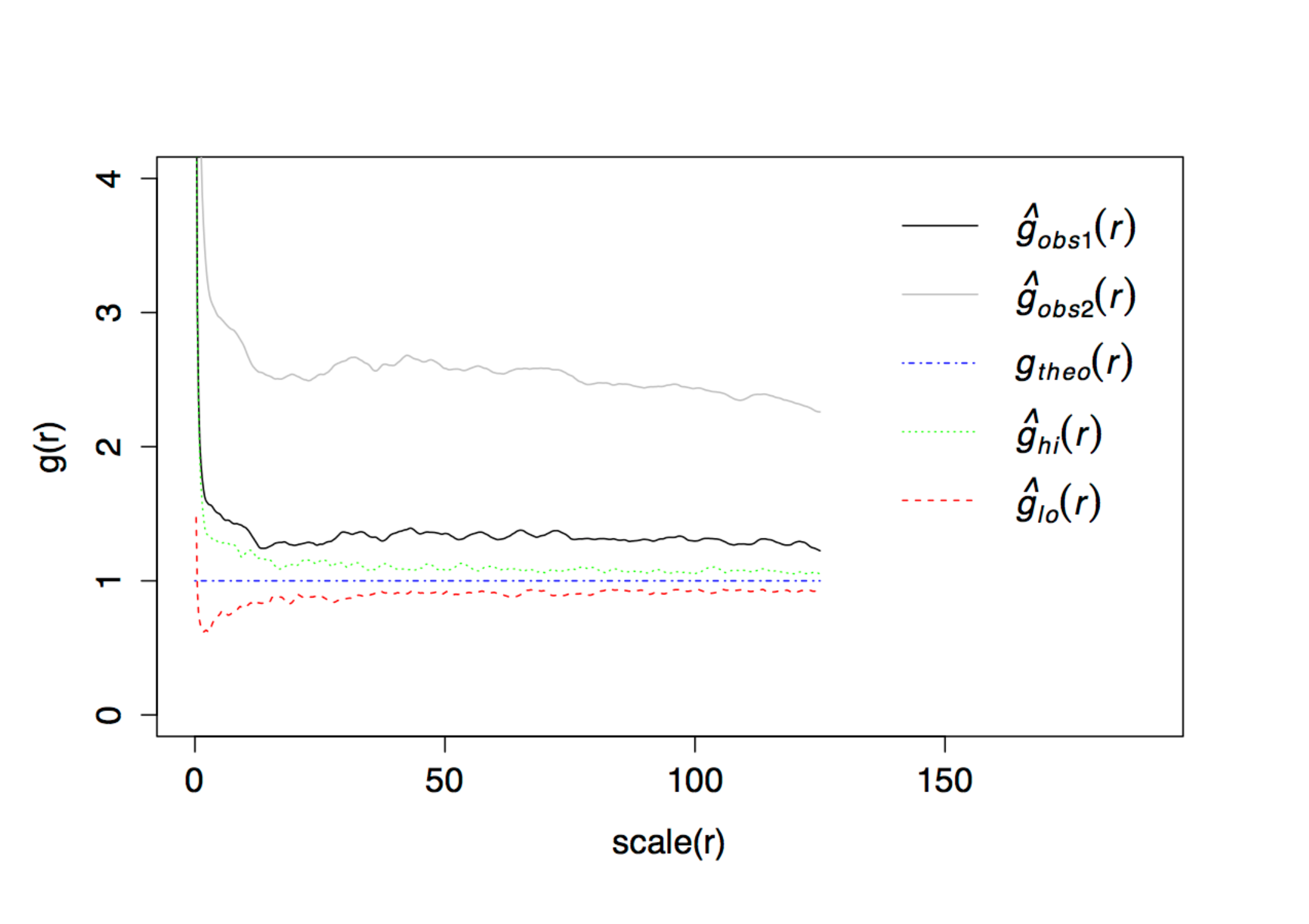}
	\\
	\vspace*{-0.9cm}
	\subfloat[{}\label{fig:Thomas_50_10_10}]
	{\includegraphics[width=55mm]{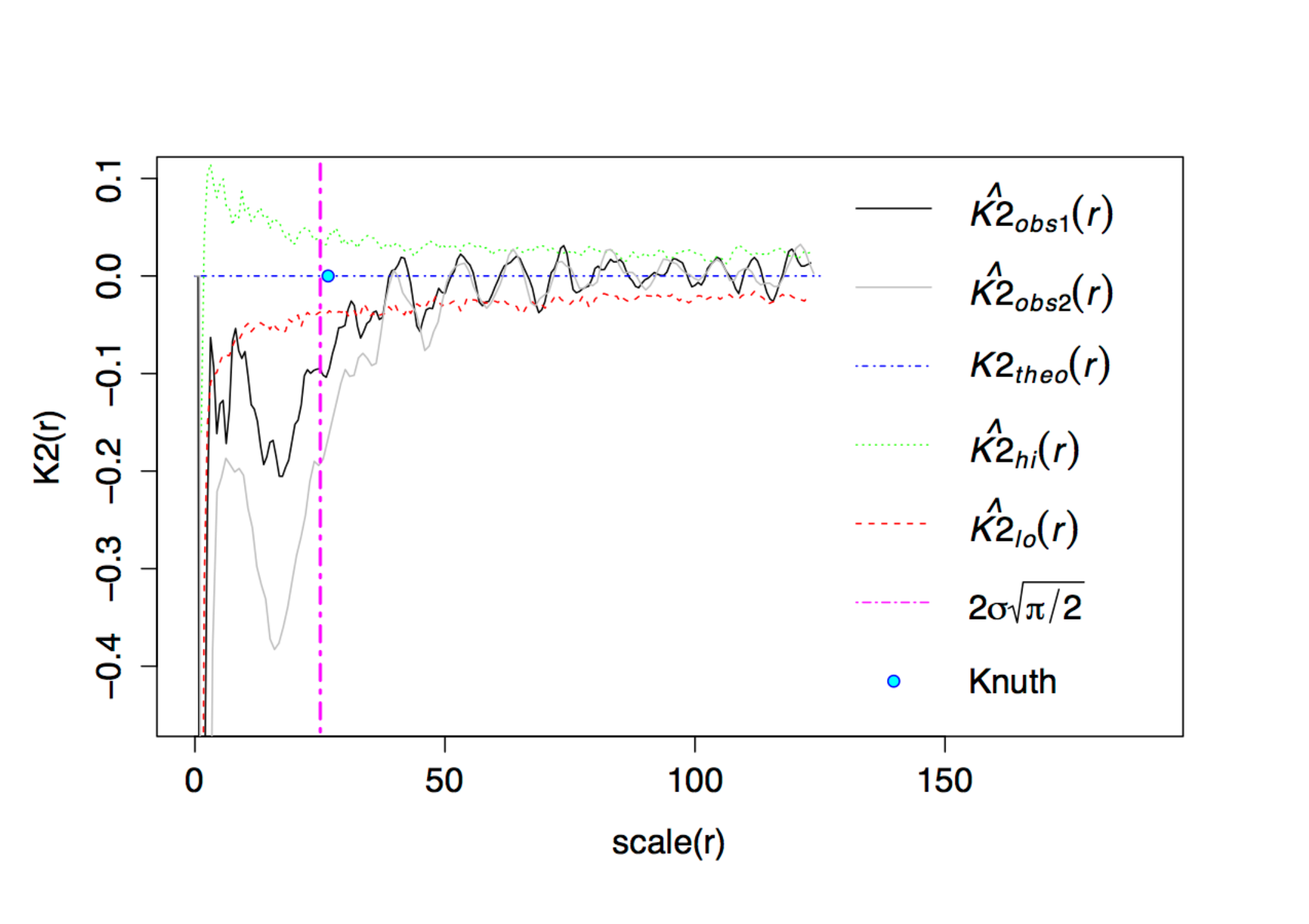}}\hspace{-0.3cm}
	\subfloat[{}\label{fig:Disp10}]
	{\includegraphics[width=55mm]{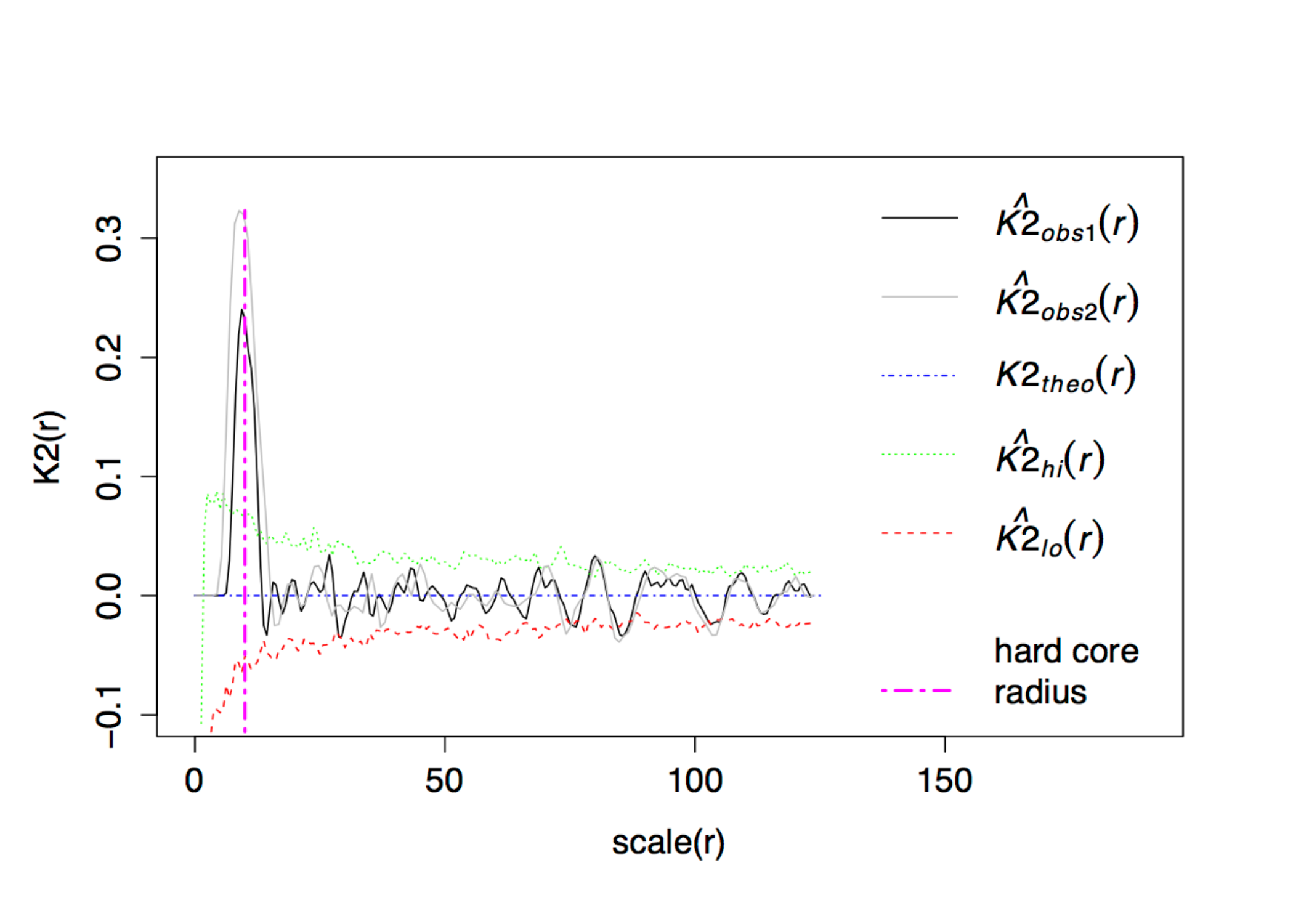}}\hspace{-0.3cm}
	\subfloat[{}\label{fig:inPoisx}]
	{\includegraphics[width=55mm]{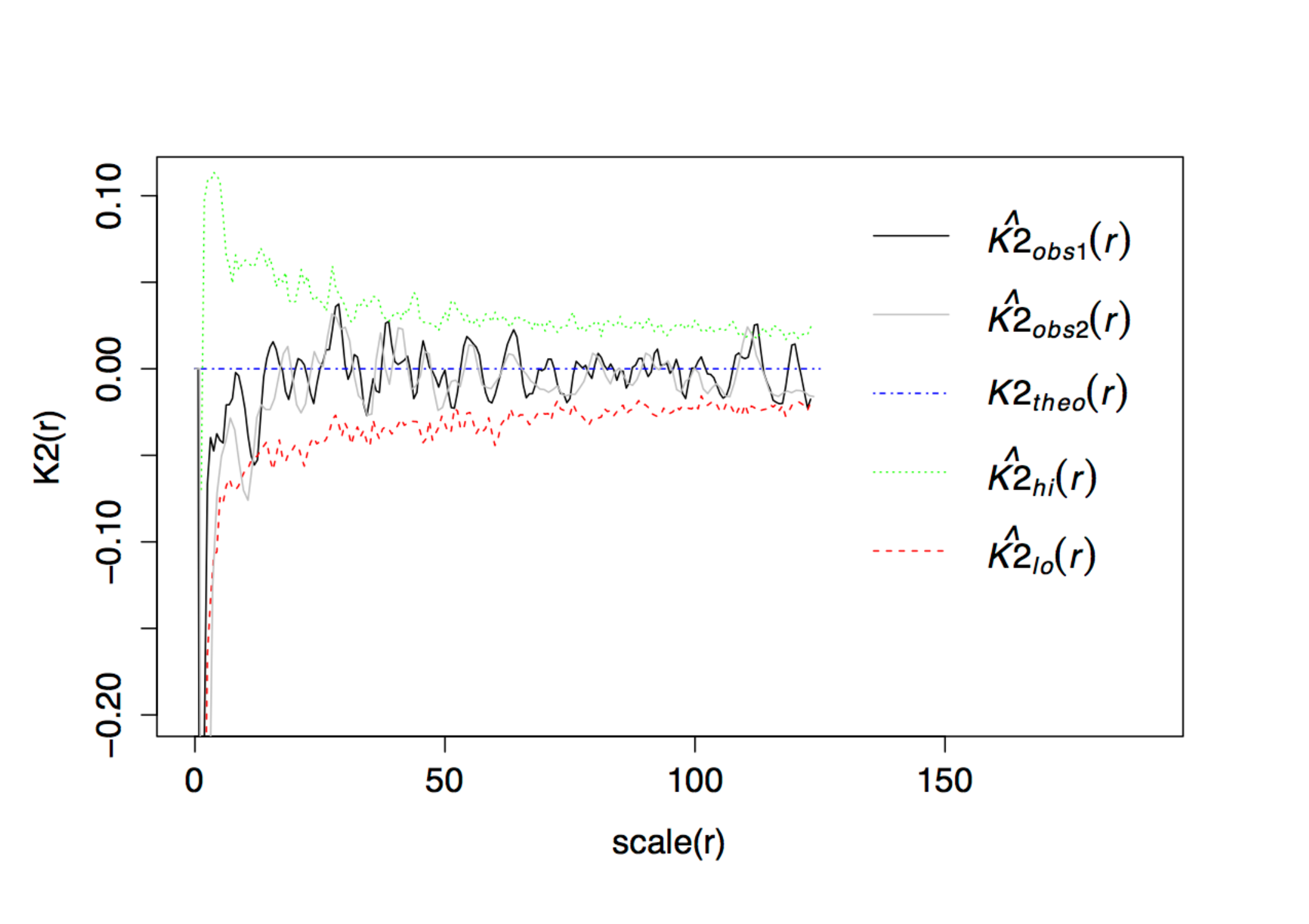}}
	\vspace*{0.3cm}
	\caption{\small Second-order statistics for three generated patterns. From top to bottom: point patterns with superimposed Knuth grid, pair correlation function and Shiffers' \emph{K2}-index. Significance departures from the CSR model are tested by Monte Carlo simulations (blue lines are theoretical CSR curves, green and red are the higher and lower band of the 99$\%$ confidence envelopes). Black lines in the graphs of $g$ and $K2$ refer to the original datasets, plotted within a $500\times500$ units window. Grey lines are obtained by considering the same data points but within an enlarged window, created by adding an empty $500\times500$ units square at the bottom of the previous window. For the $mTp$ process we also inserted the value of Knuth binning diameter $2\sqrt{a/\pi}$.
	}\label{fig:second_statistics_Knuth}
\end{sidewaysfigure}

To generate the virtual aggregation, they are firstly considered within a $500\times500$ units window (black curves of \figurename~\ref{fig:second_statistics_Knuth}), and then within a larger one, obtained from the previous by adding at its bottom an empty square box of the same area (grey curves). In \figurename~\ref{fig:second_statistics_Knuth}\subref{fig:Thomas_50_10_10} we generate a point pattern according to a \emph{mTp} with parameters $\rho=2\cdot10^{-4},\ \sigma=10,\ \mu=10$, where $\rho$ is the intensity of the Poisson process from which parents are generated, $\sigma$ is the standard deviation of the Gaussian distribution of offspring around each parent and $\mu$ is the mean number of offspring per parent. The two second-order statistics $g$ and \emph{K2} well capture the clustered structure of the pattern, either in the original window and in the expanded one. In particular, they reveal an average clump diameter around 37-40~units, which overestimate the true value $2\sigma\cdot \sqrt{\pi/2}\approx25$~units. Looking at the grey curves, we can observe the phenomenon of the \emph{virtual aggregation}. While the black pair correlation function intercepts the line $y=1$, the grey one sees the species as clustered at any scale. By contrast, their corresponding \emph{K2}-indexes are much closer one another, meaning the addition of an empty box slightly affects this last statistic. By applying the Knuth method we get a $22\times20$ grid which results in a clump size of $26$ units circa, which is therefore the closest to the real one. Notice that the Knuth method automatically restricts to the data span $V$ (see Sec.~\ref{Ksec}) therefore it results to be insensitive to voids in the pattern and hence does not suffer from the virtual aggregation effect. \\
In \figurename~\ref{fig:second_statistics_Knuth}\subref{fig:Disp10} we carry out the analysis for 500 points sampled from a uniform distribution with the additional constraint that if a point comes to fall within a fixed hard core distance from a pre-existing one, it is rejected. Once again all statistics are able to capture the overdispersion at small scale: they return an hard core radius around the true value of $10$~units, with a slight difference between black and grey curves. Applying the Knuth method we obtain a $1\times1$ grid as in the CSR case: the estimated density is therefore correctly detected by the optimal histogram which sees the homogeneity of the pattern. Here we see a limitation of the Knuth method which is unable to reveal second-order information such as the dispersion of the process and the hard core radius.\\
The last pattern shown in \figurename~\ref{fig:second_statistics_Knuth}\subref{fig:inPoisx} is sampled from a Poisson process with intensity $\lambda$ increasing with the $x$-coordinates, thus presenting a small gradient along the axis. Here the phenomenon of virtual aggregation is well visible looking at the graph of $g$: the pattern is detected as clustered at all scales by the statistic. Instead, the \emph{K2} index is not affected by this, although it cannot distinguish the process from a CSR one. By contrast, the Knuth method arrange data in a $4\times1$ grid, which permits us to capture the homogeneity along the $y$-axis and the gradient along the $x$-axis.\\
In conclusion, combining the Knuth method to the Shiffers' \emph{K2} index leads to a better understanding of the underlying process from which a sample is observed: the former permits to detect homogeneity against gradient in densities and gives a measure of how structured it is in the sense that small bins indicate clumping while large bin indicate Poisson-like distributions. The latter, on the other hand, allows to give quantitative information on the scale of both clumping and dispersion.

\subsection{Detection of anisotropy}
\begin{figure}[p]
	\centering
	\includegraphics[width=11cm]{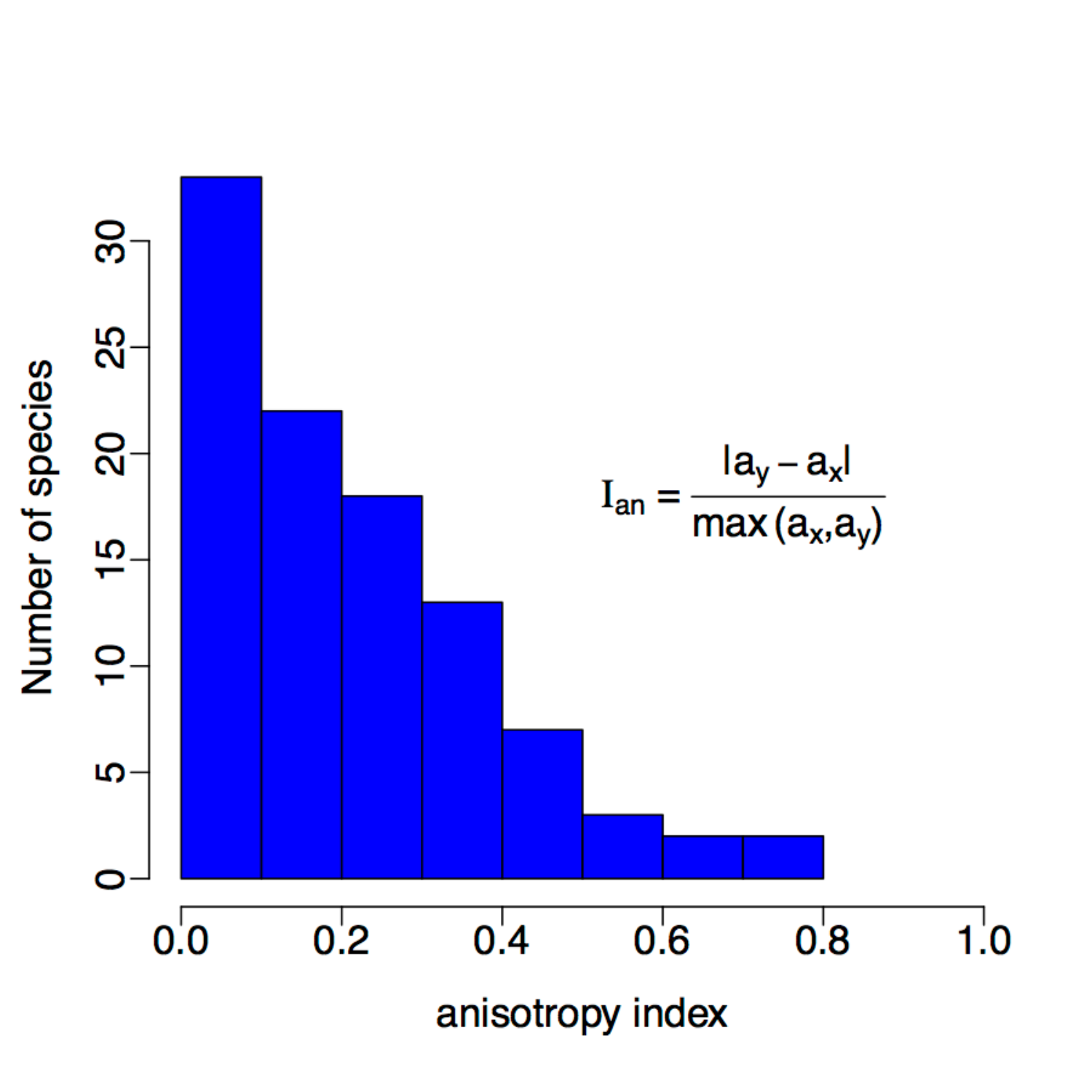}
	\caption{Knuth anisotropy index for the database consisting of 100 datasets generated from a Poisson cluster process with $N=3000$, $5$ clusters and $\sigma$ varying from 1 to 100. In the formula, $a_x$ is the bin width along the $x$ axis and $a_y$ the bin width along the $y$ axis.}\label{fig:anisotropy_Poisson}
\end{figure}

We tested the Knuth method's capability of detecting another relevant characteristic of a process: the anisotropy. By construction, Knuth algorithm is sensitive to the inversion of the orthogonal axes: if the two-dimensional pattern is tilted by a multiple of $\ang{90}$, also the resulting optimal grid is rotated of the same angle.
In Supplementary Information we applied the Knuth method on a generated dataset aggregated in a unique anisotropic Gaussian cluster distributed according to a bivariate Gaussian with standard deviation along the $x-$axis twice than along the $y-$axis; it resulted sensitive to the inversion of variance's direction between $\ang{0}$-$\ang{90}$ and $\ang{45}$-$\ang{135}$ cases, which leads to the inversion of the optimal bin sizes.\\
These results confirm that the Knuth method is very efficient at detecting the anisotropy of the cluster structure. However, an obvious limitation of the method is that it returns a unique bin size for the whole plot. Therefore, when dealing with real patterns, where anisotropic clusters may be structured at different scales or oriented in many directions, the optimal bin size is the result of a compromise between these different sub-structures of the dataset. Notice that while different bin sizes denote that the Knuth method classifies the pattern as anisotropic, the contrary does not hold.\\
However, previous remarks suggest that the difference in bin width along two orthogonal axes is an index of how anisotropic is the spatial pattern. We define the anisotropy index of a pattern as $I_{an}=|a_y-a_x|/\max(a_x,a_y)\in [0,1]$, where $a_x$ is the bin width along the $x$ axis and $a_y$ the bin width along the $y$ axis in order to be invariant with respect to inversion of the axes. To test the usefulness of the anisotropy index as a tool for detecting the anisotropy of a spatial pattern, we computed $I_{an}$ for a set of patterns generated from the \emph{mTp} process with parameters $N=3000,\ \rho\cdot A_0=5$ and $\sigma\in{0,...,100}$. The obtained frequency histogram (\figurename~\ref{fig:anisotropy_Poisson}) is highly picked around zero, meaning that the Knuth method detects the isotropic structure of the data distribution along the principal axes. In Sec.~\ref{sec:Application to the BCI dataset} below we computed the anisotropy index for a dataset containing the location of the trees of around $180$ species showing that the Knuth method can be efficiently used to test hypothesis on the underlying process from which real data are sampled.

\section{Application to the BCI ecological database}
\label{sec:Application to the BCI dataset}
We have considered an open access ecological dataset consisting of the spatial coordinates of individuals belonging to 300 different species of plants located in a 50~ha rectangle of the Barro Colorado Island (BCI) rainforest in Panama. Our goal is to show 
 that the choice of modelling BCI species' distribution through a \emph{mTp}, which has been proven to be efficient in capturing some important biological curves but not others (\cite{Plotkin, Morlon, Azaele}), in many cases is not supported by the Knuth method.\\
 Notice that checking the goodness-of-fit of a fitted model using a minimum contrast method when the theoretical form of the summary function (usually Ripley's $K$) is not known is a difficult task. For a $mTp$ process the form of Ripley's $K$ is known (see Supplementary Information) hence the fitted pattern is already optimal with respect to a minimum contrast goodness-of-fit criterion. Nevertheless, the analysis of the difference of the optimal bin size for the real and the $mTp$-generated pattern may reveal a strong departure form the real data (see below).\\
To find the species that are suitable to be described by a clumped pattern, 
we have selected all species with abundance between 20 and 3000 individuals (204 species) and we have then computed the \textit{mTp} parameters $(\rho,\sigma,\mu)$, for each species by fitting the empirical Ripley's \emph{K} function (minimum contrast estimation, see \cite{Diggle}, Ch.~7, and Supporting Information). As in \cite{Morlon}, we have discarded the species whose distribution is not quite distinguishable from a random one --in this case the Poisson cluster process cannot capture correctly the underlying structure of the data-- according to the following criteria:
i) the mean cluster diameter $\sigma\sqrt{2\pi}$ is smaller than $500$~m and ii)
 the number of clusters is smaller than the number of the individuals.
 
There are 183 of the 204 species satisfying both criteria.
For each of the 183 species we have computed $\sqrt{a/\pi}$ which is the radius of the circle equivalent to a rectangular Knuth optimal bin of area $a$. We know this is a measure of the size of the underlying minimal structure of the dataset. 
The frequency histogram (see Supporting Information) indicates that there is no preferred choice for a common optimal bin area for all species, since the values of the optimal binning areas span from $100$~m$^2$ to $ 5\cdot10^5$~m$^2$ circa. This is not surprising, since each species has its own distribution due to myriad of factors such as seed dispersal, gap recruitment or adaptation to the surrounding soil (\cite{Augspurger,Plotkin}). 

\subsection{Difference index for BCI}
\begin{figure*}[p]
	\centering
	\includegraphics[width=12cm]{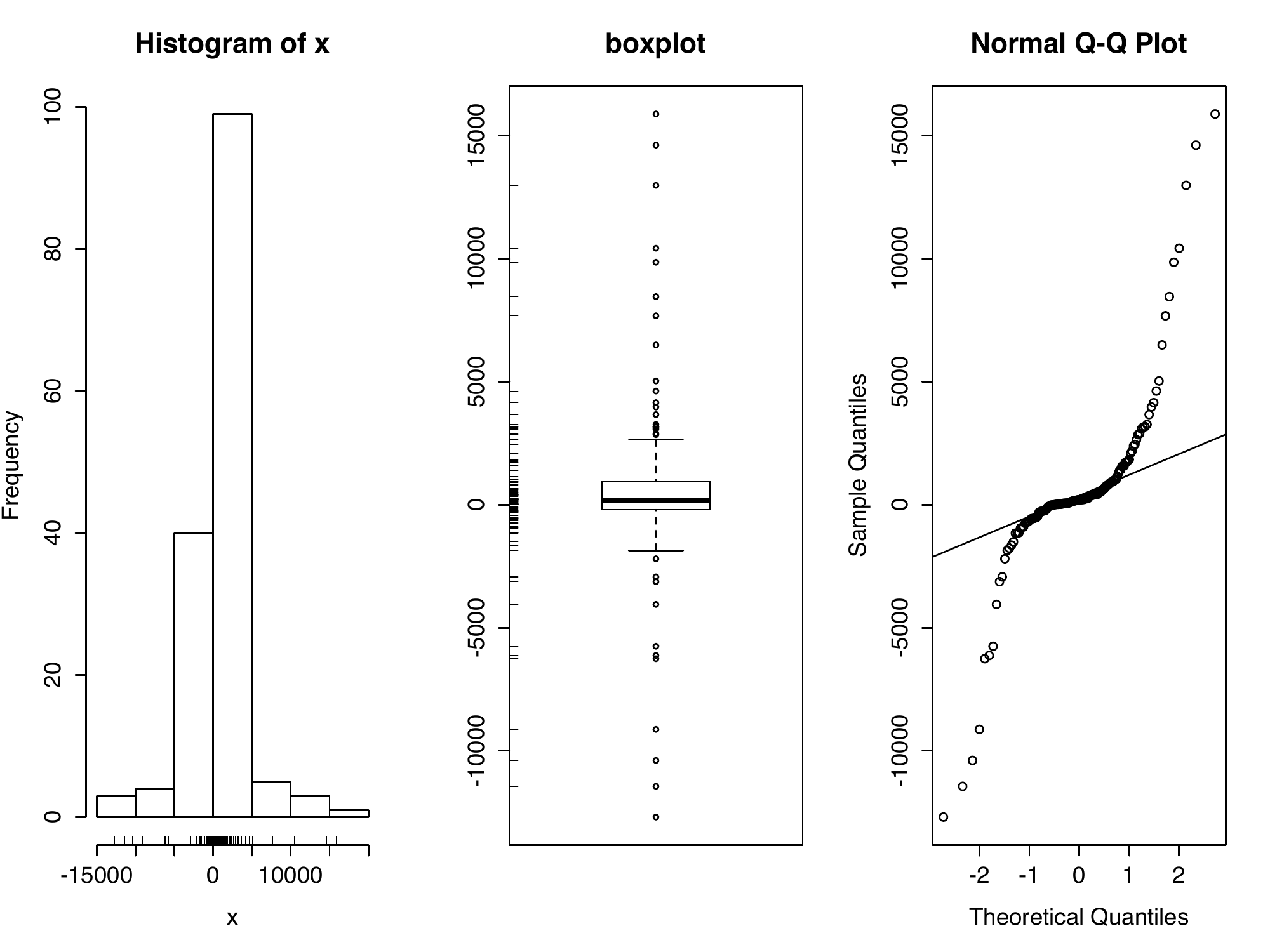}
	 \caption{Exploratory analysis of difference data $x=\Delta = a(mTp) -a(real)$: histogram, boxplot and relation with quartiles of a normal distribution. The mean is $546.4$, the median $202.9$ and the standard deviation equals $3458.9$.
	 	 }
	 	 \label{fig:qqplot}
\end{figure*}

\begin{figure*}[p]
	\centering
	\vspace*{-0.8cm}
	\includegraphics[width=60mm]{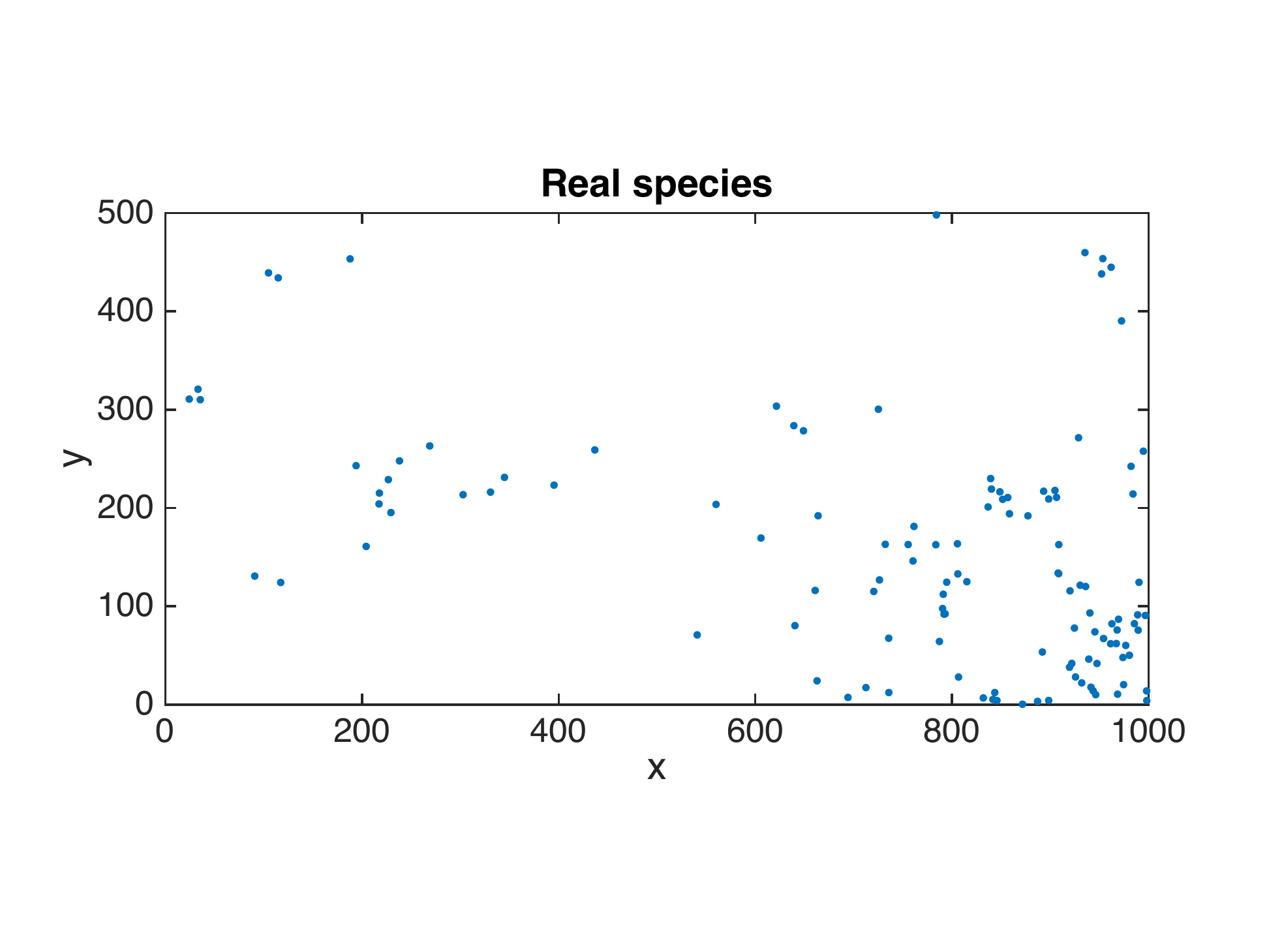}\hspace{-0.3cm}
	\includegraphics[width=60mm]{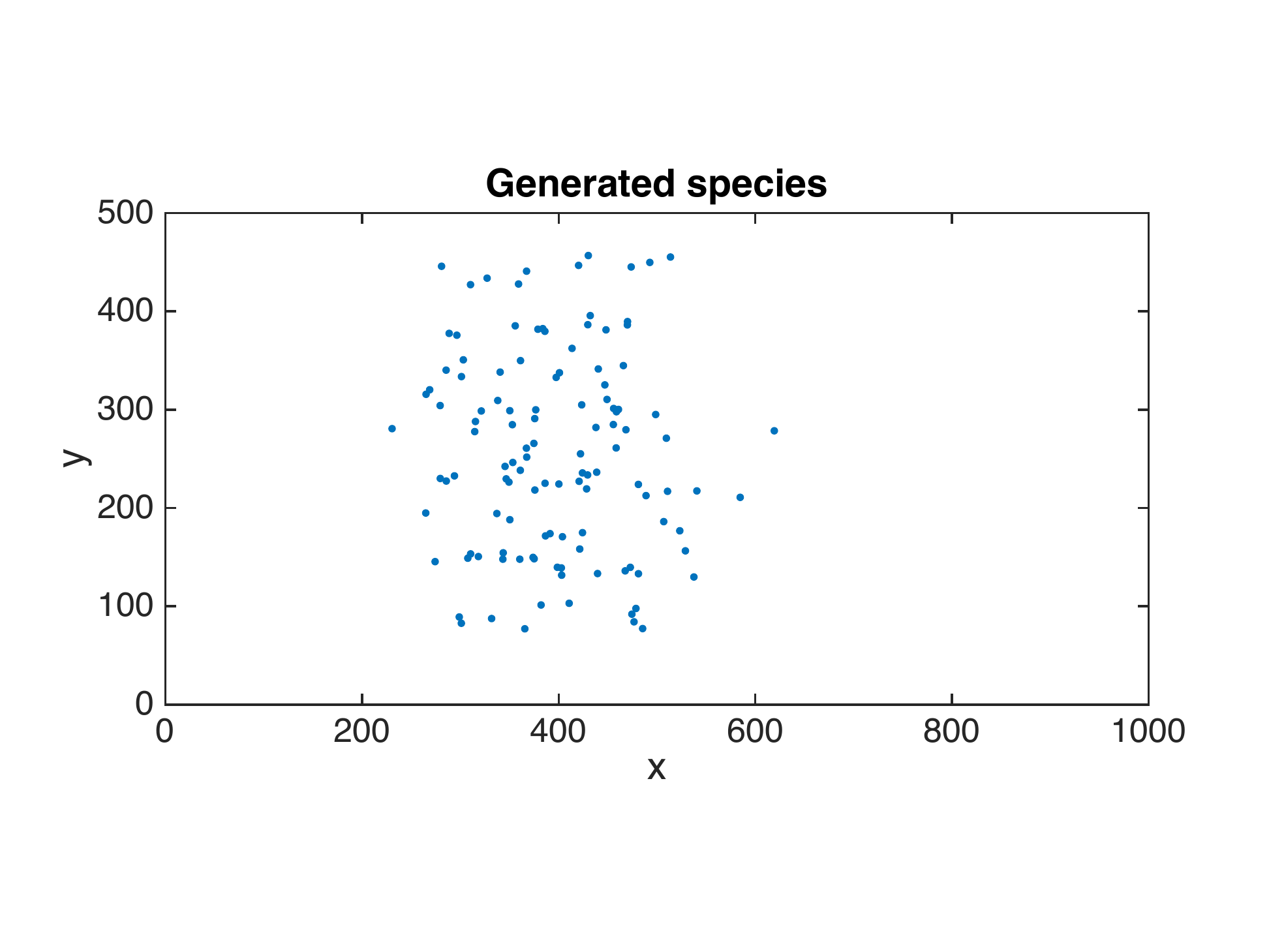}\\
	\vspace*{-0.55cm}
	\includegraphics[width=60mm]{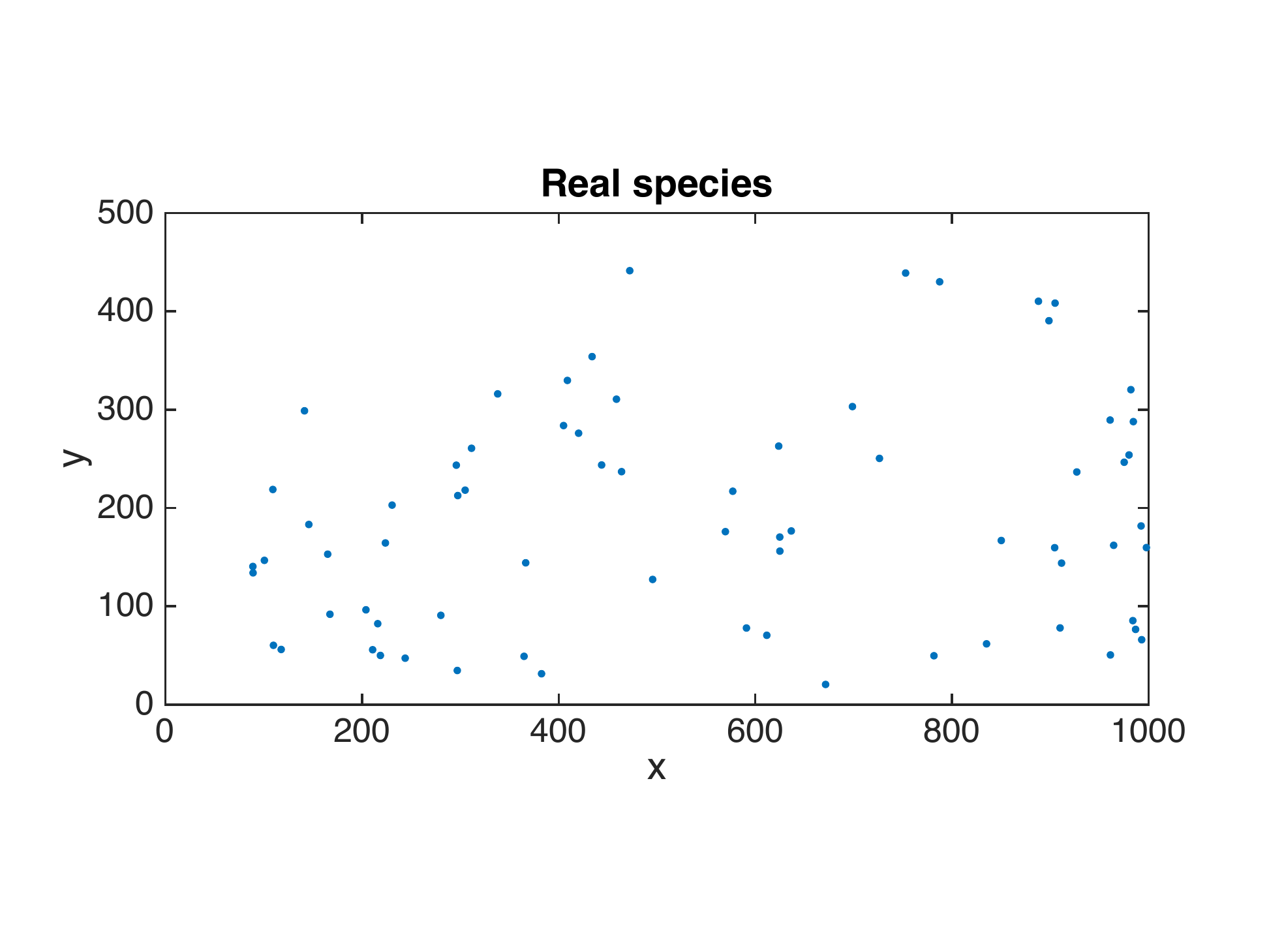}\hspace{-0.3cm}
	\includegraphics[width=60mm]{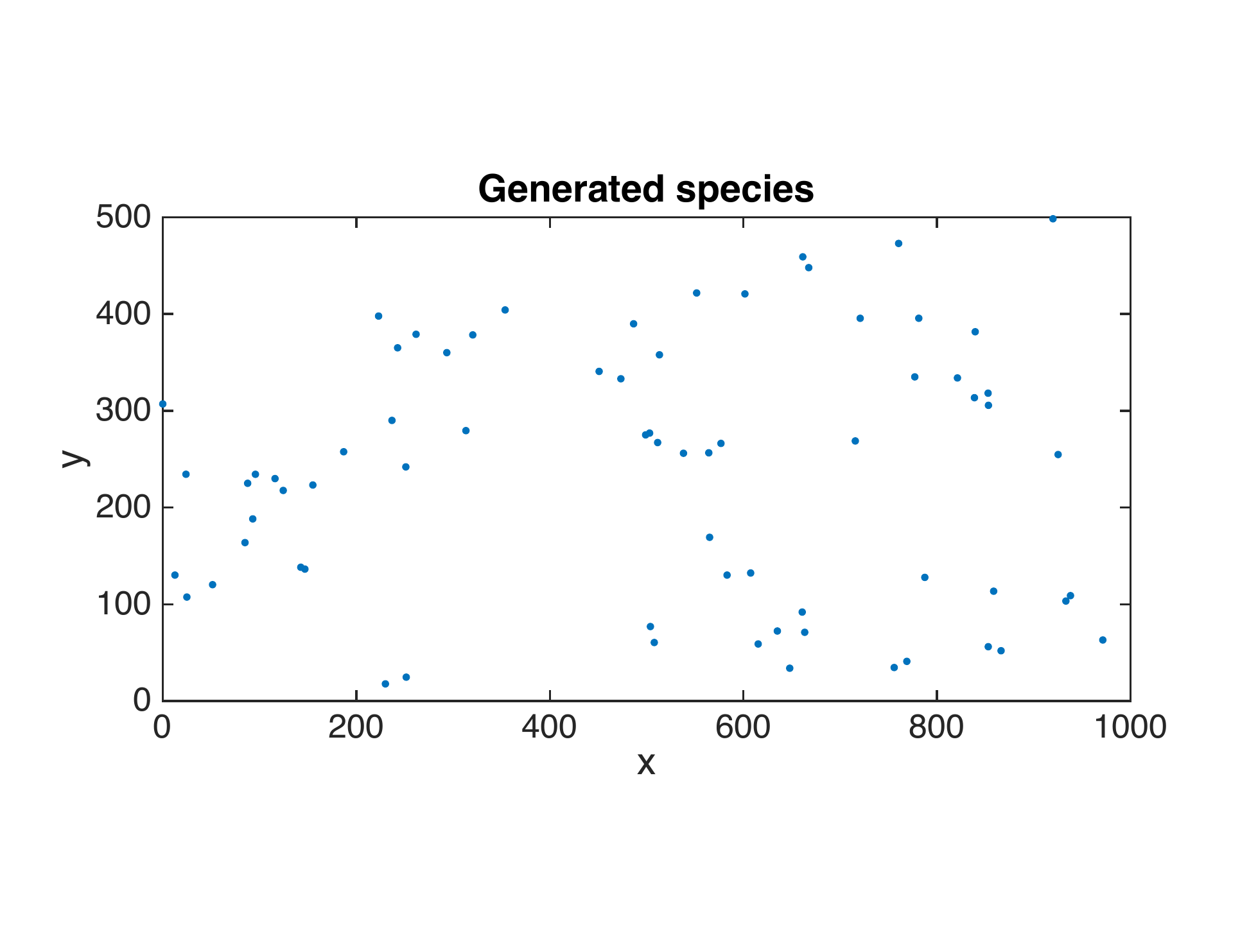}\\
	\vspace*{-0.55cm}
	\includegraphics[width=60mm]{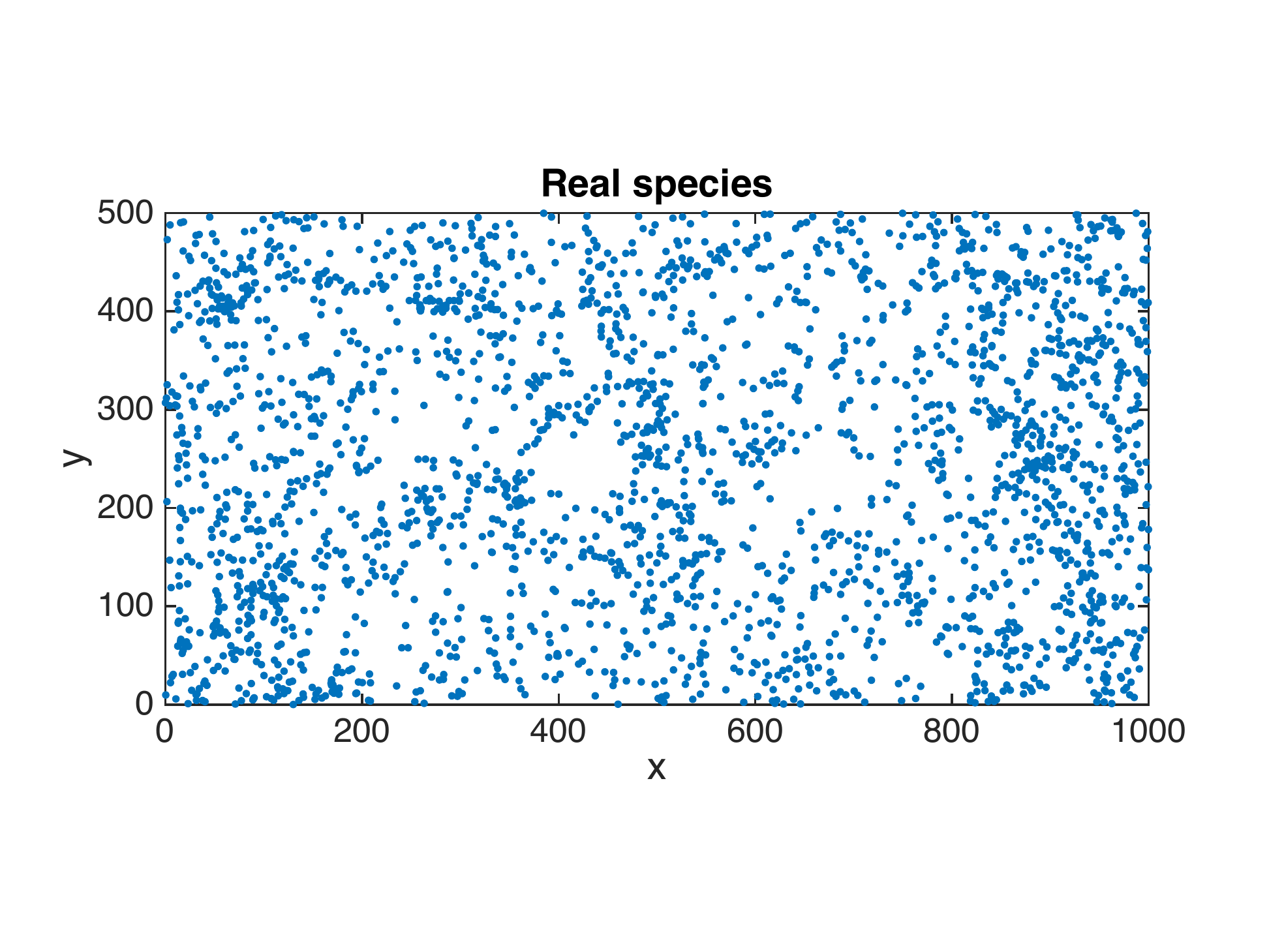}\hspace{-0.3cm}
	\includegraphics[width=60mm]{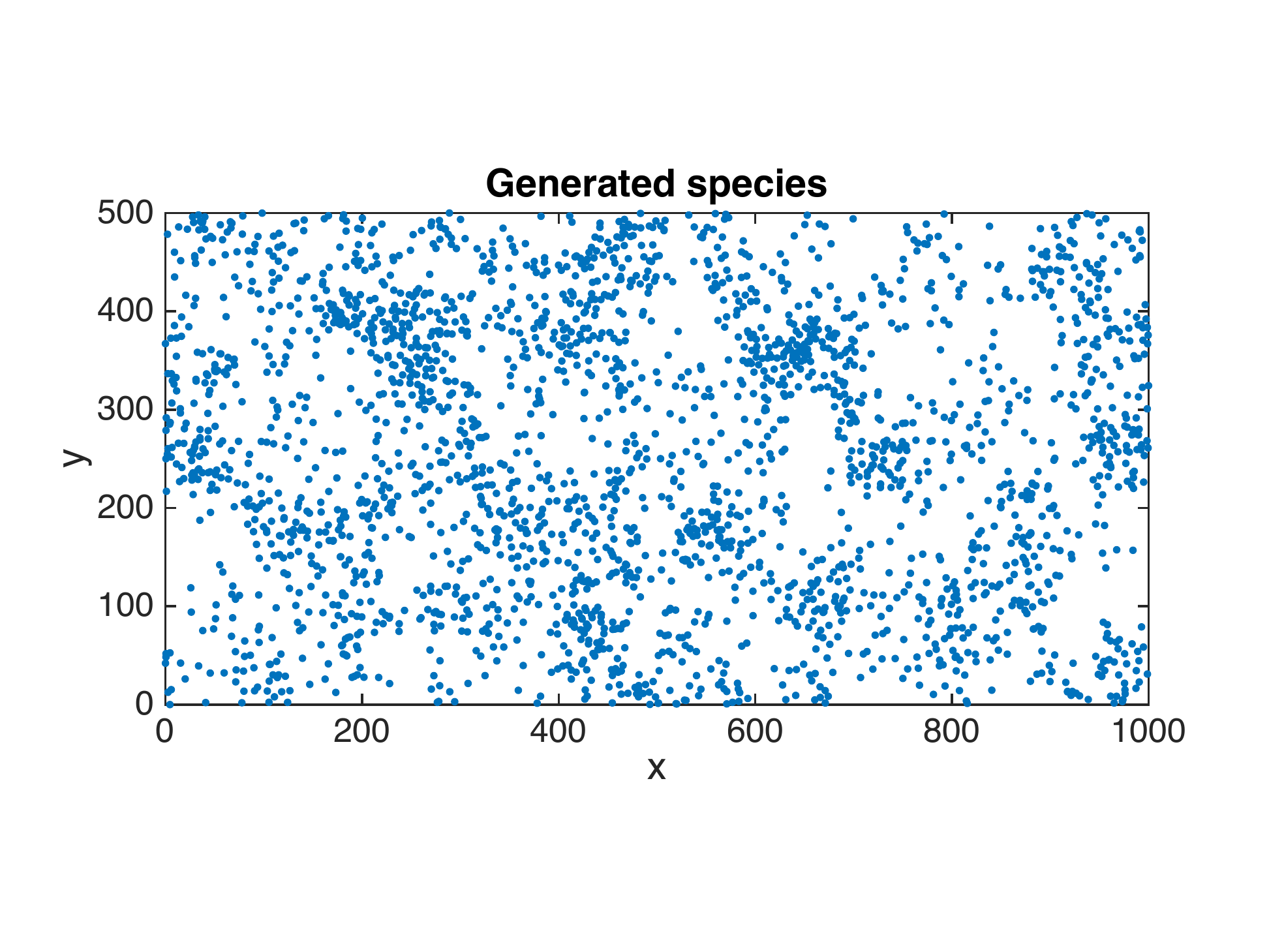}
	\caption{\small 
		Plot of three species distribution from the BCI surveyed area (left column) and plot of the same species distribution generated according to the $mTp$ with parameters fitted from the real data (right column). Species were selected to display the three cases : i)(\textit{Quassia amara}, top) $a(mTp) \gg a(real)$, 	 meaning that the Knuth method detects a finer structure for the real species compared to the generated one; 
		ii) (\textit{Posoqueria latifolia}, middle) $a(mTp)\approx a(real)$ the Knuth method recognises as similar the two spatial structures and therefore the hypothesis that the species is distributed according to an $mTp$ is not rejected.
		iii) (\textit{Soroacea affinis}, bottom) $a(mTp) \ll a(real)$, the Knuth methods detects a finer structure for the generated species with respect to the real one.}\label{fig:Knuth_mTp_realspecies}
\end{figure*}

\begin{figure}[p]
	\centering
	{\includegraphics[width=11cm]{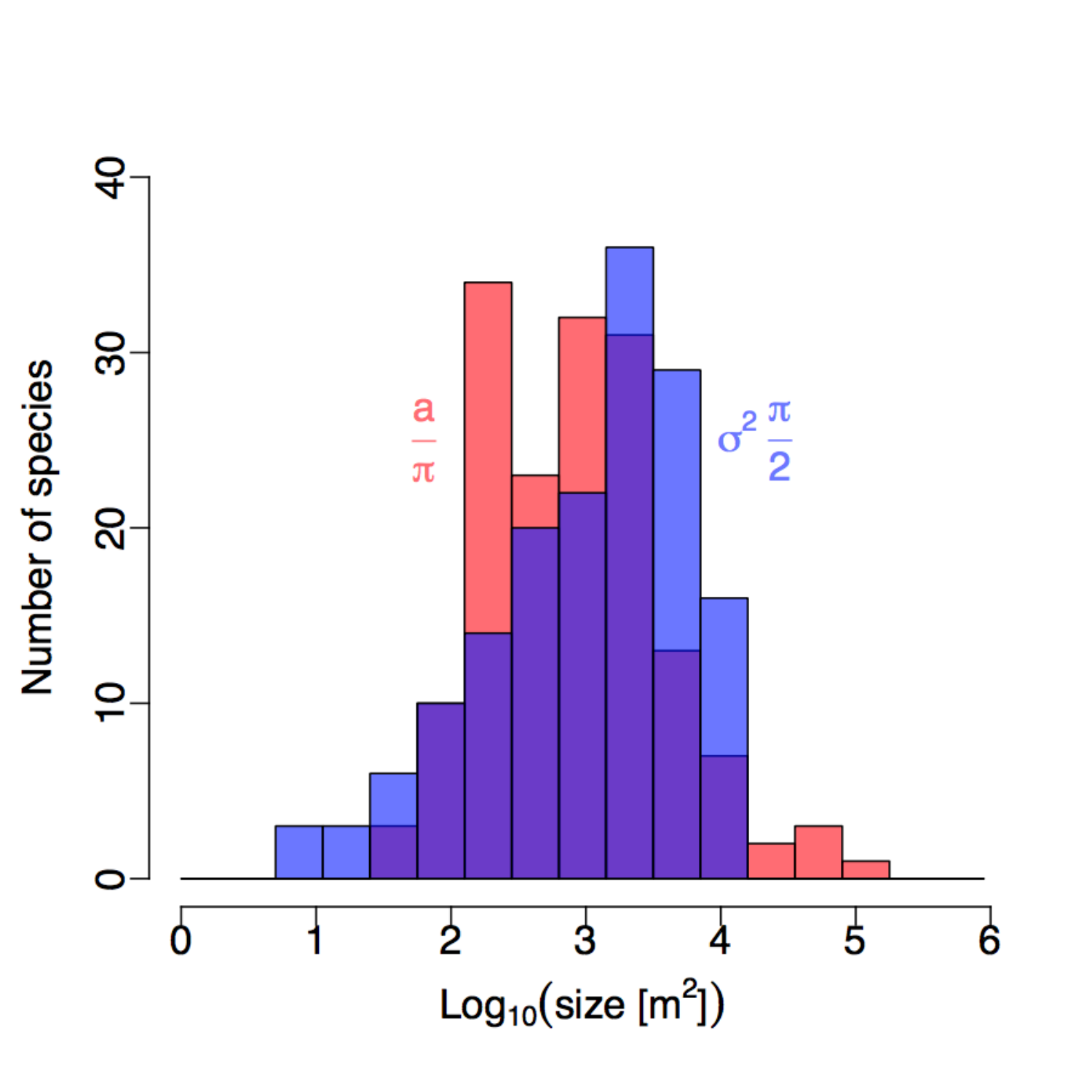}}
	\caption{Comparison between histograms of the clump area of \textit{mTp} (blue) and Knuth optimal bin area (red) for BCI species.}\label{fig:cmp_a_sigma}
\end{figure}

We computed the Knuth optimal bin size $a$ both for BCI real species with $\sigma^2\pi/2<10^{4}$~m$^2$ and for their correspondent \emph{mTp} generated counterparts and their difference 
 $ \Delta =  a(mTp)-a(real)) $. 
 In \figurename~\ref{fig:qqplot} we display the exploratory analysis (histogram, boxplot and QQ plot) of the difference data. We see that the histogram is right-skewed with fat tails. For $ 50 \%$ of the species the difference of the bin area $\Delta$ is bigger than $2$ times the smaller of the two. For these species which are in the tails of the distributions, the $mTp$ process fails at reproducing the real pattern either because the generated clustered pattern has a coarser scale than the real one losing details of the original fine structure ($a(real) \ll a(mTP)$, right tail) or because it introduces an artificial clustered structure on a more uniform real pattern ($a(real) \gg a(mTp)$, left tail). \\
To have a visual inspection of the difference between real and generated species, we have selected three species respectively i) in the right tail, ii) in the centre and iii) in the left tail. In \figurename~\ref{fig:Knuth_mTp_realspecies} we display the real and the $mTp$ generated patterns with parameters fitted from the data. In the first case the $mTp$ reproduces a coarser pattern than the real one while in the latter it introduces an artificial finer structure.\\ 
In \figurename~\ref{fig:cmp_a_sigma} the frequency histogram of the $mTp$ clump area $\sigma^2\cdot\pi/2$ is superimposed to the one of $a/\pi$. In this cumulative plot the two histograms are respectively right and left-skewed showing that  globally Knuth method assigns a finer structure to the real patterns with respect to the generated ones.
 
\subsection{Anisotropy index for BCI}
\begin{figure}[p]
	\centering
	{\includegraphics[width=11cm]{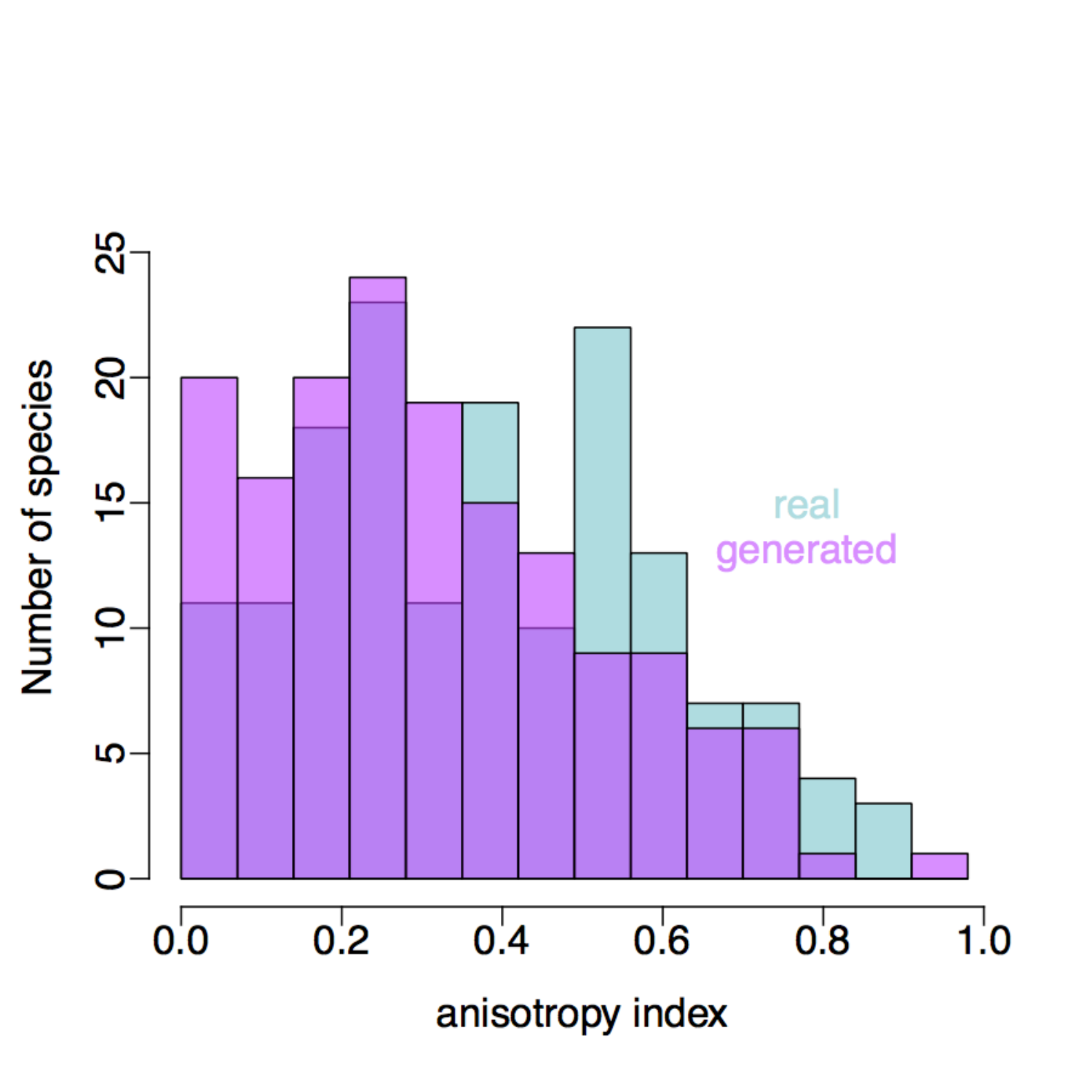}}\\
	\vspace*{0.3cm}
	\caption{Frequency histograms of anisotropy index for real BCI species and the ones generated by a modified Thomas process with parameters fitted by data. }\label{fig:anisotropy_index}
\end{figure}

We computed the anisotropy index $|a_y-a_x|/\max(a_x,a_y)$ both for BCI real species with $\sigma^2\pi/2<10^{4}$~m$^2$ and for their correspondent \emph{mTp} generated counterparts. 
 \figurename~\ref{fig:anisotropy_index} shows the frequency histogram of anisotropy index for real species superimposed on the one for $mTp$ generated species. In the first case there is a high number of species whose index is far from 0, meaning that their underlying density function is not recognised as isotropic by the Knuth method. In the second case, Knuth histogram is more shifted against the $y-$axis, meaning that the Knuth method sees the new generated species' process more isotropic than before. This fact suggests that a reason for which \textit{mTp} process fails in capturing some important ecological curves is the fact that the hypothesis of isotropic clusters is too strong and is not supported by real data. 

\subsection{Relation with abundance}
The relation between species abundance and degree of aggregation is still debated as an important issue of ecological theory (\cite{He,Plotkin,Condit,Morlon}). Moreover, as pointed out in \cite{Morlon}, the correlation between these two quantities strongly depends on how they are measured. For example, for Pasoh forest, in (\cite{He}) the proposed \textit{Donnelly clumping index} based on nearest-neighbour distance shows a slightly positive correlation between abundance and aggregation. By contrast, the \textit{relative neighbourhood density} $\Omega_{0-10}$ of \cite{Condit} and the Cramer-von Mises-type $k$ statistic of \cite{Plotkin} are negatively correlated to abundance.\\ 
To investigate if \textit{mTp} and Knuth aggregation parameters are similarly correlated with species abundance or not, 
we have computed on the one hand the correlation between the abundance and the following $mTp$ quantities: the mean number of parents $\rho\cdot A_0,$ the mean clump radius $\sigma\cdot\sqrt{\pi/2},$ the mean number of offspring per parent $\mu$ and the relative neighbourhood density $\Omega_{0-10}$ (see \cite{Morlon}) and on the other hand the correlation of species' abundance with Knuth optimal bin area and index of anisotropy. These latter, as we have seen, give us information on how structured is the data density function and how far it is from a uniform or isotropic one (see Supporting Information).\\
From the determination coefficients, only the relative neighbourhood density shows a negative correlation with abundance, while the other \textit{mTp} parameters result to be slightly positive correlated with it. This is in accordance with the literature (\cite{Morlon}). Knuth optimal bin area (\figurename~\ref{fig:num_parameters}) and index of anisotropy are, instead, insignificantly correlated with the abundance with respect to the determination coefficient. This is quite reasonable because Knuth optimal grid depends only on data distribution, and not on their abundance.

%------------------------------------------------
\phantomsection
\section*{Results and Discussion}
 Knuth optimal binning method, based on Bayes' Theorem and Maximum a-posteriori Estimation allows to infer the least biased estimate of the underlying density function of a point pattern. The optimal bin size sets the most informative scale at which to observe the data. \\
We showed how to use the Knuth optimal bin size and shape to estimate the intensity of a spatial process and infer characteristic spatial features as anisotropy and clusterization. We tested it against most currently used kernel method for two-dimensional datasets (Epanechnikov) and one-dimensional (Stone binning rule) ones and it resulted to be more efficient in detecting CSR processes and avoiding sample fluctuations in both cases. It is therefore a reliable method for determining the intensity function of a pattern. Moreover it does not need any a priori assumption about the phenomena that generated the data.\\
Since the Knuth method is based on a maximization procedure it does not contain adjustable parameters and it is not subject to the virtual aggregation phenomenon. When used in conjunction with the \emph{K2}-index, it allows to infer qualitative (it correctly detects homogeneity or the presence of a gradient in the density function, even if it is not strong) and quantitative (cluster size) information on both first and second-order statistics.\\
Our analysis indicates that Knuth bin size is a good indicator of how finely structured is the dataset and the relative difference of the rectangular bin sides is a measure of the anisotropy of the pattern. For the above reasons it can be used as a trusted tool for the preliminary statistical analysis of a spatial dataset.\\
Finally, we tested our findings on the BCI ecological dataset to have information about distribution of the size of cluster-like structure of plants, anisotropy of plant distribution and existence of uniformly distributed species. We found evidence that the choice of modelling a species' distribution through a modified Thomas process, which has been proven to be efficient in capturing some important biological curves but not others, is not always supported by the Knuth method. Moreover, we found that cluster size, measured by optimal bin area is insignificantly correlated with the abundance of a species. 
Globally, it provides a reliable method to test whether an hypothesis made on the underlying process of a pattern is justified or not. We are confident that this survey of Knuth algorithm' performance can be of help for the scientific community.
%------------------------------------------------
\phantomsection
\section*{Acknowledgments} 
We are grateful to Samir Suweis, Sandro Azaele, Matteo Adorisio, Jacopo Grilli and Jayanth R. Banavar for many insightful discussions.\\ We also thank the Center of Tropical Research Science (R.~Condit, S.~Hubbell, R.~Foster) for providing the empirical data of the BCI forest. We thanks the anonymous referees for their valuable criticisms and suggestions.

%----------------------------------------------------------------------------------------
%	PICTURES
%----------------------------------------------------------------------------------------

\clearpage
\clearpage
\clearpage
\clearpage
\clearpage
\clearpage
\clearpage
\clearpage
\begin{figure}[p]
	\centering
	\includegraphics[width=11cm]{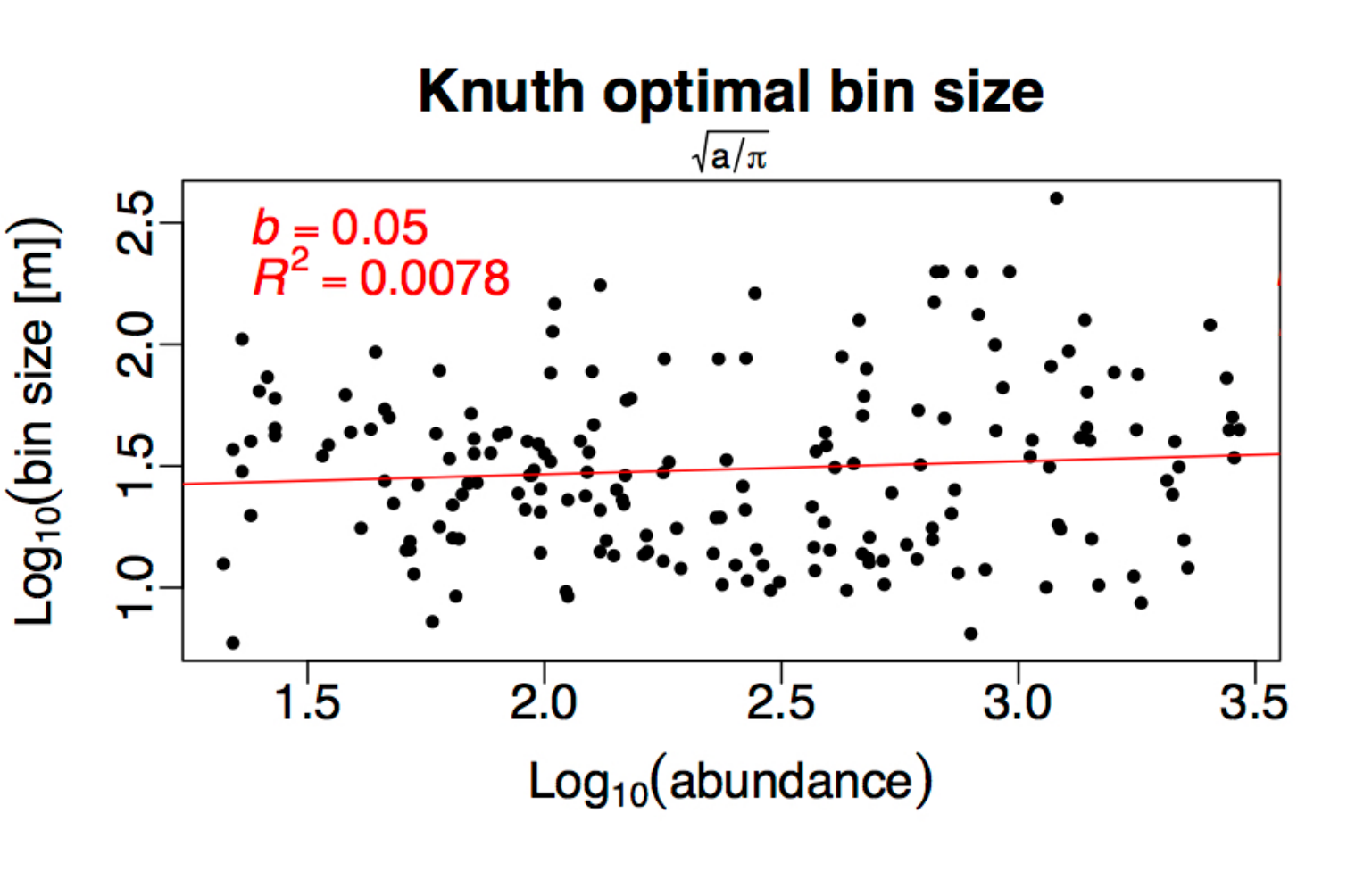}
	\caption{Correlation between species' abundance with Knuth optimal bin radius $\sqrt{a/\pi}$. From the determination coefficient the two quantities result insignificantly correlated. In the figure $b$ is the slope of the fitting line.}\label{fig:num_parameters}
\end{figure}

\part*{Supplementary Information}

\section{Application of Knuth method on one-dimensional datasets}
\label{sec:Knuth_method}
\begin{figure*}
	\centering
	\vspace{-0.9cm} 
	\subfloat[][{\label{fig:RelLog_uniform}}]
	{\includegraphics[width=.35\textwidth]{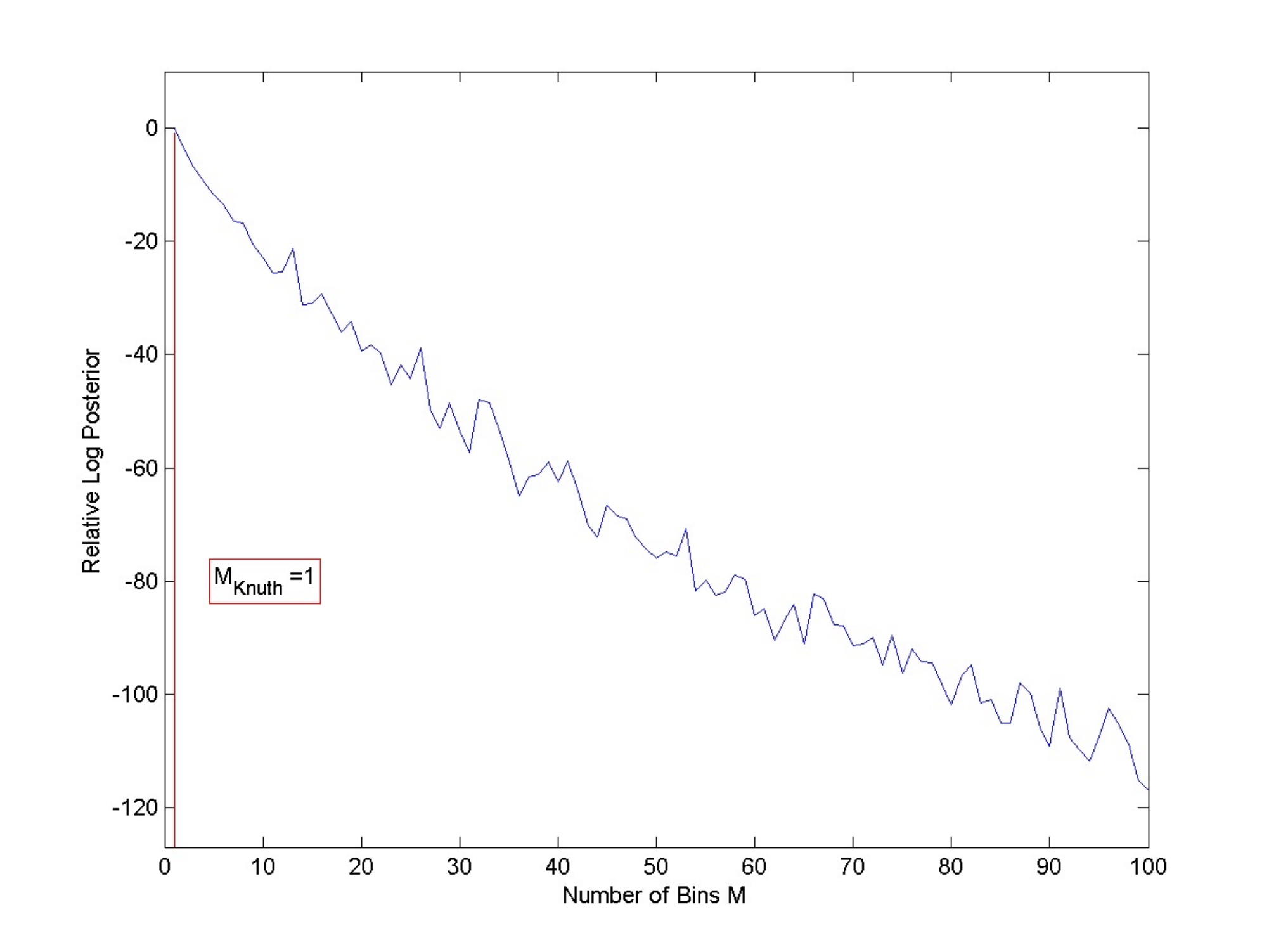}} \quad
	\subfloat[][{\label{fig:Knuth_uniform}}]
	{\includegraphics[width=.35\textwidth]{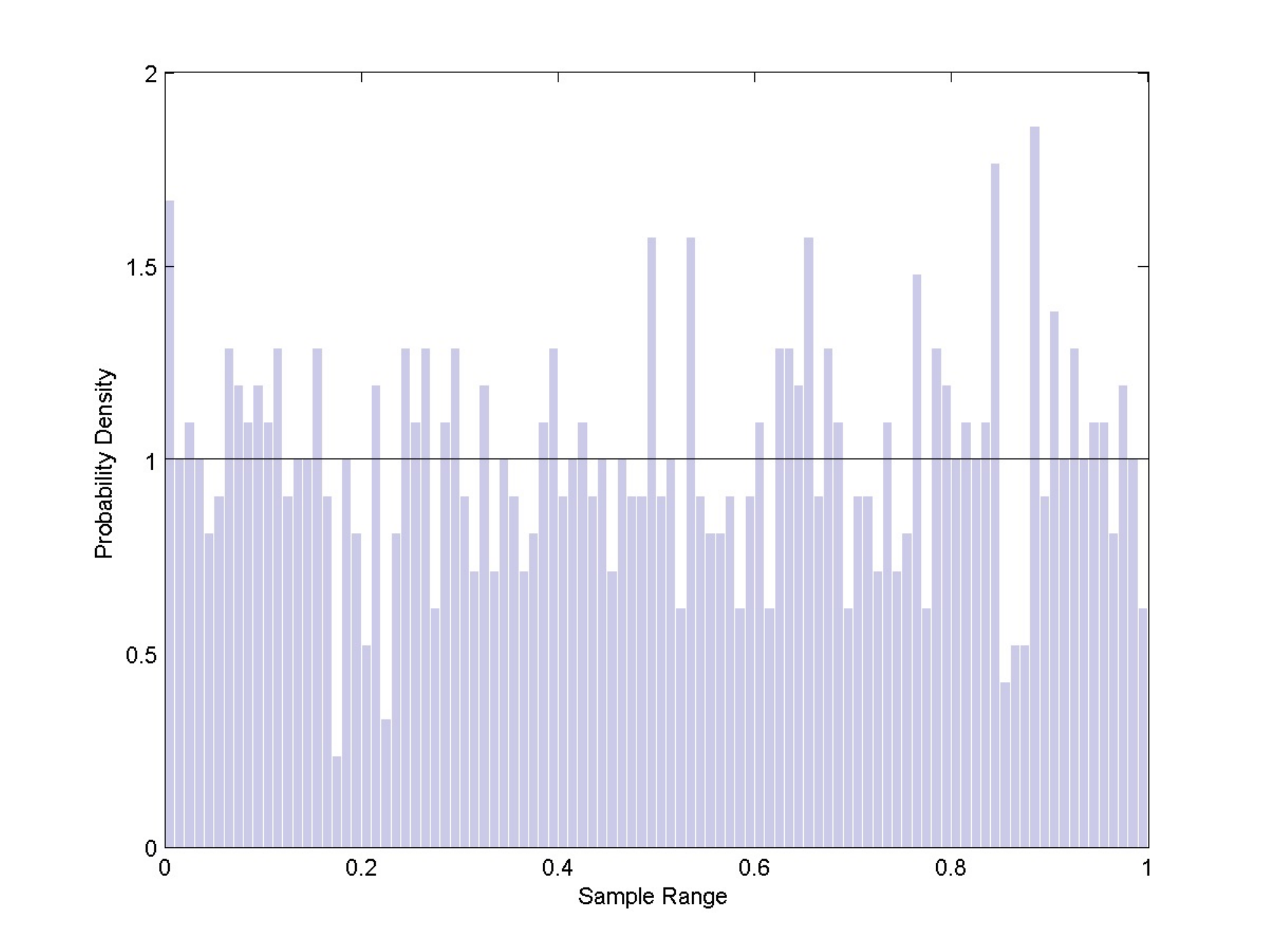}} \\
	\vspace{-0.6cm}
	\subfloat[][{\label{fig:RelLog_steps}}]
	{\includegraphics[width=.35\textwidth]{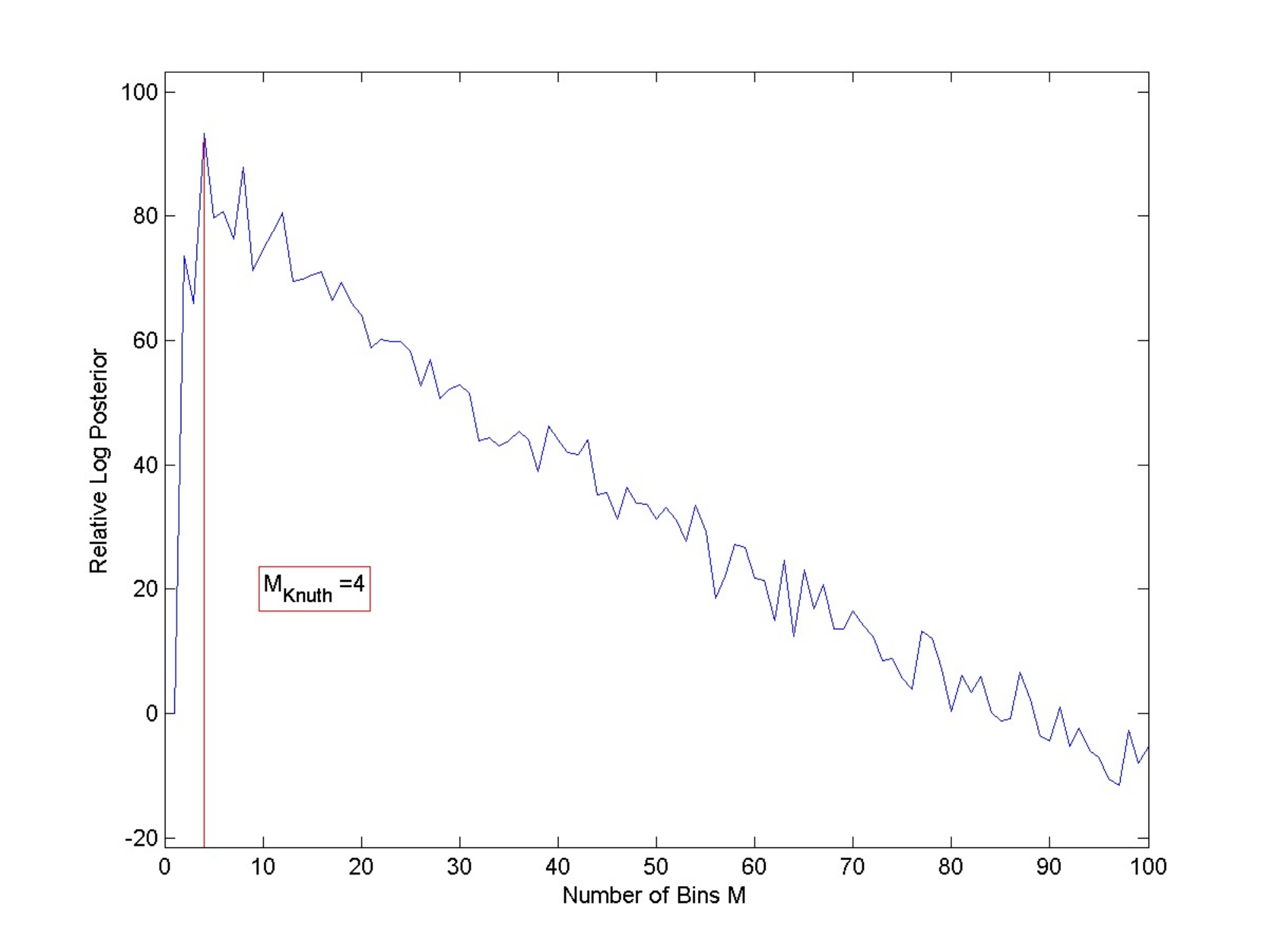}} \quad
	\subfloat[][{\label{fig:Knuth_steps}}]
	{\includegraphics[width=.35\textwidth]{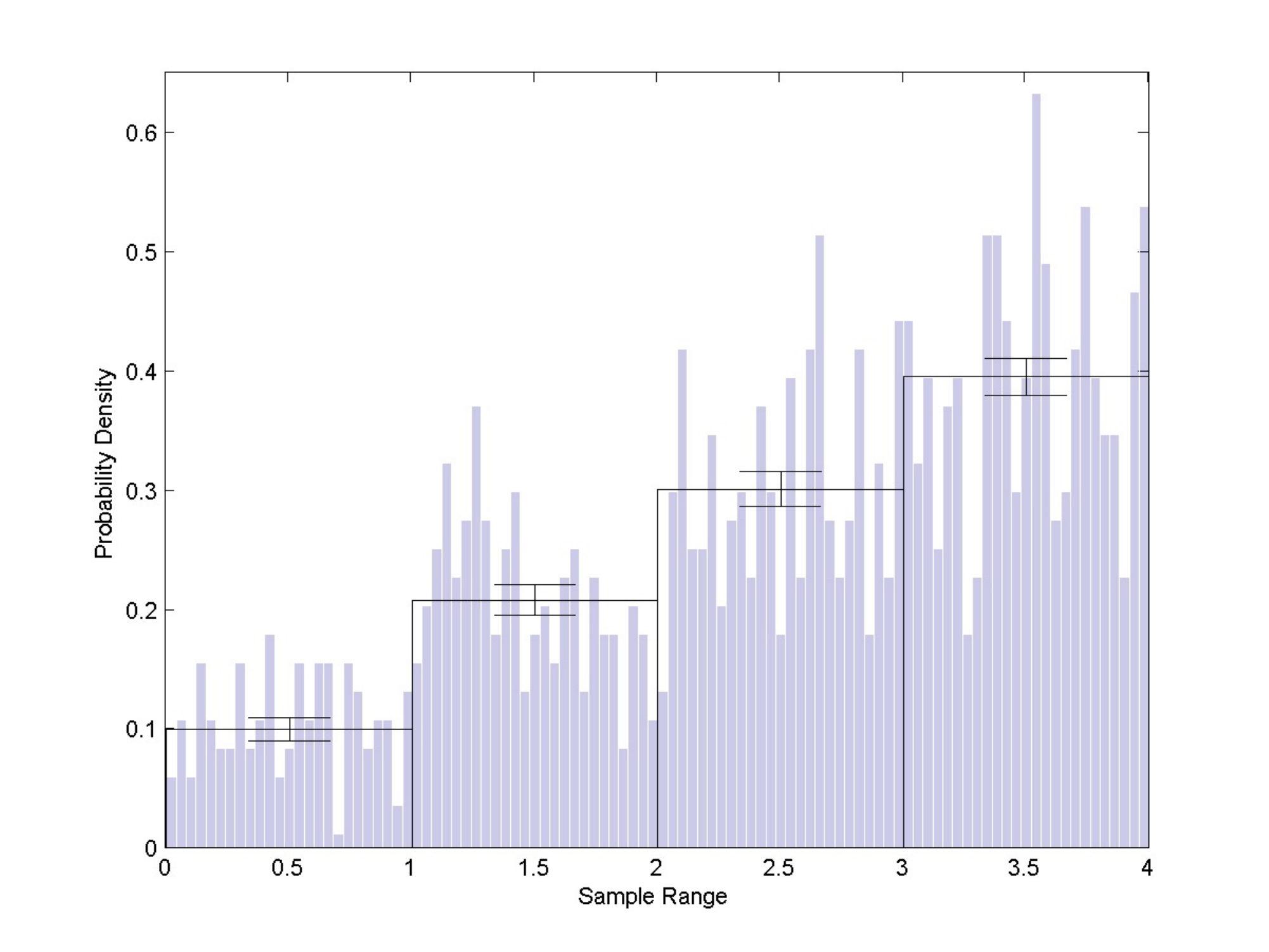}} \\
	\vspace{-0.6cm}
	\subfloat[][{\label{fig:RelLog_gauss}}]
	{\includegraphics[width=.35\textwidth]{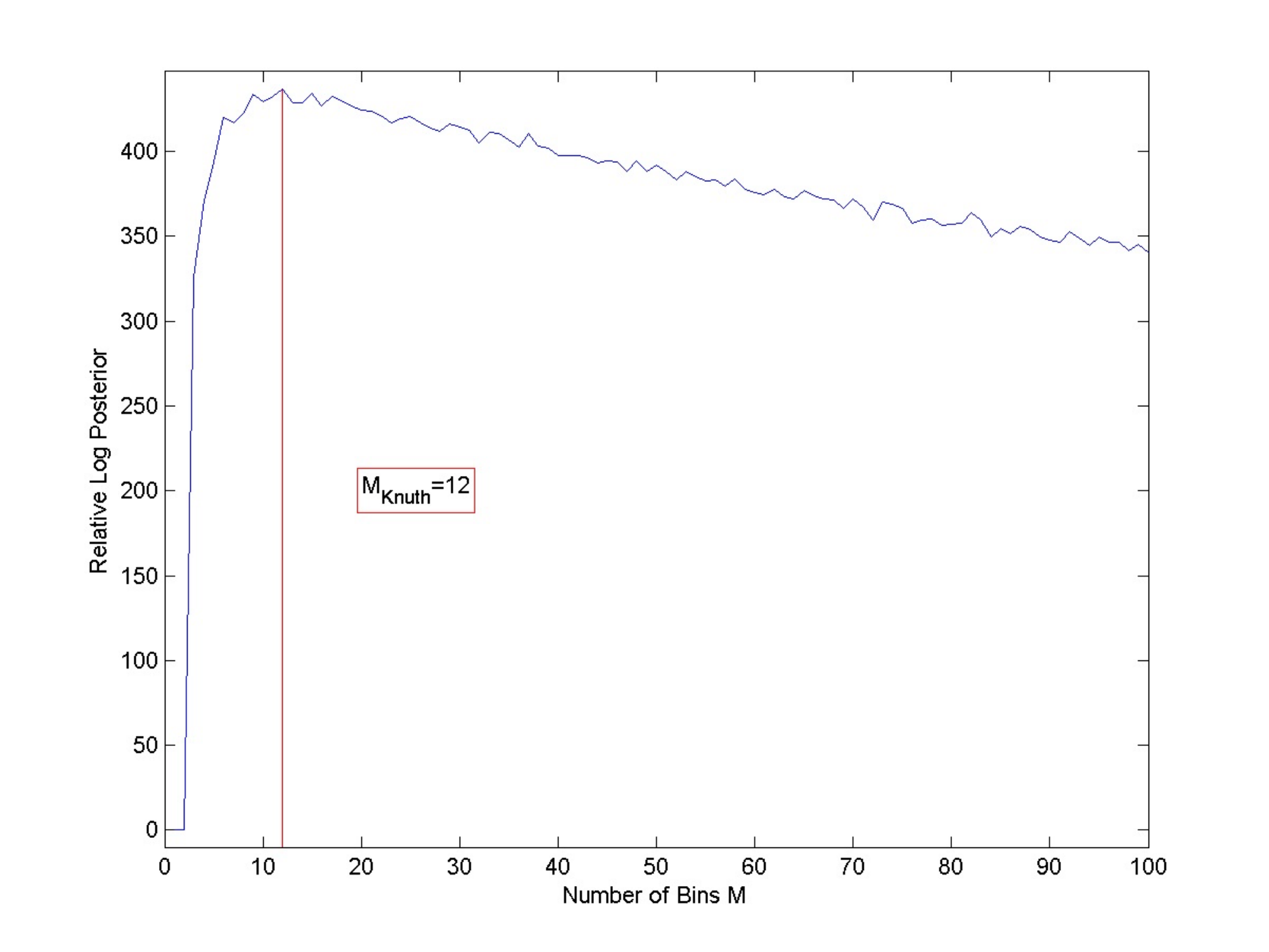}} \quad
	\subfloat[][{\label{fig:Knuth_gauss}}]
	{\includegraphics[width=.35\textwidth]{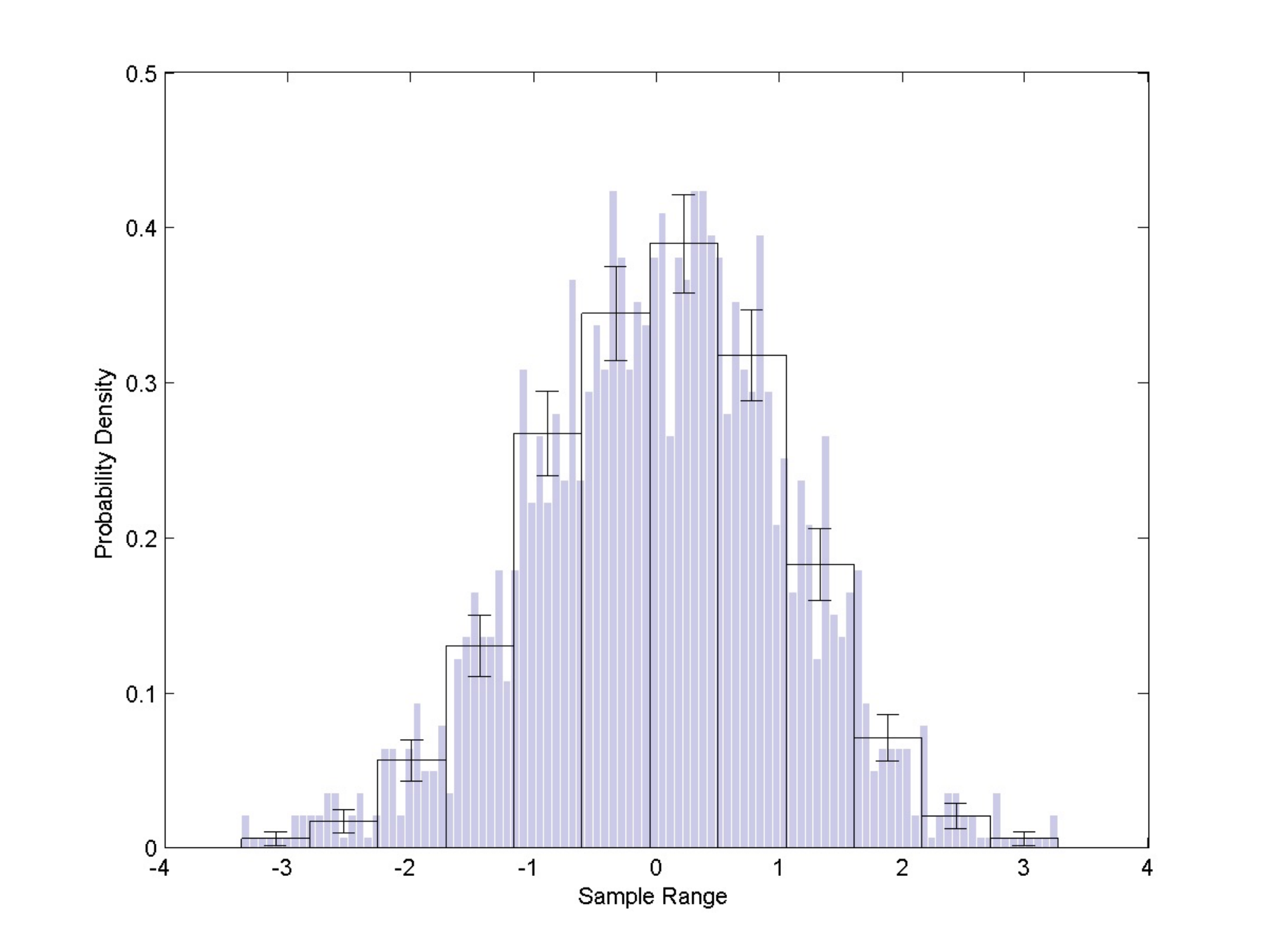}} \\
	\vspace{-0.6cm}
	\subfloat[][{\label{fig:RelLog_threegauss}}]
	{\includegraphics[width=.35\textwidth]{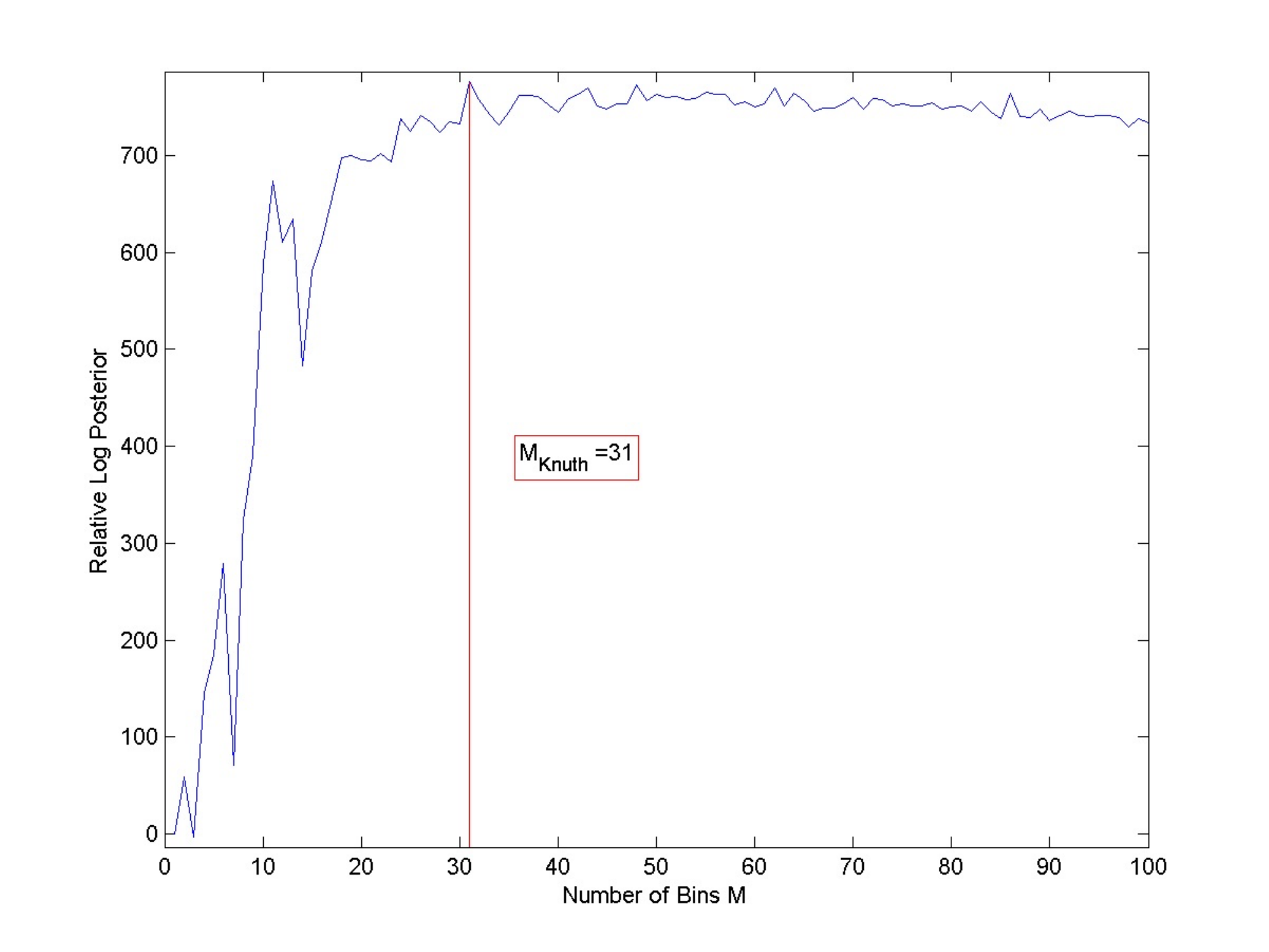}} \quad
	\subfloat[][{\label{fig:Knuth_threegauss}}]
	{\includegraphics[width=.35\textwidth]{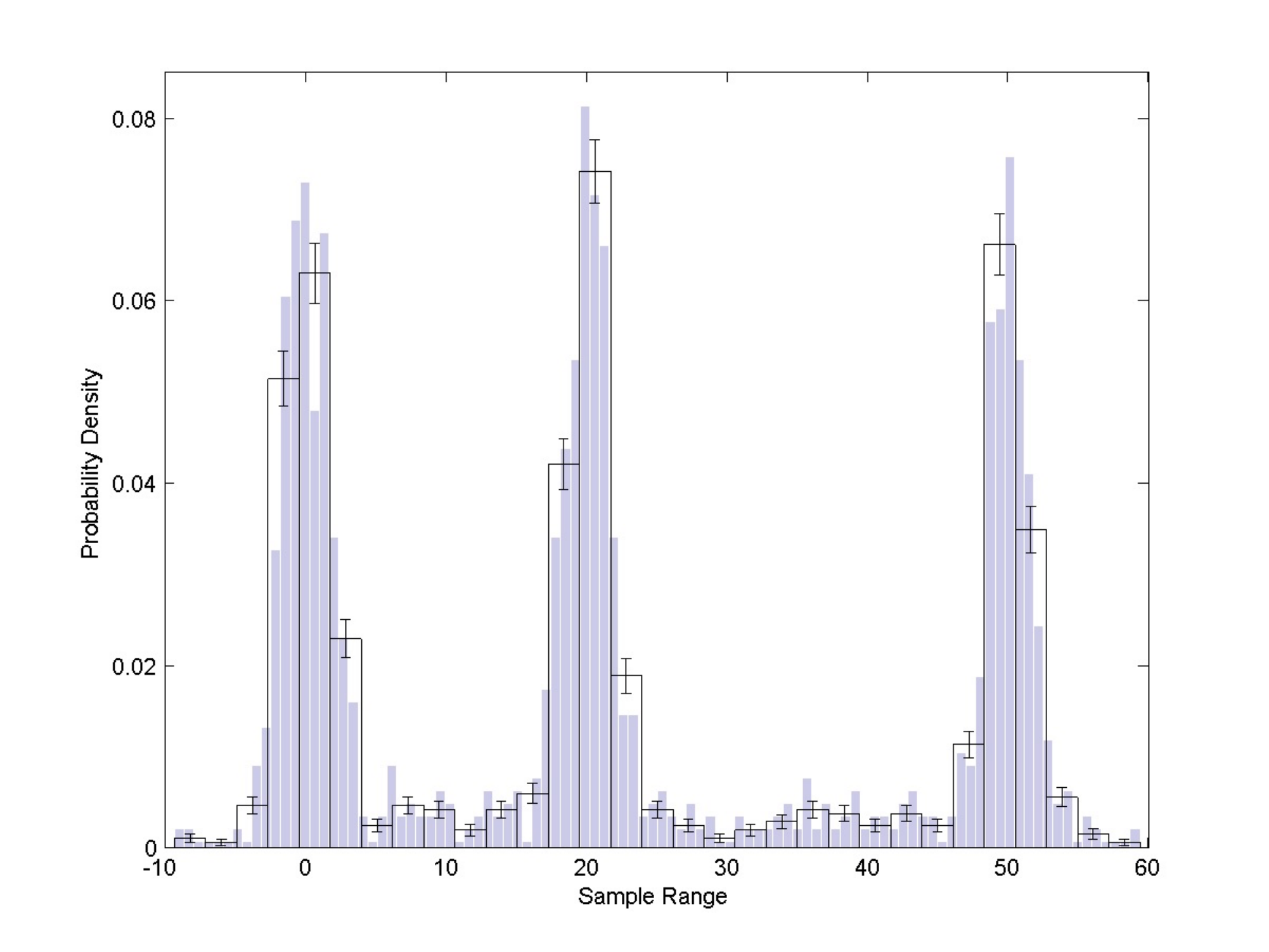}}\\
	\vspace{-0.3cm} 
	\caption{Application of Knuth method to four datasets sampled from known distributions: uniform (a,b), four-steps (c,d), Gaussian (e,f), three Gaussian peaks on a uniform background (g,h). On the left column we plot the graph of the relative log posterior and the point $ \hat{M}$ where it attains the maximum (Knuth optimal binning number). On the right column we arrange the data in an histogram with that number of bins. To highlight the fact that the method is able to avoid sampling fluctuations we plot, under each optimal histogram, the one we would get using 100 bins instead of $ \hat{M} $, which therefore better shows the spatial distributions of the data sets.}
	\label{fig:Knuth1D}
\end{figure*}

As in \citep{Knuth} we test Knuth's method with four sets of data, each consisting of 1000 one-dimensional points sampled from  known probability density functions. In order to do this we resort to $ MATLAB $ software and Knuth's \textit{OPTBINS binning package v1.0}. 
The results are shown in \figurename~\ref{fig:Knuth1D}.\\
For each set of data we plot the logarithm of the relative posterior given by eq.(8) of the main article and we then arrange data in an histogram with the found optimal bin number $ \hat{M} $. It represents the piece-wise constant density function of our data we wished to obtain. To highlight the fact that the method is able to avoid sampling fluctuations while it captures the main characteristics of the underlying density function, we plot, under each optimal histogram, the one we would get using 100 bins instead of $ \hat{M} $, which therefore better shows the spatial distributions of the data points.\\
Let us notice that the optimal binning numbers that we obtain in the four cases depend on the particular realisation of the process from which the data we are analysing are sampled. Anyway, as we will show later for CSR processes e.~g., Knuth's method resulted to be very stable, so that the results it gives slightly differ from a sample to the other.\\ 
The first case we consider is the sampling of 1000 data points from a uniform density. From \figurename~\ref{fig:Knuth1D}\subref{fig:RelLog_uniform}, we see that the relative log posterior reaches its maximum at $ \hat{M}=1 $. Knuth's histogram therefore coincides perfectly with the underlying density function.\\
In the second example, the set of data is sampled from a four-steps  density function. Again we can obtain it by using Knuth's algorithm, whose relative log posterior peaks at four bins.\\
In Figures \ref{fig:Knuth1D}\subref{fig:RelLog_gauss} and \subref{fig:Knuth_gauss} we show the results we obtain applying the method to a set of data sampled from a standard Gaussian density function $ \mathcal{N}(0,1) $.
As we can see from \figurename~\ref{fig:Knuth1D}\subref{fig:RelLog_gauss} the relative log posterior reaches its maximum at 12 bins, which capture very well the structure of the underlying density function.\\
Finally, in the last example, we sample 1000 data points from a density function consisting of three Gaussian peaks over a uniform background.
In this case the relative log posterior has a maximum for $ \hat{M}=31 $. Looking at \figurename~\ref{fig:Knuth1D}\subref{fig:RelLog_threegauss} we can notice that the algorithm gives an optimal binning number much higher than the ones obtained in the previous examples. This is due to the fact that, since we are imposing that the bins have equal width $ v $, we need an high number of bins in order to capture the real structure of the underlying density function and to identify the three peaks, in spite of the uniform background, usually arranged within only one bin, as we have seen in the first example.\\
This fact suggests that the number of bins given by the method could be a good parameter to distinguish clustered data from uniform ones, since the more structured is a species, the higher the number of bins we need to capture it correctly.\\
Moreover, we can notice that the relative log posterior decreases much slower than the previous ones, so that each value $ M\geq\hat{M} $ could be taken as the number of bins of our histogram in order to faithfully approximate the underlying density function. This is a general feature that we can observe when we are dealing with data distributions characterised by the presence of clusters, such as the three Gaussian peaks of the last example.   	
		
\section{Comparison with Stone}
\label{sec:Stone}
We now introduce a non-kernel method to select the optimal number of bins developed by Charles~J.~Stone in 1983 \citep{Stone}. We will then compare the two methods to show how Knuth one is more efficient in avoiding sample fluctuations.

\paragraph{Stone's optimal selection rule}
Let us assume we have a set of $ N $ data $ (\underline{d_i})\in \R^{n\times N}$, sampled from a known probability density function $ p $.\\
Let us denote with $ \underline{a}=(a_1,...,a_n)\in \R^n $ the point in the space where we wish to start building our histogram and with $ \underline{b}=(b_1,...,b_n)\in \R_+^n $ the dimension of the $ n $-dimensional bin we set. Our histogram will therefore depend on the pair $ (\underline{a},\underline{b}) $.  We wish now to compute the value of the piece-wise constant probability density of the histogram in the $ \underline{k}^{th} $ bin, whose integer coordinates, starting from $ \underline{a} $, are $\underline{k}= (k_1,...k_n)\in\mathbb{Z}^n $ and which therefore occupies, in $ \R^n $, the place
\begin{linenomath}
\begin{equation}
	\begin{split}
		I_{(\underline{a},\underline{b}),\underline{k}}=(a_1+(k_1-1)b_1,\ & a_1+k_1b_1)\times...\\
		&\times (a_n+(k_n-1)b_n,\ a_n+k_nb_n).
	\end{split}
\end{equation}
\end{linenomath}
We have obtained a partition of $ \R^n $ into identical bins of dimension $ v_{(\underline{a},\underline{b})}=\prod_{i=1}^{n} b_i$.
The probability mass of the $ \underline{k}^{th} $ bin, i.e. the volume of the column over it, is given by the empirical distribution
\begin{linenomath}
\begin{equation}
	\pi_{(\underline{a},\underline{b}),\underline{k}}=\dfrac{1}{N}\left\vert\{i| 1\leq i\leq N,\:\underline{d_i}\in I_{(\underline{a},\underline{b}),\underline{k}}\}\right\vert.
\end{equation} 
\end{linenomath}
From this we can compute the height of the column, which is simply
\begin{linenomath}
\begin{equation}
	h_{(\underline{a},\underline{b}),\underline{k}}=\frac{\pi_{(\underline{a},\underline{b}),\underline{k}}}{v_{(\underline{a},\underline{b})}}.
\end{equation}
\end{linenomath}
The probability density function of the histogram is therefore given by 
\begin{linenomath}
\begin{equation}
	h_{(\underline{a},\underline{b})}(x)=\sum_{\underline{k}} h_{(\underline{a},\underline{b}),\underline{k}}\chi_{I_{(\underline{a},\underline{b}),\underline{k}}}(x), 
\end{equation}
\end{linenomath}
where $ \chi_{I_{(\underline{a},\underline{b}),\underline{k}}} $ is the characteristic function of $ I_{(\underline{a},\underline{b}),\underline{k}} $, defined as
\begin{linenomath}
\begin{equation}
	\chi_{I_{(\underline{a},\underline{b}),\underline{k}}}(x)=
	\begin{system}
		0\ $ if $\ x\notin I_{(\underline{a},\underline{b}),\underline{k}} \\
		1\ $ if $\ x\in I_{(\underline{a},\underline{b}),\underline{k}}
	\end{system}
\end{equation}
\end{linenomath}
Similarly we can define the probability mass as follows:
\begin{linenomath}
\begin{equation}
	\pi_{(\underline{a},\underline{b})}(x)=\sum_{\underline{k}}\pi_{(\underline{a},\underline{b}),\underline{k}}\chi_{I_{(\underline{a},\underline{b}),\underline{k}}}.
\end{equation}
\end{linenomath}
In order to best approximate the density function from which the data were sampled, Stone's idea relies on minimising the integrated squared error of  $ h_{(\underline{a},\underline{b})} $
\begin{linenomath}
\begin{equation}
	\label{error_stone}
	E_{(\underline{a},\underline{b})}=\int_{\R^n}(h_{(\underline{a},\underline{b})}(\underline{x})-p(\underline{x}))^2 d\underline{x}.
\end{equation}
\end{linenomath}
Under some conditions, (see \citep{Stone}), minimising (\ref{error_stone}) is asymptotically equivalent to minimising 
\begin{linenomath}
\begin{equation}
	K_{(\underline{a},\underline{b})}=\frac{1}{v_{(\underline{a},\underline{b})}}\biggl(\frac{2}{N}-\sum_{\underline{k}}\pi_{(\underline{a},\underline{b}),\underline{k}}^2\biggr).
\end{equation}
\end{linenomath}

\begin{figure*}
	\centering
	\vspace{-0.8cm}
	\subfloat[{}\label{fig:KvsS_RelLog}]
	{\includegraphics[width=70mm]{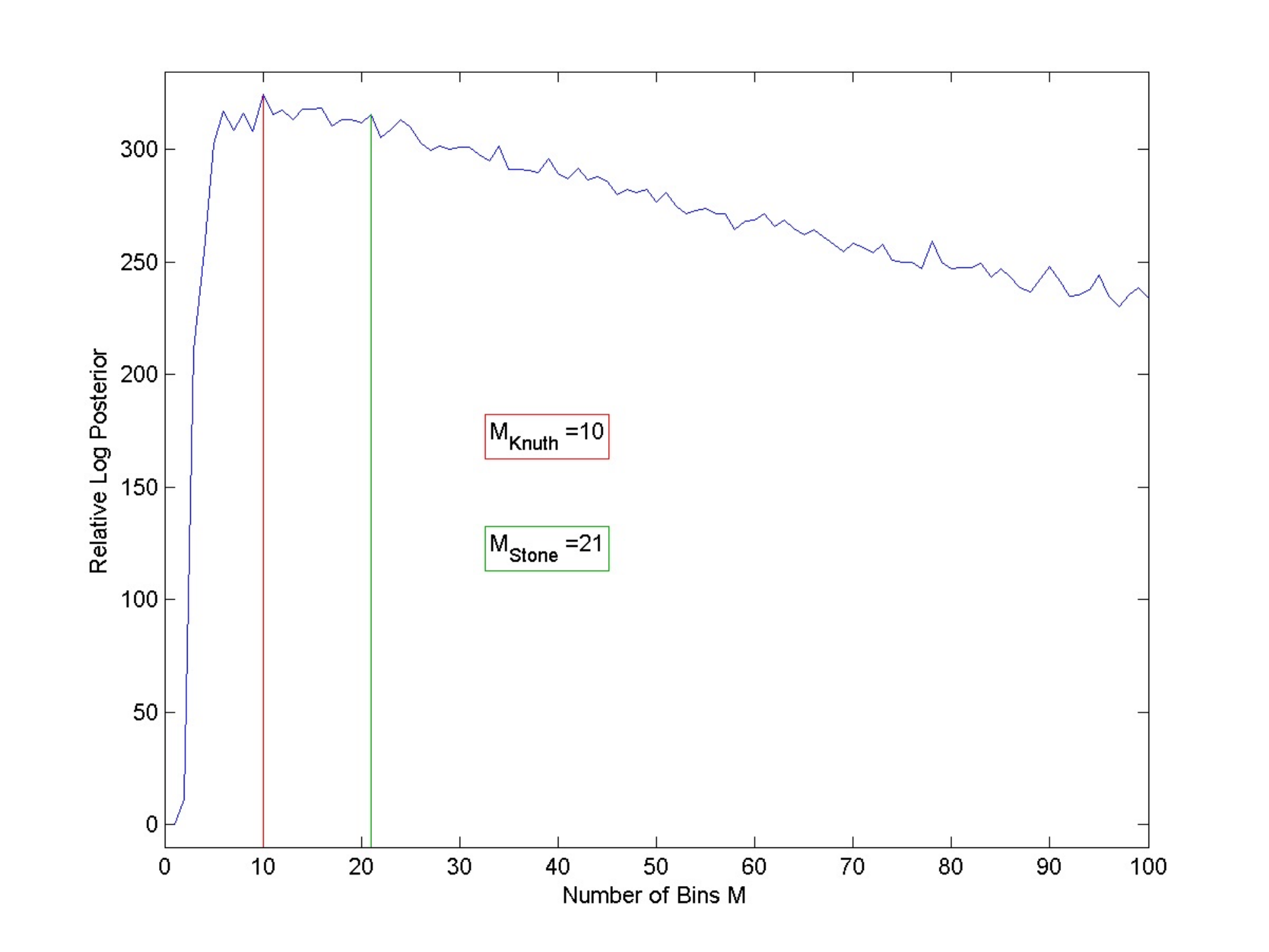}}
	\subfloat[{}\label{fig:KvsS_100bin}]
	{\includegraphics[width=70mm]{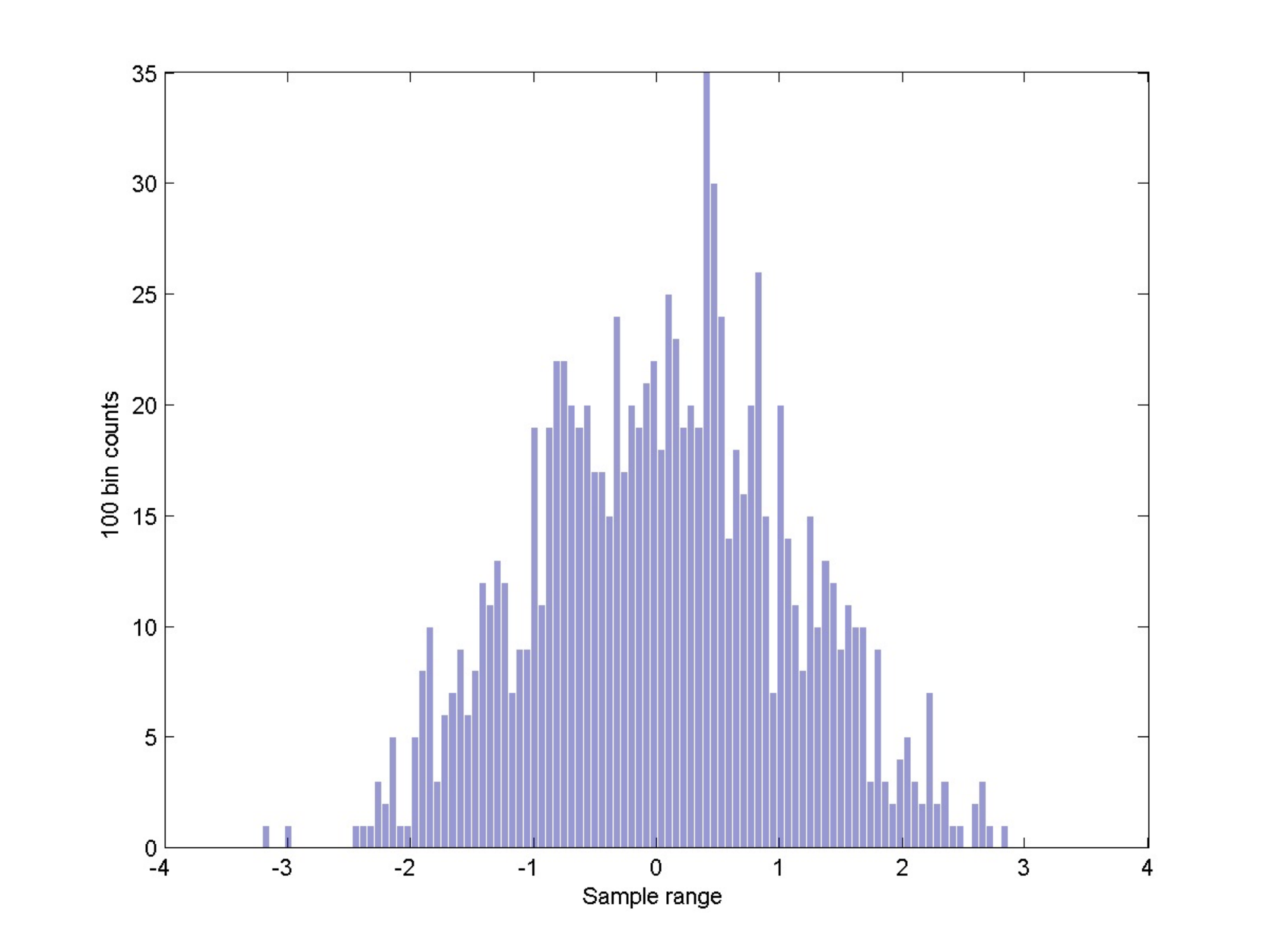}}\\
	\vspace{-0.5cm} 
	\subfloat[{}\label{fig:KvsS_Stone}]
	{\includegraphics[width=70mm]{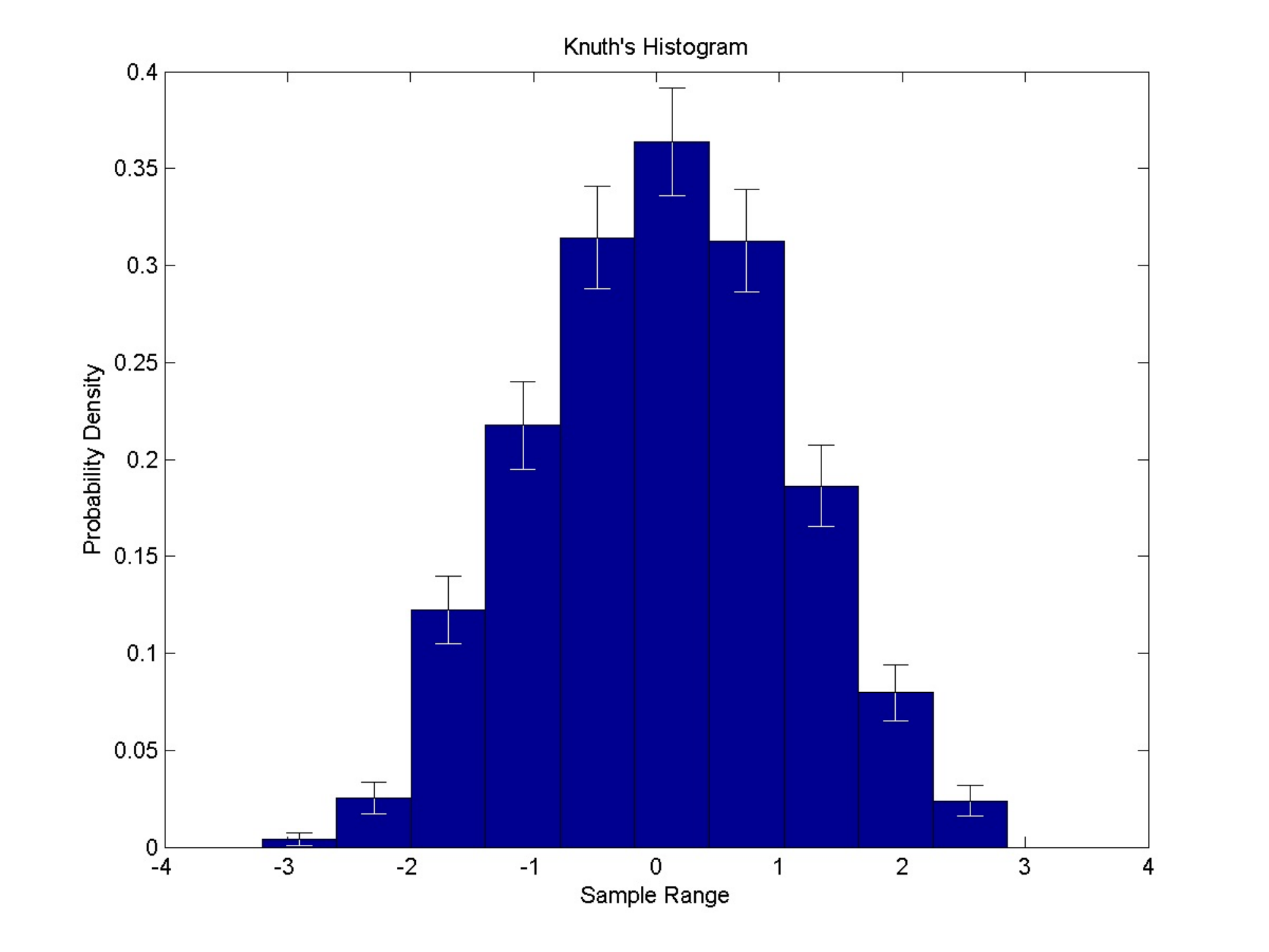}}
	\subfloat[{}\label{fig:KvsS_Knuth}]
	{\includegraphics[width=70mm]{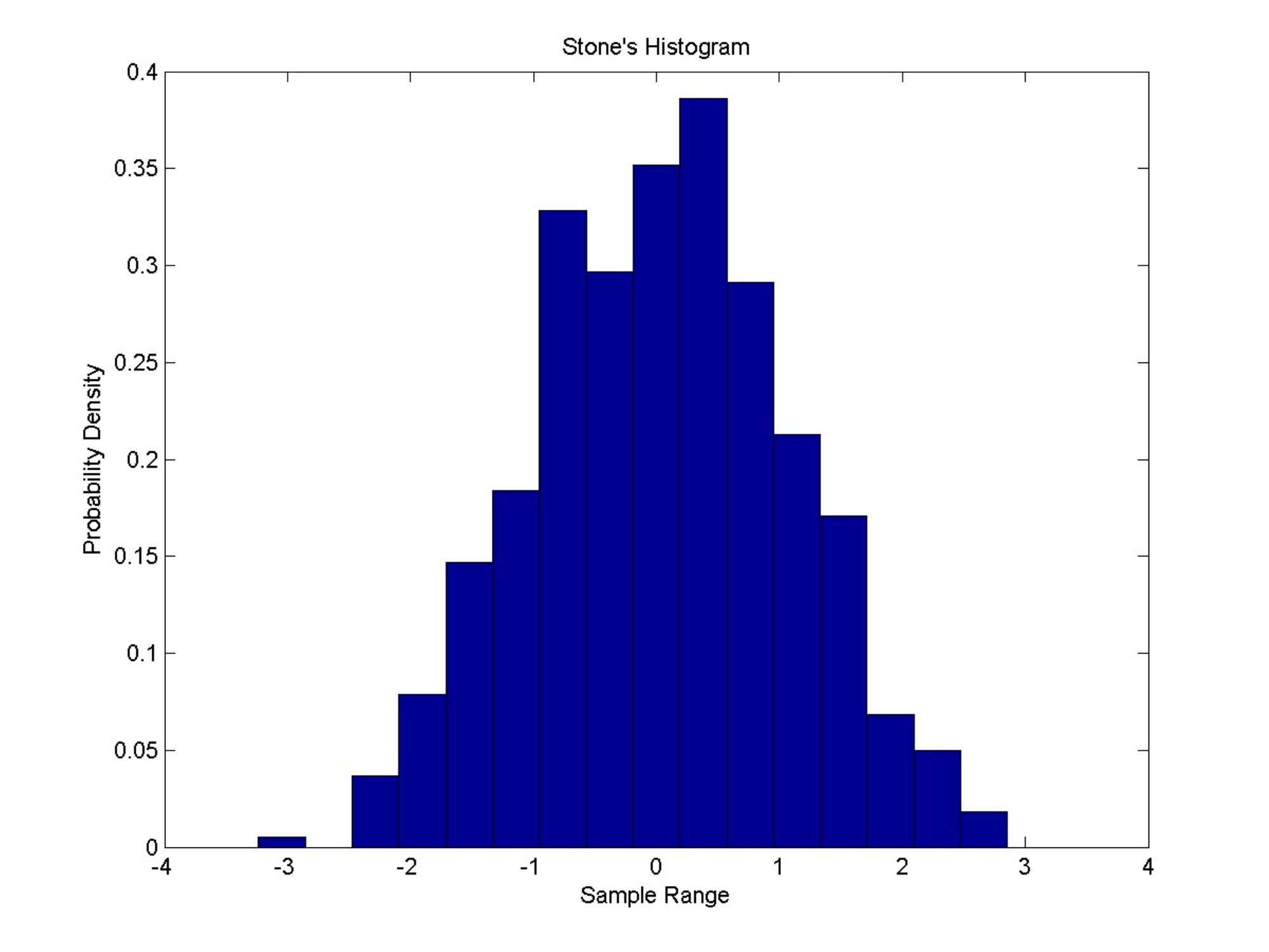}}\\
	\vspace{-0.5cm} 
	\subfloat[{}\label{fig:dist_L1}]
	{\includegraphics[width=70mm]{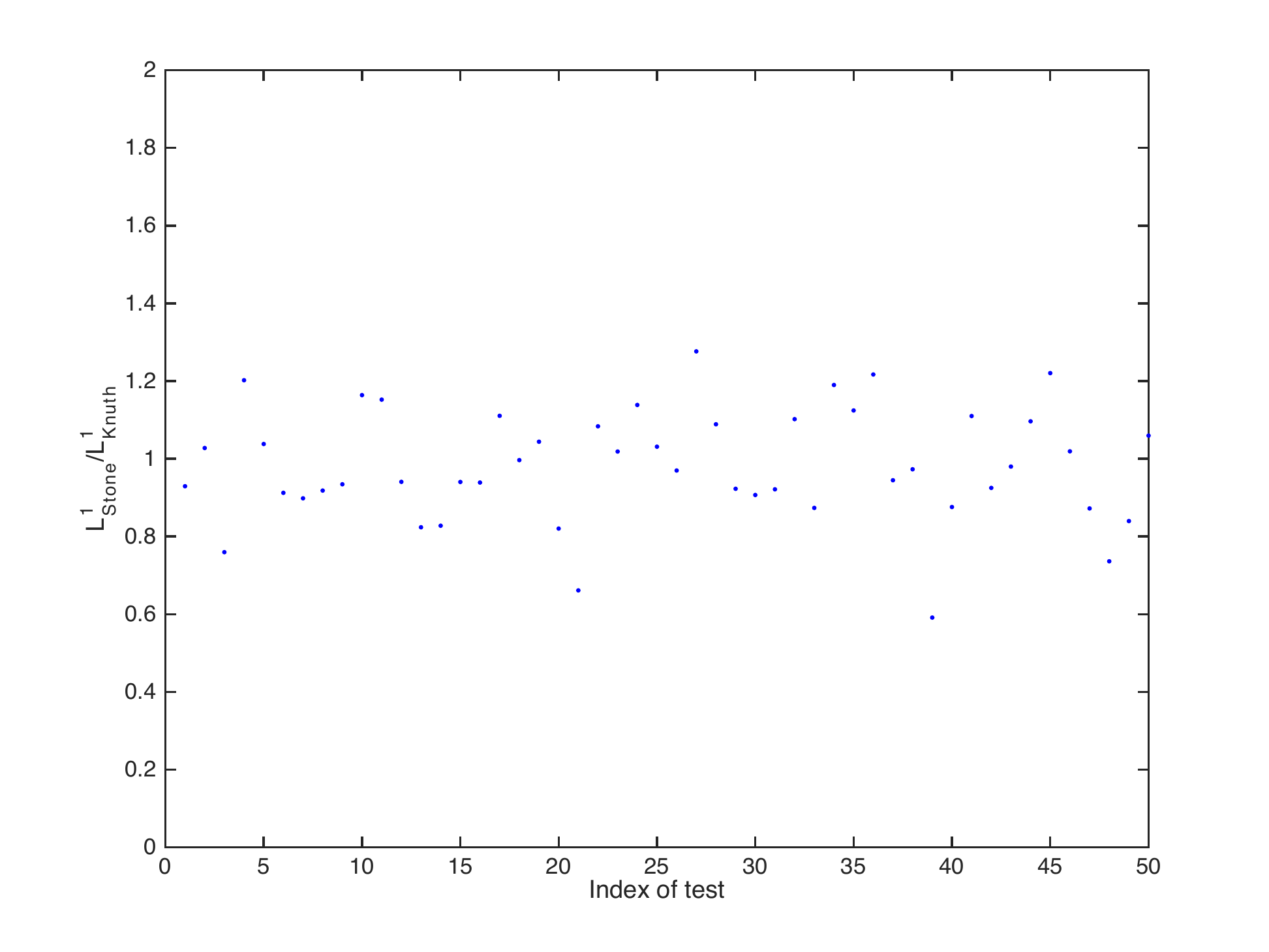}}
	\subfloat[{}\label{fig:dist_L2}]
	{\includegraphics[width=70mm]{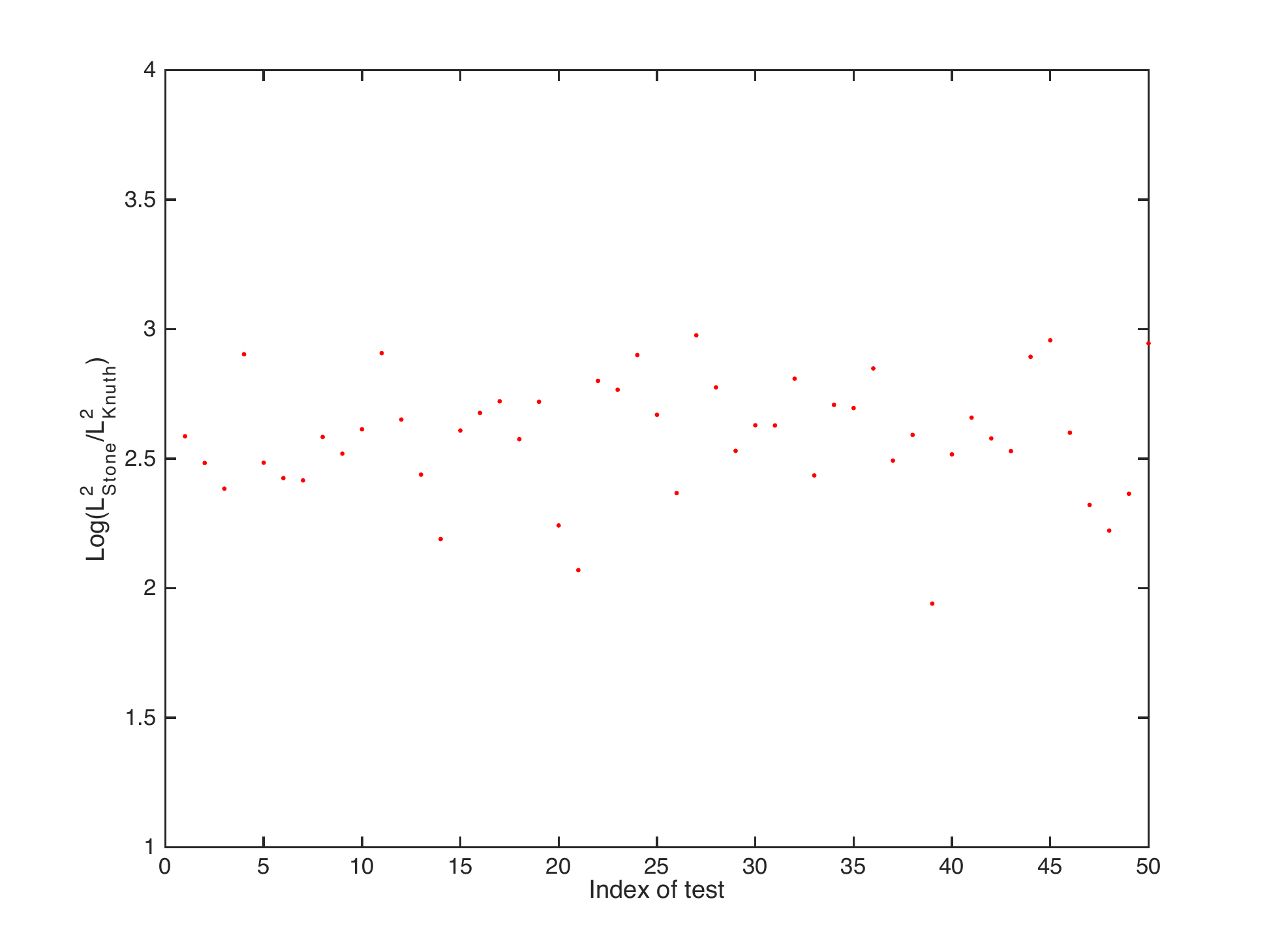}}
	\vspace{-0.3cm} 
	\caption{{$ L^1 $ and $ L^2 $-comparisons between Stone's and Knuth's method on datasets generated from a Gaussian distribution. In (a-d) we see an example of Knuth and Stone answer to the same dataset. As we can see from the histograms, contrarily to Stone,  Knuth is able to avoid sample fluctuation. In (e-f) we see the ratio between Stone and Knuth's histogram distances from the underlying density function.}\label{fig:L1_L2_distance}}
\end{figure*}

\paragraph{Comparison to one-dimensional datasets}
Let us firstly remark that the condition above is satisfied, for example, if there is some non-empty open subset of $ \R $ on which the derivative of $ p $ exists, it is continuous and non-zero \citep{Stone}. As a consequence, if we are dealing with uniform or step density functions, such as we did in the previous section with Knuth, we cannot apply Stone's optimal selection rule, which has therefore a smaller range of applications.
This is a drawback especially when working with a real dataset sampled from an unknown density function, since we cannot exclude a priori such densities.\\
We have compared the two methods in the one-dimensional case using $ L^1 $ and $ L^2 $ distance method \citep{distances}.\\
We have generated 50 datasets each consisting of 1000 one-dimensional points sampled from a standard Gaussian density function $\mathcal{N}(0,1) $. For each test, we have computed Knuth and Stone optimal binning number and then we have constructed the correspondent histograms. We finally have compared the $ L^1 $ and $ L^2 $ distances between these and the underlying density function.\\
In \figurename~\ref{fig:L1_L2_distance}\subref{fig:dist_L1} we plot the ratio between Stone's and Knuth's $ L^1 $ distance, while in \figurename~\ref{fig:L1_L2_distance}\subref{fig:dist_L2} we plot the logarithm of the ratio between Stone's and Knuth's $ L^2 $ distance.\\
As we can see in \figurename~\ref{fig:L1_L2_distance}\subref{fig:dist_L1}, while the two methods are practically equivalent in the first case, Knuth's $ L^2 $ distance is about ten times smaller than Stone's one (\figurename~\ref{fig:L1_L2_distance}\subref{fig:dist_L2}), due to the fact that the latter is more sensitive to sample fluctuations Knuth is able to avoid instead. The result may be surprising, since Stone's optimal selection rule relies on minimizing exactly the $ L^2 $ distance between the underlying density function and the one obtained by the histogram. The matter is that the rule only assures that this minimum is reached when the number of data tends to infinity. In fact, it is in this asymptotic case that the sampling fluctuations diminish and thus the histogram recalls quite faithfully the distribution from which the data are sampled.\\
For the sake of completeness, we also insert the graphics we obtain testing the two methods on one dataset such as the ones described above (a Gaussian). From \figurename~\ref{fig:L1_L2_distance}\subref{fig:KvsS_Knuth} and \subref{fig:KvsS_Stone}, it is clear that while Knuth's histogram correctly captures the main characteristics of the underlying probability density function, Stone is more sensitive to sampling fluctuations, which results in a higher optimal binning number.

\section{Knuth sensitivity to uniform distribution}
\begin{figure}
	\centering
	\includegraphics[width=7cm]{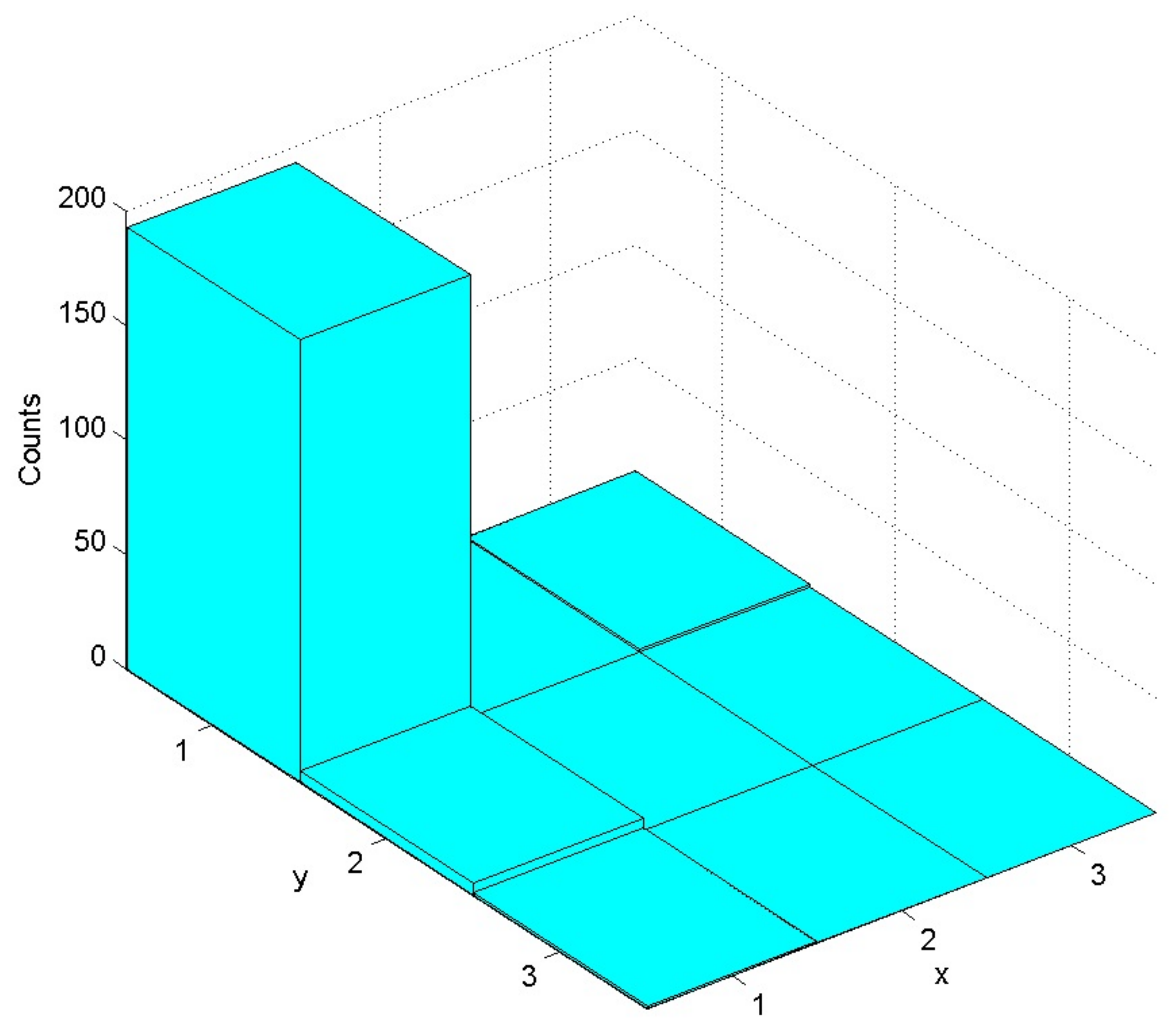}\vspace*{3mm}
	\caption{Histogram of the optimal binning number obtained with Knuth's algorithm tested 200 times on a uniformly distributed dataset}\label{fig:hist_200_uniform}
\end{figure}

We firstly test how sensitive and robust is the algorithm with respect to the dataset in order to see if the answer that it gives depends from it rather than from the distribution from which it has been sampled.\\
With this goal in mind, we test Knuth algorithm's stability by seeing if, generating more times a dataset from a known probability density function, the result is always the same. In particular, we generate a dataset consisting of 1000 points within a $ 500\times500 $~units sampled from a uniform density function. We therefore computed, for each generated dataset, Knuth optimal binning number.\\
From Section~\ref{sec:Knuth_method}, we know that for a one-dimensional set of data uniformly distributed, the answer Knuth's algorithm gives is $ \hat{M}=1 $, which correctly captures the underlying distribution. We wish to see if the answer for the two-dimensional datasets is the same for almost every test, which would imply that the algorithm is stable with respect to the density function from which we sample the data. We arrange our results in a $ 3\times3 $ histogram, since the answers given by Knuth algorithm lie in this range.\\
From \figurename~\ref{fig:hist_200_uniform} we can see that for almost all of the 200 tests we performed the optimal binning number results to be $ M=[1\ 1], $ so that the probability density function we get thanks to Knuth's  histogram coincides perfectly with the one from which the data were sampled. This fact provides that the optimal binning number that we get with Knuth's algorithm for each species strictly depends on their distribution rather than on the data itself.

\section{Poisson Cluster Process}
Let us recall the definition of the modified Thomas process:
\begin{itemize}
	\item \emph{Parents} are distributed according to a Poisson process with intensity $ \rho $. 
	\item To each \emph{parent} a random number of \emph{offspring} is assigned, drawn from a Poisson distribution of intensity $\mu$.
	\item For each \emph{parent}, its \emph{offspring} is located according to a two-dimensional Gaussian distribution centred at the location of the \emph{parent} in the plot and with standard deviation $ \sigma. $
	\item The \emph{parents} are removed from the plot, so that the clumped pattern is formed by \emph{offspring} only.
\end{itemize}
Let us notice that with this model we are assuming that the formation of clusters is due to an isotropic local propagation of \emph{offspring} from the \emph{parent}, which is clearly an oversimplification of the complex natural mechanisms which actually determines their rise.\\ In order to compute the process parameters $(\rho,\sigma,\mu)$ from a dataset, we computed the empirical $K$-Ripley function, which, for a spatial Poisson point process with intensity $ \lambda $, is defined by
\begin{linenomath}
\begin{equation}
\label{Kd}
K(d)=\dfrac{1}{\lambda}\langle \text{number of extra events within a distance \textit{d} from an arbitrary event}\rangle.
\end{equation}
\end{linenomath}
In our case, the arbitrary event is the location of any individual belonging to the species. If this latter is randomly distributed, then, since the mean number of stems within a circle of radius $ d $ is $ \lambda\pi d^2 $, we get that for a random placement model (\emph{RPM}) $ K_{RPM}(d)=\pi d^2. $\\
Let us therefore consider a species with $n$ individuals whose location in a plot of area $A_0$ are $ s_1,...,s_{n} $ and let us define with
\begin{linenomath}
\begin{equation}
\hat{\lambda}=\dfrac{n}{A_0},
\end{equation}
\end{linenomath}
the estimator of the intensity $ \lambda $ for the  species. Let us then denote with $ w(s_i,s_j) $ the proportion of the circumference of the circle with centre $ s_i $ passing through $ s_j $ which lies in the plot. Lastly, let us define the indicator function 
\begin{linenomath}
\begin{equation}
II(||s_i-s_j||\leq d)=
\begin{system}
1\ \ \ \text{if}\ ||s_i-s_j||\leq d\\
0\ \ \ \text{otherwise}
\end{system}
\end{equation}
\end{linenomath}
where $ ||\cdot|| $ indicates the euclidean distance between the individuals located in $ s_i $ and $ s_j $ respectively. The canonical edge-corrected estimator of Ripley's $ K $ function is then \citep{Plotkin}
\begin{linenomath}
\begin{equation}
\hat{K}(d)=\dfrac{1}{\hat{\lambda}}\sum_{i=1}^{n}\sum_{j=1, j\neq i}^{n}\dfrac{1}{w(s_i,s_j)}\dfrac{II(||s_i-s_j||\leq d)}{n}.
\end{equation}
\end{linenomath}
Defining with $ d_{max} $ the larger distance at which we measure a cluster within the area $ A_0 $ of our rainforest, we evaluate the edge-corrected estimator between $ 0 $ to $ d_{max} $.\\ 
In order to get the empirical pairs $ (\rho,\sigma) $ for a generated dataset, we need to compute the Ripley's $ K $ function for the modified Thomas process with parameters $ (\rho,\sigma) $, which is given by \citep{Cressie}
\begin{linenomath}
\begin{equation}
K(d)_{mTp}=\pi d^2+\dfrac{1}{\rho}\bigg(1-\exp\bigg(-\dfrac{d^2}{(2\sigma)^2}\bigg)\bigg).
\end{equation}
\end{linenomath}
We therefore have to choose the pair $ (\rho,\sigma) $ such that the above function best fits the empirical values $ \hat{K}(0),..., \hat{K}(d_{max}), $ where we set $d_{max}=300$ m as in \citep{Morlon}.\\
We do this by the \emph{Method of Minimum Contrast} \citep{Diggle}
\begin{linenomath}
\begin{equation}
\int_{0}^{d_{max}}(\hat{K}(h)^{\frac{1}{4}}-K(h)_{mTp}^{\frac{1}{4}})^2 \text{d}h
\end{equation}
\end{linenomath}
Once obtained the parameters $ (\rho,\sigma) $ from the dataset, we generated a Poisson cluster model by the following four steps:
\begin{itemize}
	\item{We simulate the Poisson cluster process by placing $ \lfloor \rho\times A_0+1/2\rfloor $} \emph{parents} randomly distributed within the plot. 
	\item{We randomly assign each of the $ n $ individuals of the species to one of the previously generated parents.}
	\item {For each parent, we locate the associated stems according to a two-dimensional Gaussian distribution centred at the location of the \emph{parent} in the plot and with variance $ \sigma. $ If the case that the \emph{offspring} falls out of the plot, we impose toroidal boundary conditions.}
	\item{We remove the parents from the plot, so that only the \emph{offspring} remains within it.}
\end{itemize}

\section{Knuth answer for different clustered structures}
\begin{figure*}
	\centering
	\vspace{-0.8cm}
	\includegraphics[width=.30\textwidth]{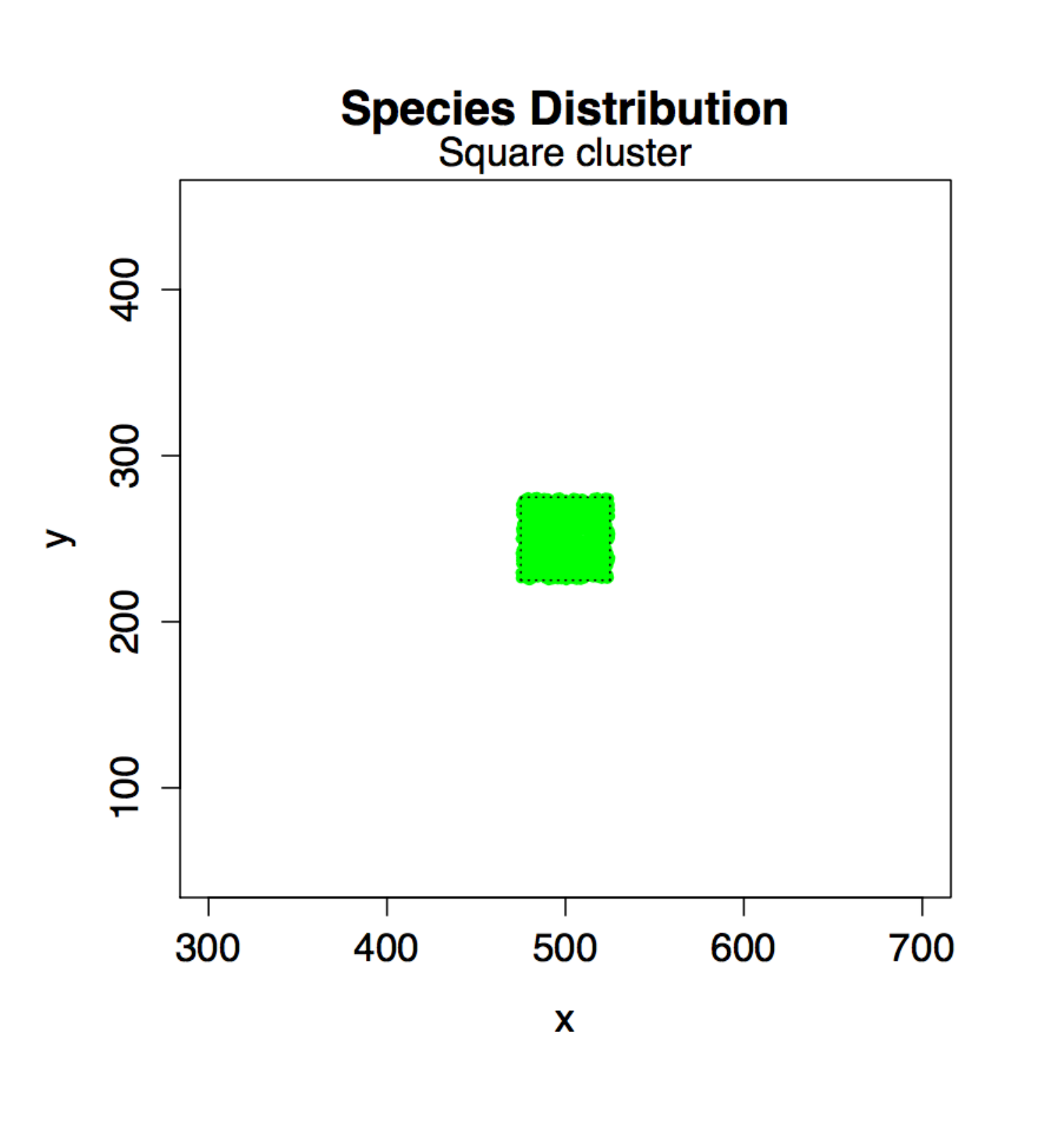}
	\includegraphics[width=.30\textwidth]{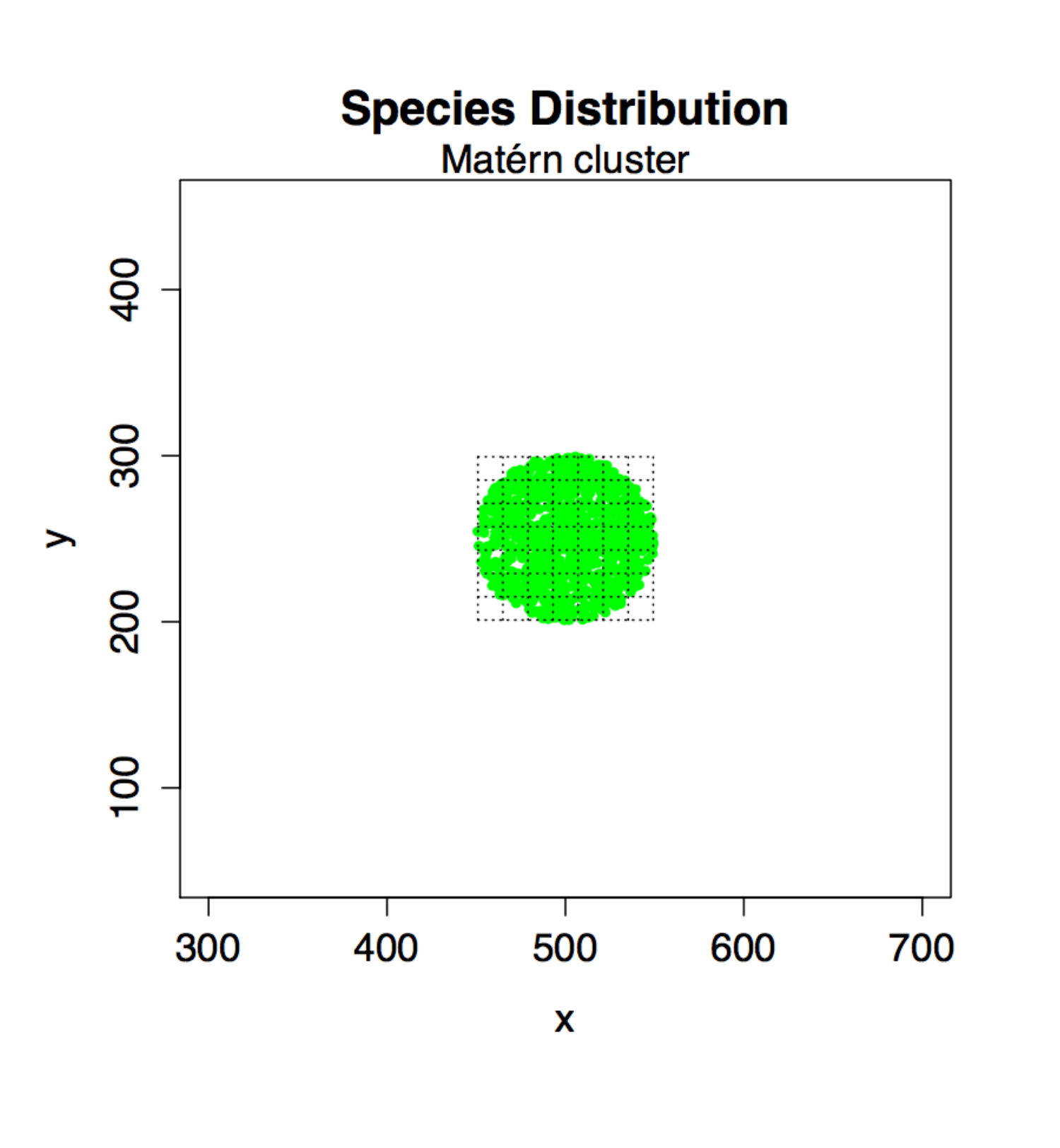}
	\includegraphics[width=.30\textwidth]{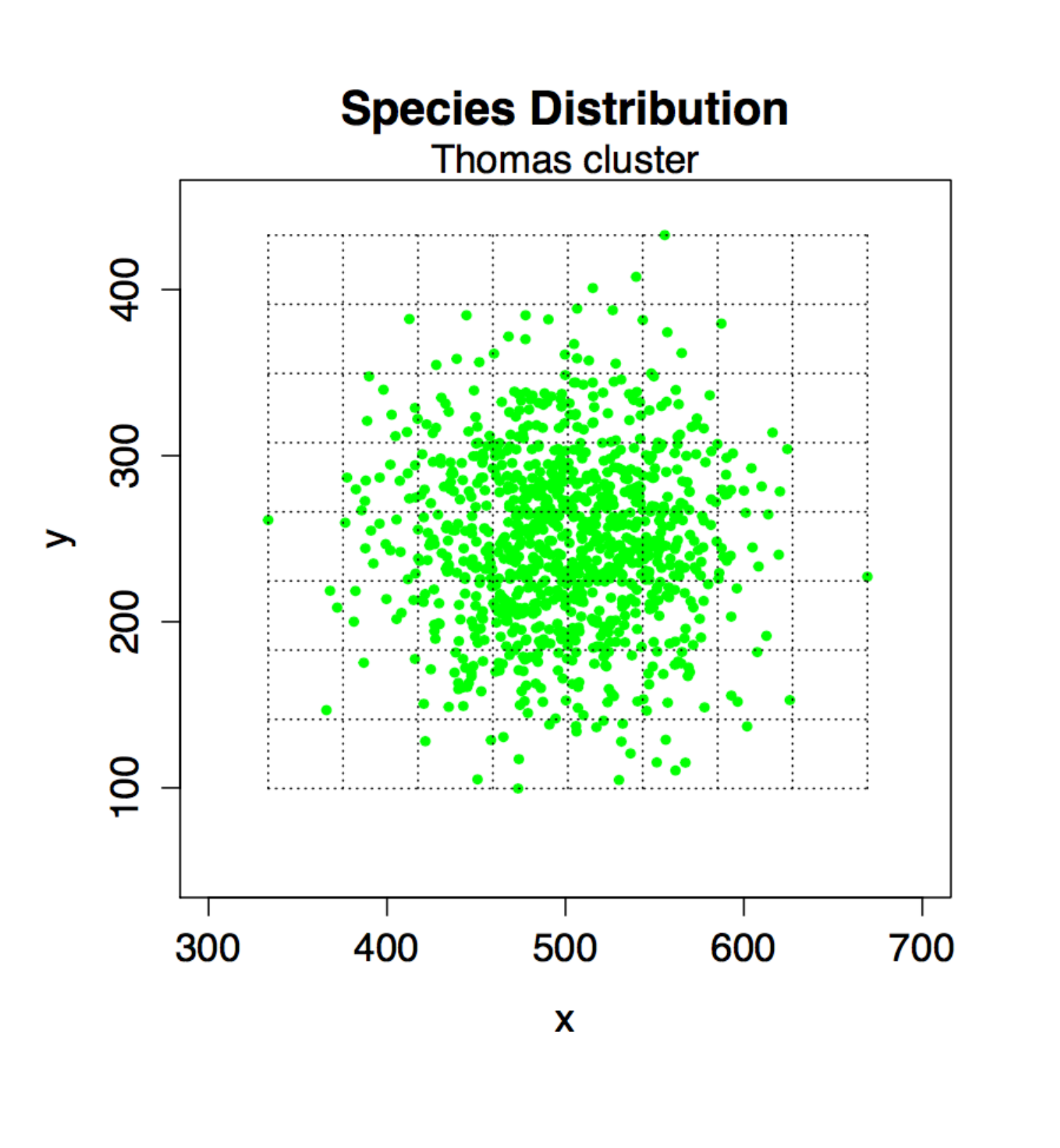}\\
	\vspace*{-0.3cm}
	\includegraphics[width=.30\textwidth]{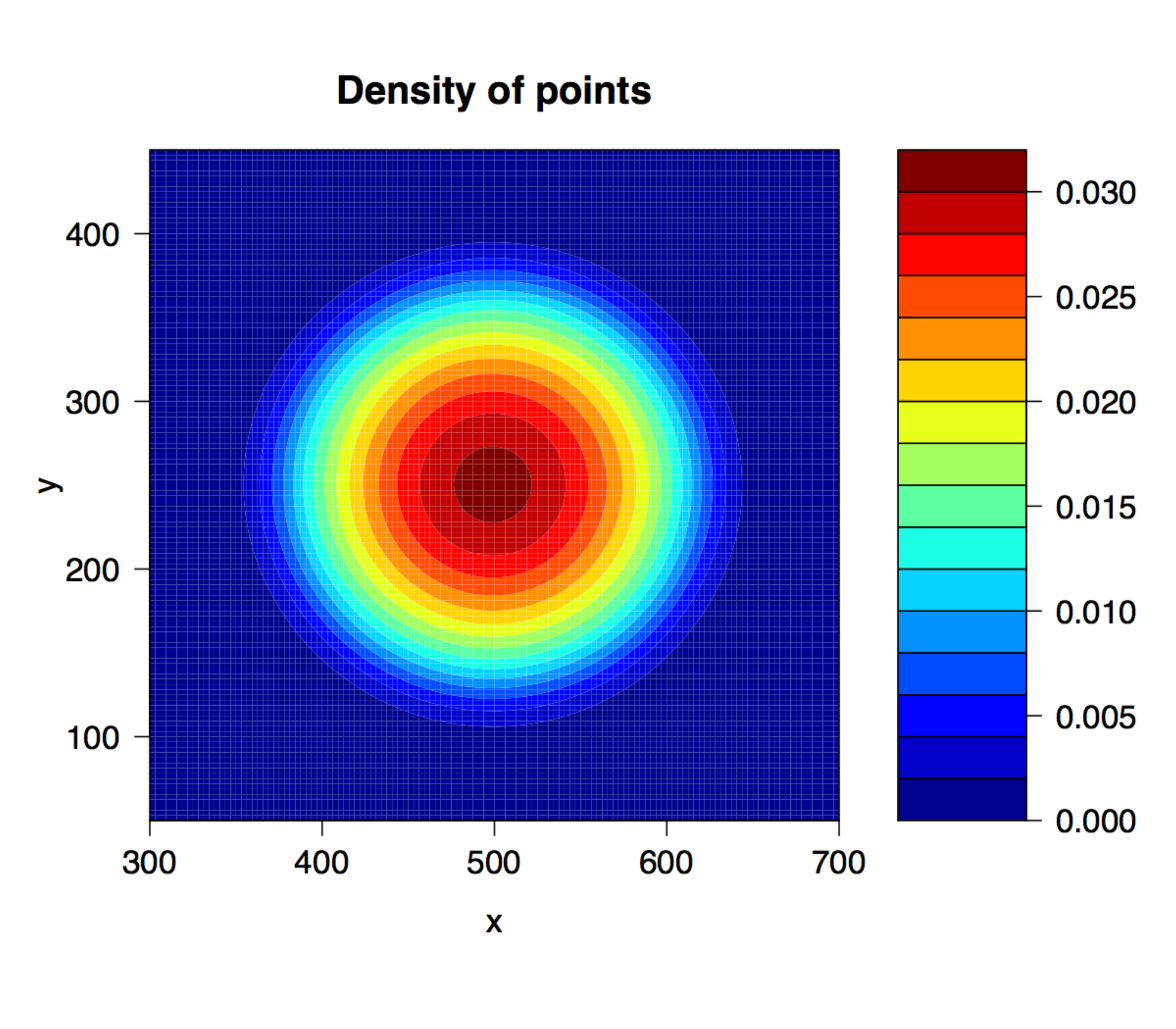}
	\includegraphics[width=.30\textwidth]{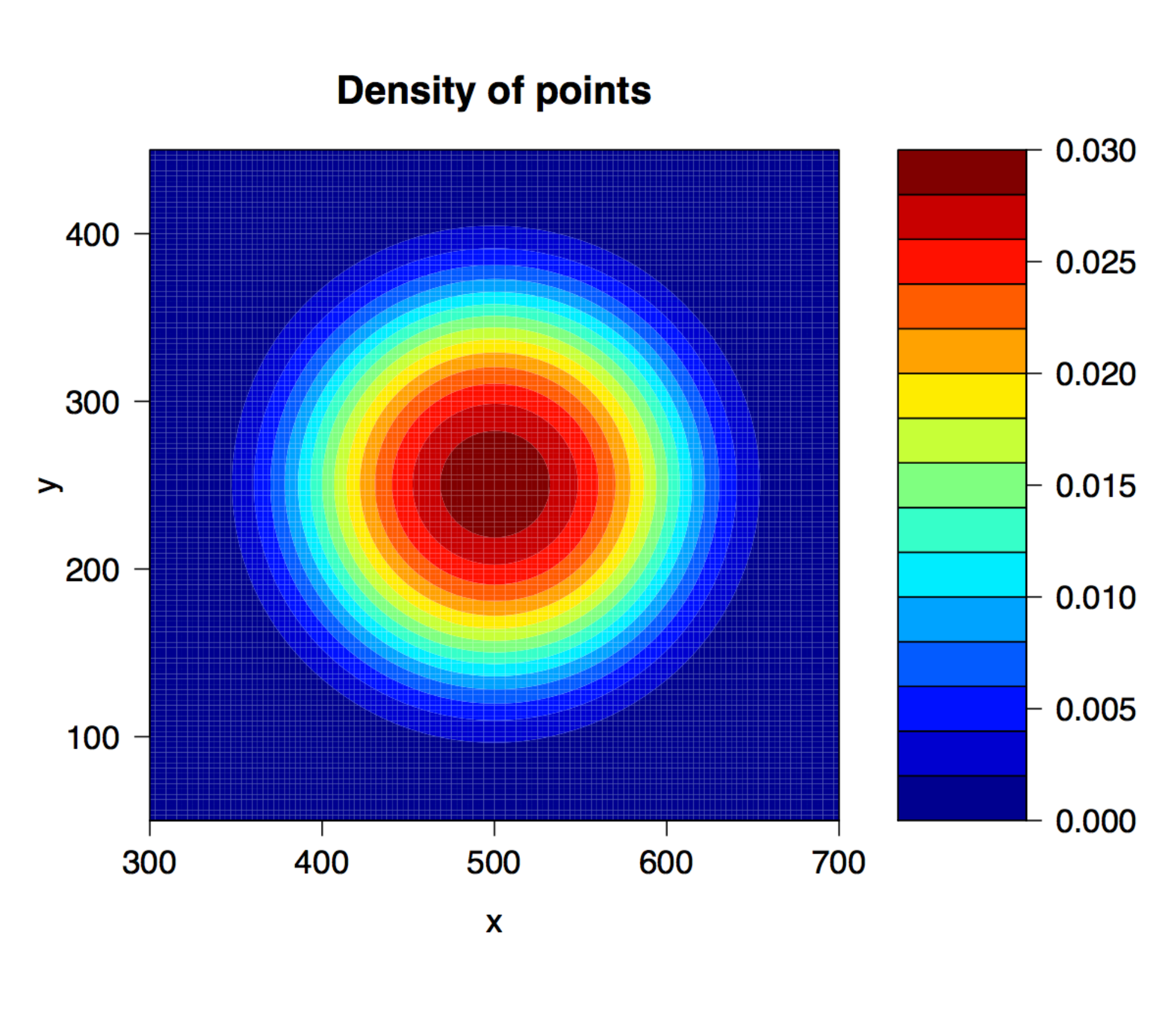}
	\includegraphics[width=.30\textwidth]{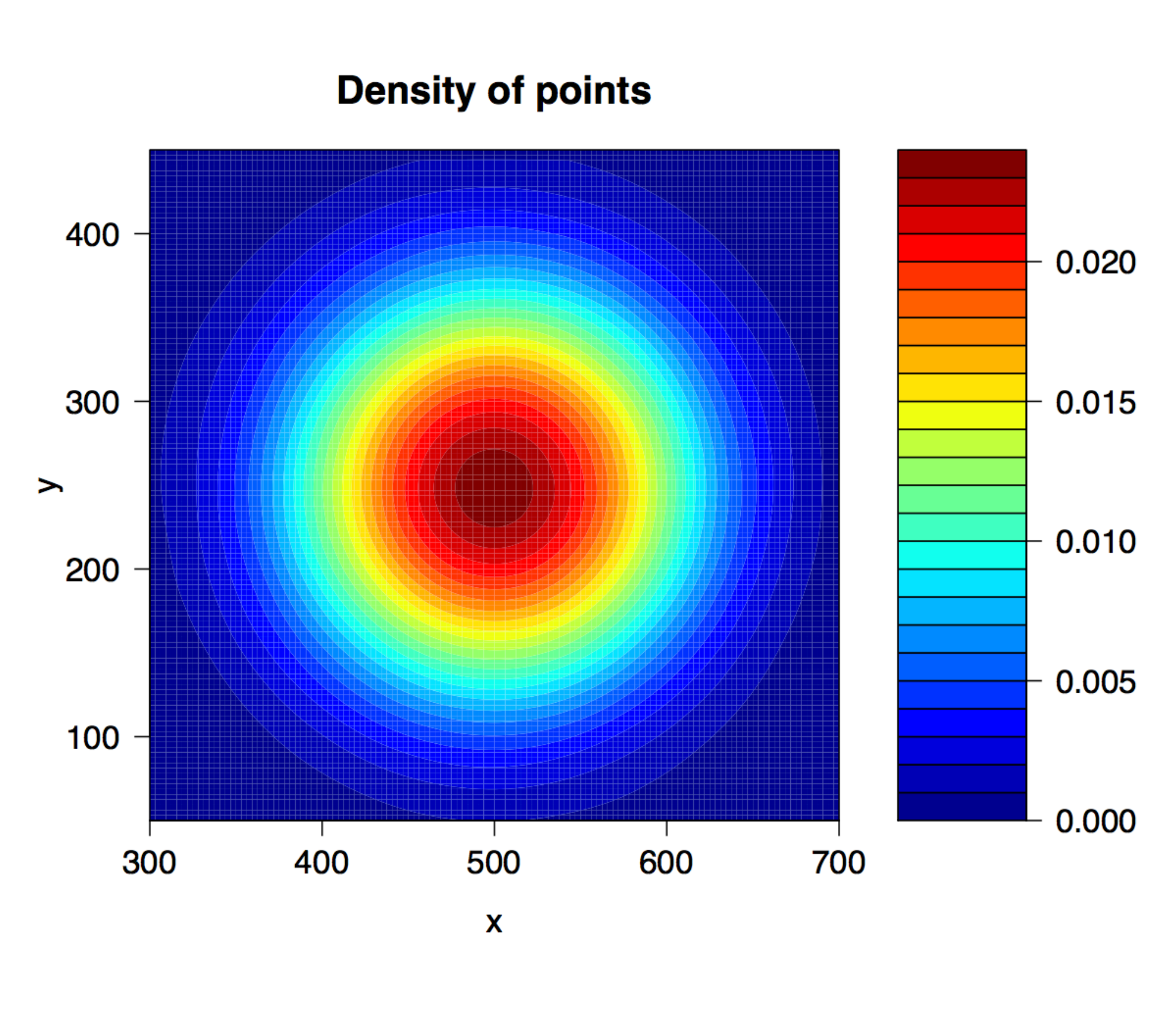}\\
	\includegraphics[width=.30\textwidth]{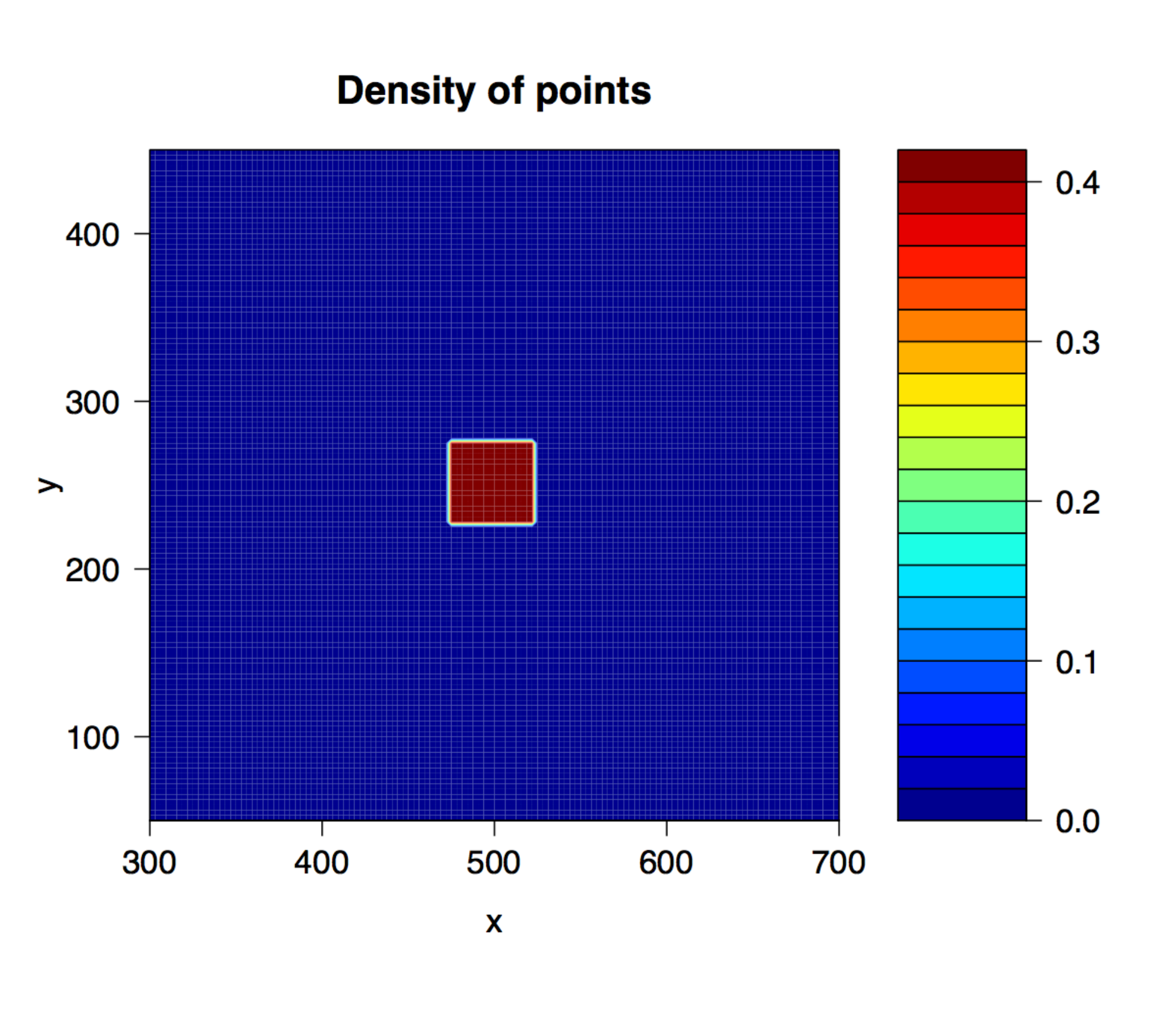}
	\includegraphics[width=.30\textwidth]{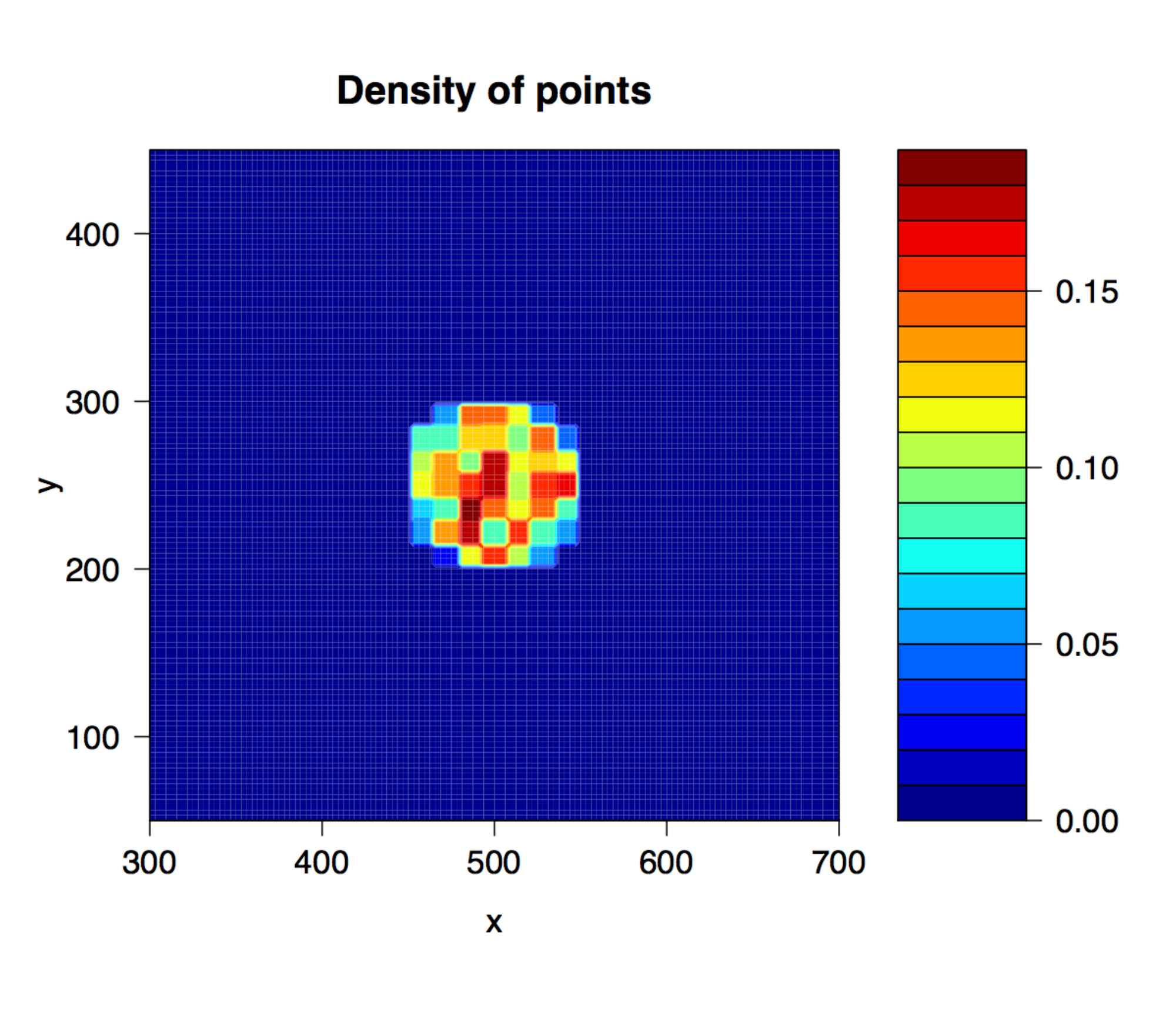}
	\includegraphics[width=.30\textwidth]{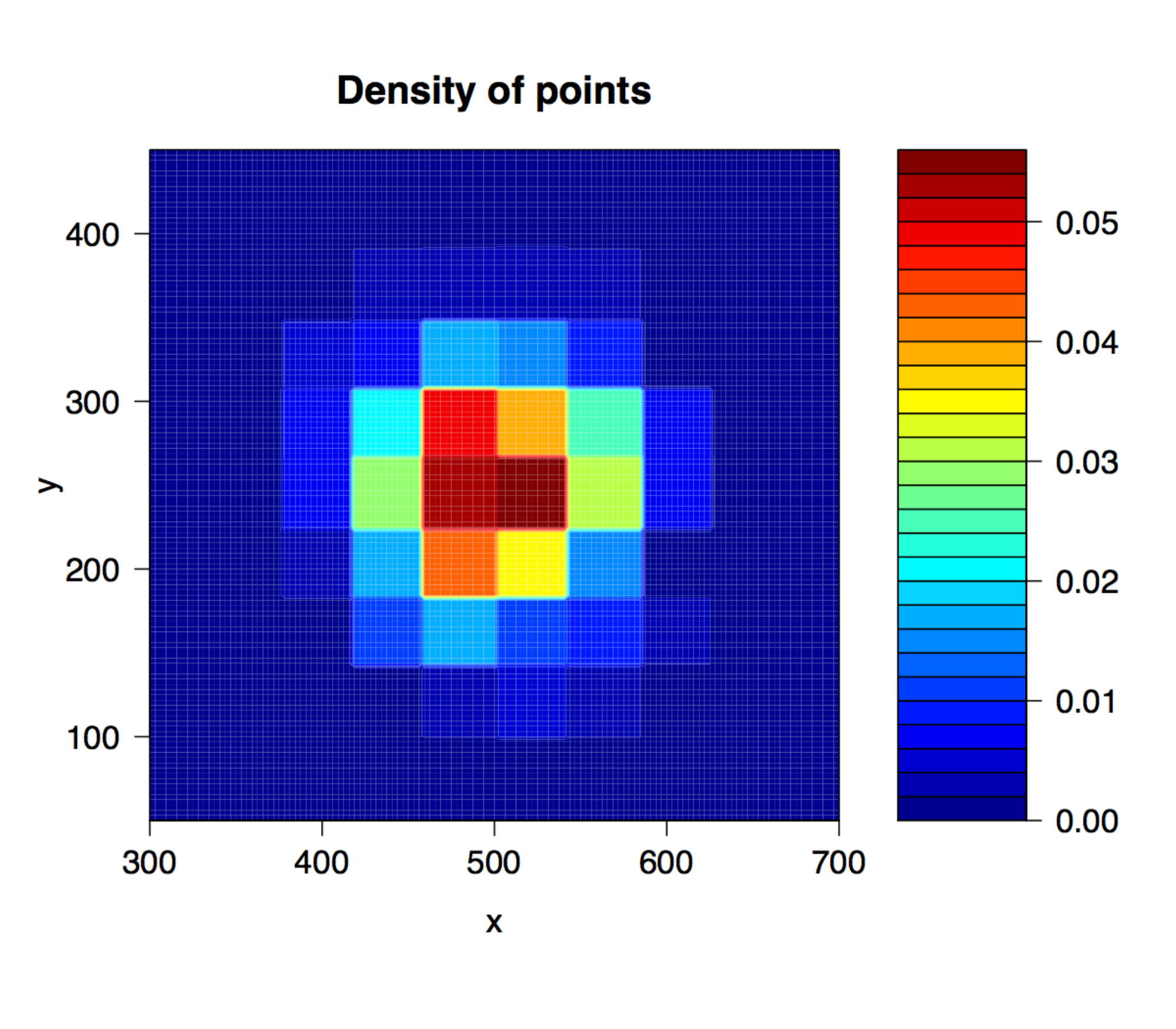}\\
	\vspace{-0.7cm}
	\hspace{-0.8cm}
	\subfloat[][{\label{fig:hist_square}}]
	{\includegraphics[width=.30\textwidth]{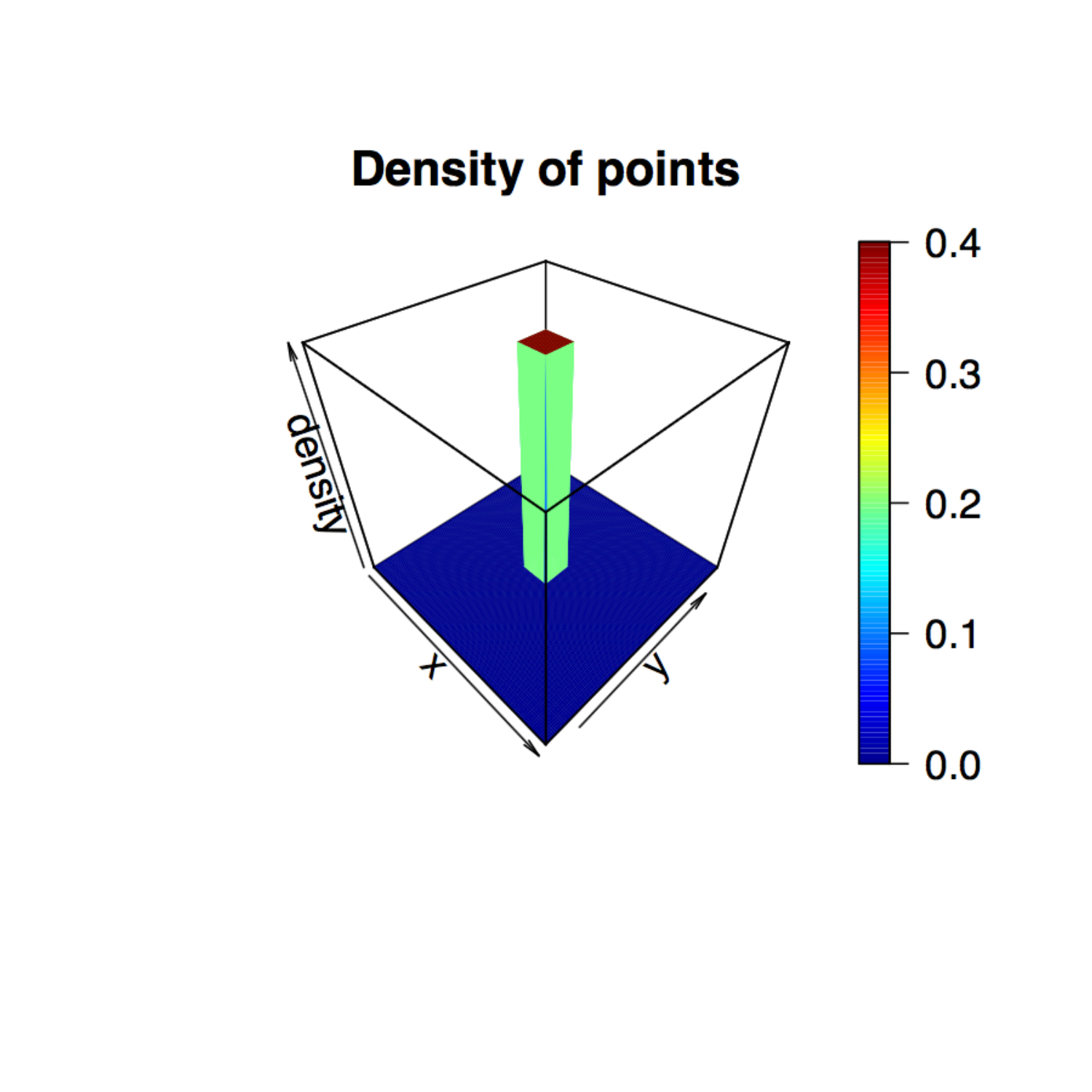}}
	\hspace{0.1cm}
	\subfloat[][{\label{fig:hist_Matern}}]
	{\includegraphics[width=.30\textwidth]{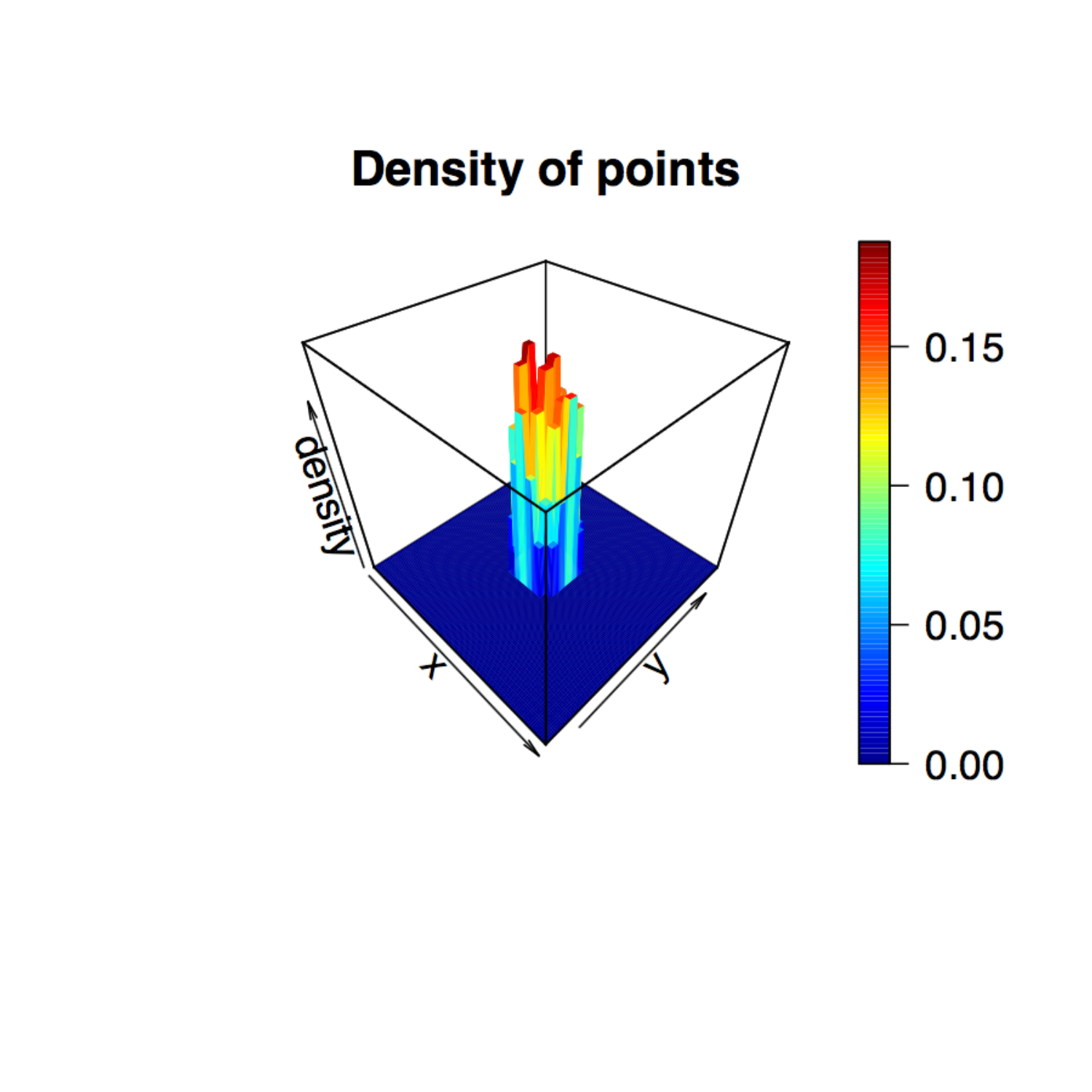}}
	\hspace{0.1cm}
	\subfloat[][{\label{fig:hist_Thomas}}]
	{\includegraphics[width=.30\textwidth]{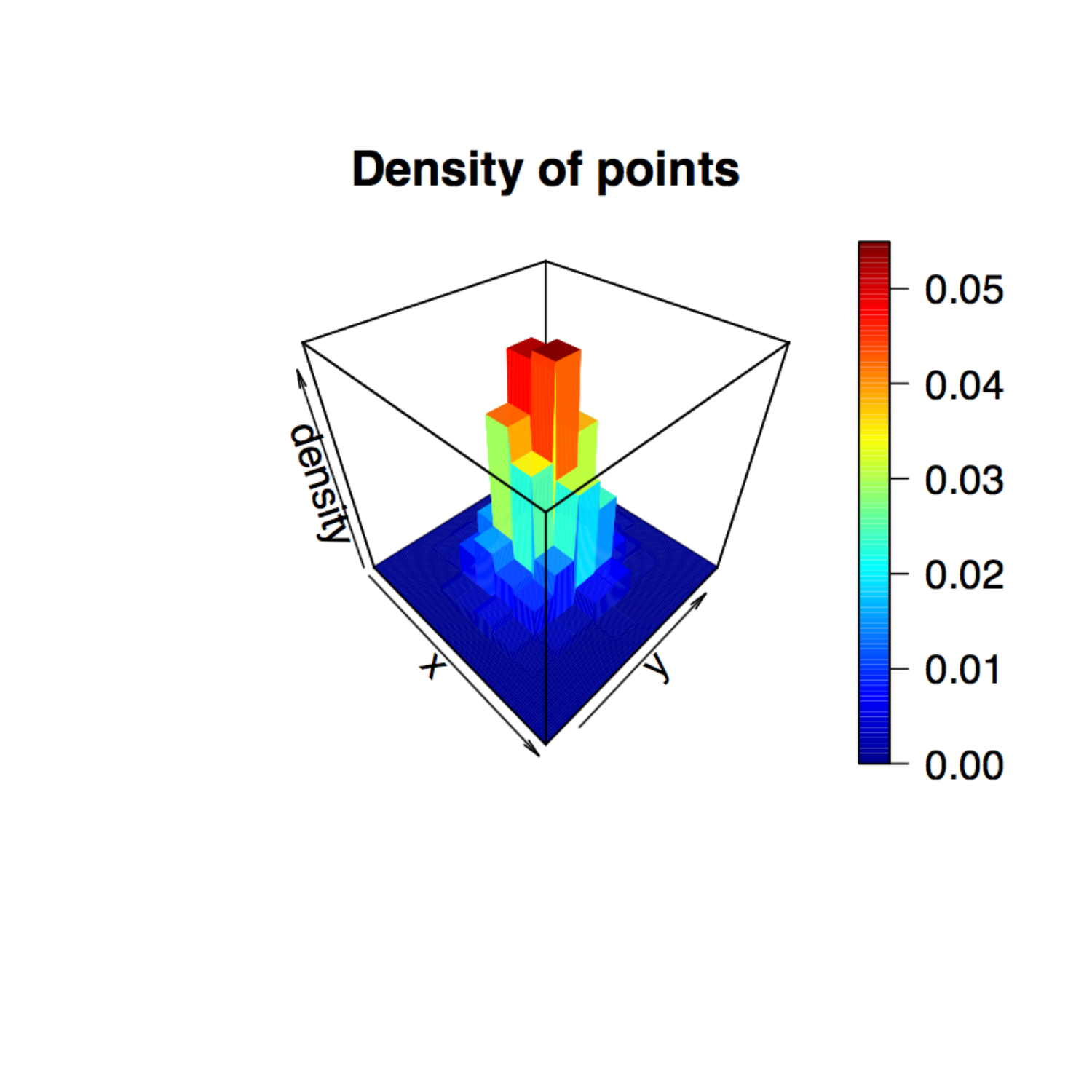}}\\
	\vspace{-0.3cm} 
	\caption{Kernel vs Knuth on three different type of clusters: (a) square with uniform density, (b) circular with uniform density as for Mat\'ern process and (c) circular with Gaussian density as for \emph{mTp} process. From top to bottom: distribution of points, kernel estimation of the intensity, Knuth estimation and Knuth histogram. In the first case Knuth collects the points in a unique cluster correctly detecting the homogeneity of the square cluster structure. In the second and in the third case it arranges data with a higher number of bins to capture the circular boundary of the clusters. On the contrary, from kernel density plots there seems to be no actual difference from the three clusters apart from their size.}
	\label{fig:one_cluster}
\end{figure*}

In the main article we found a strong correlation between Knuth optimal bin area and the characteristic size of three different type of clusters consisting of 1000 points: a square and a circular one with constant density and a circular one with Gaussian density. In \figurename~\ref{fig:one_cluster} we plot an example of three of these datasets with $\sigma=50$.\\
We compare here Knuth method with Epanechnikov kernel method in estimating the intensity function of these different cluster-structures.\\ 
In the first case Knuth collects the points in a unique cluster, as we can see from the optimal grid we insert in the plot of the data distribution. Thus the resulting estimation of the intensity perfectly coincides with the underlying one, given by the constant value $\lambda=1000/50^2=0.4$ in the $50\times50$ cluster area.\\ In the second and in the third case Knuth arranges data in a histogram with a higher number of bins in order to capture the circular boundary of the clusters. In all cases, we remark that the bin sides are practically equal, meaning Knuth detects the isotropy of the cluster structures.\\
By contrast, looking at the density plots obtained by Epanechnikov kernel method, we can see that the three clusters are not quite distinguishable one from the other if not slightly for their sizes
. 
\section{Anisotropic Gaussian clusters}
\begin{figure*}
	\centering
	\includegraphics[scale=0.4]{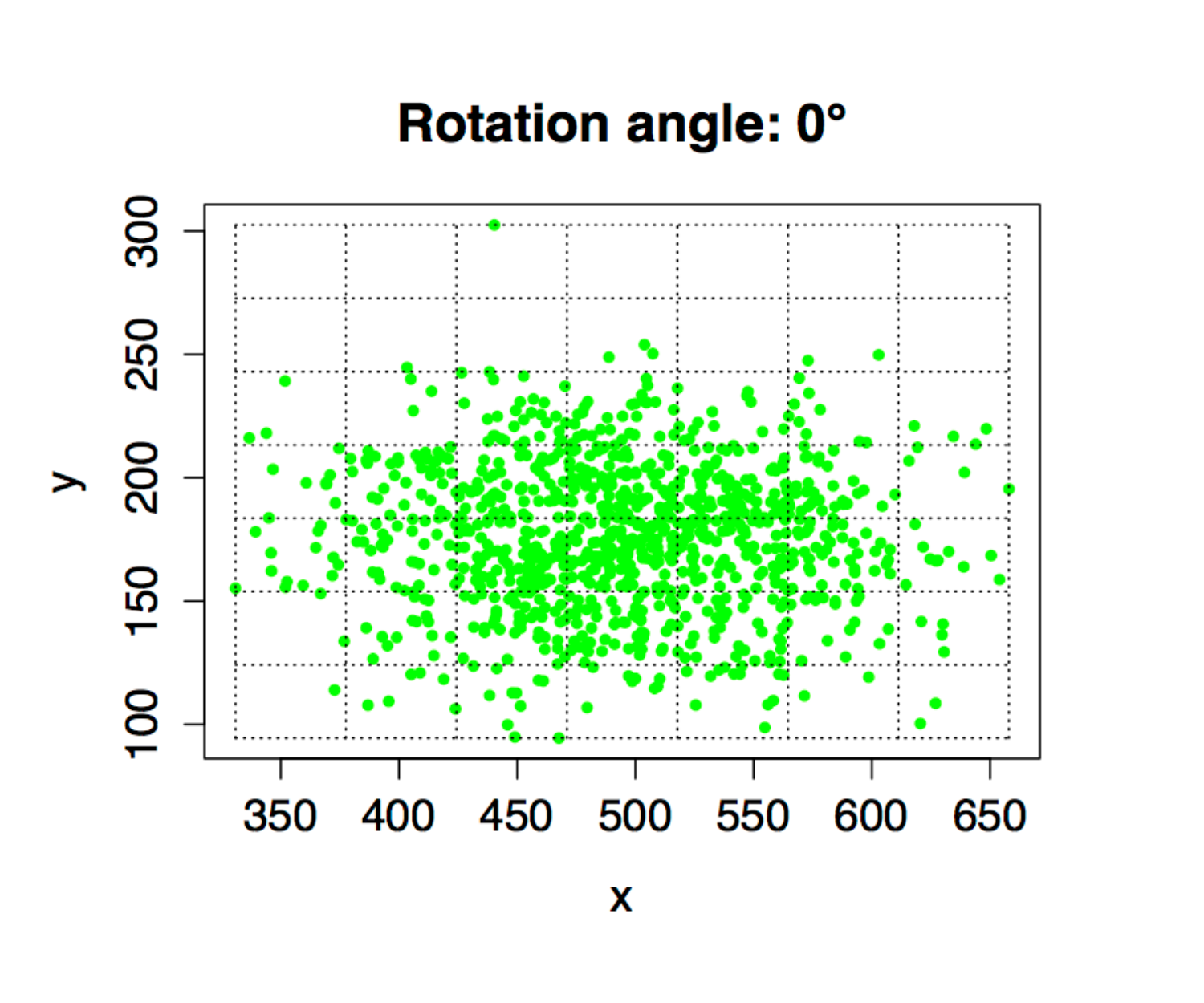}
	\includegraphics[scale=0.4]{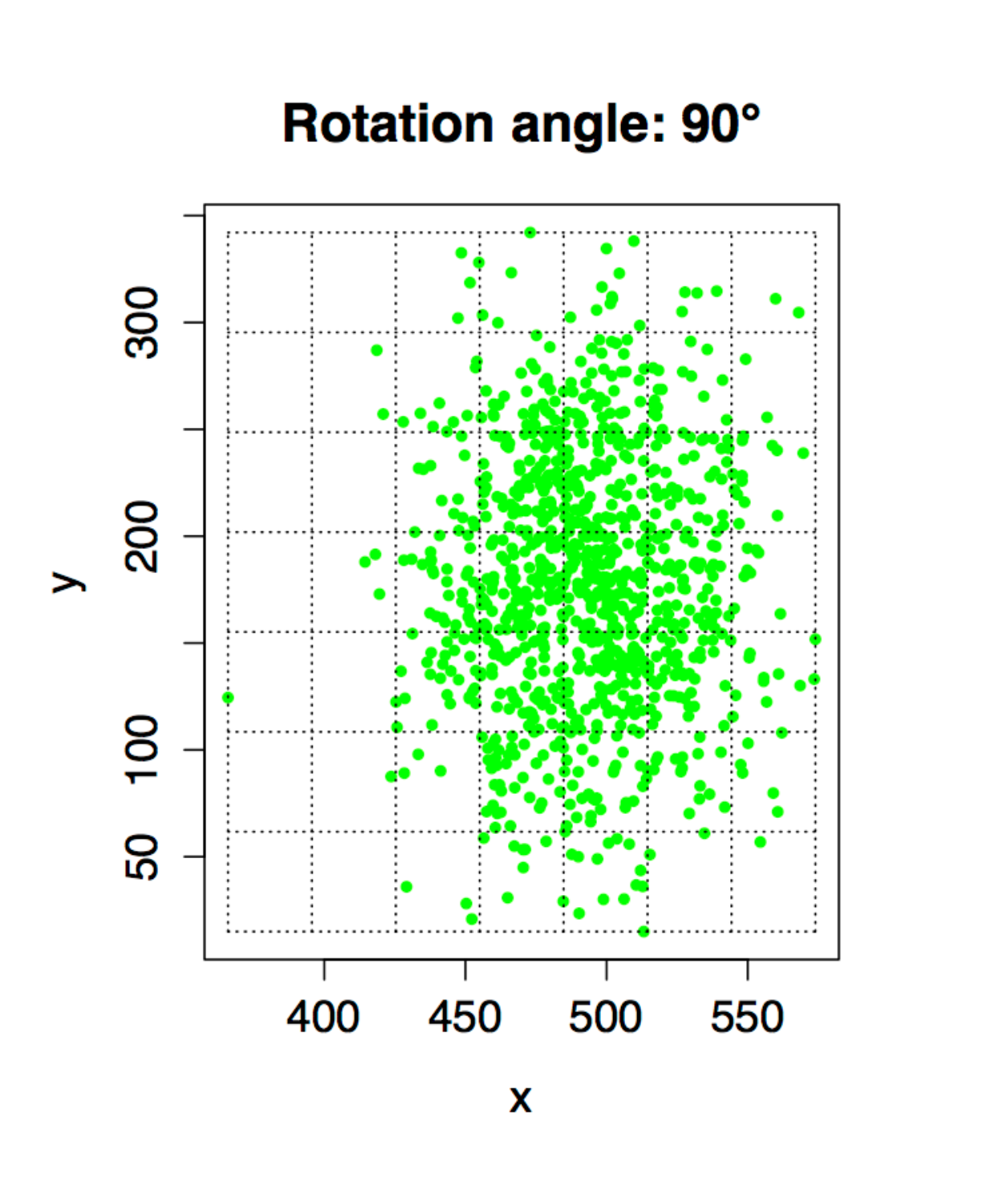}\\
	\vspace*{-0.3cm} 
	\includegraphics[scale=0.4]{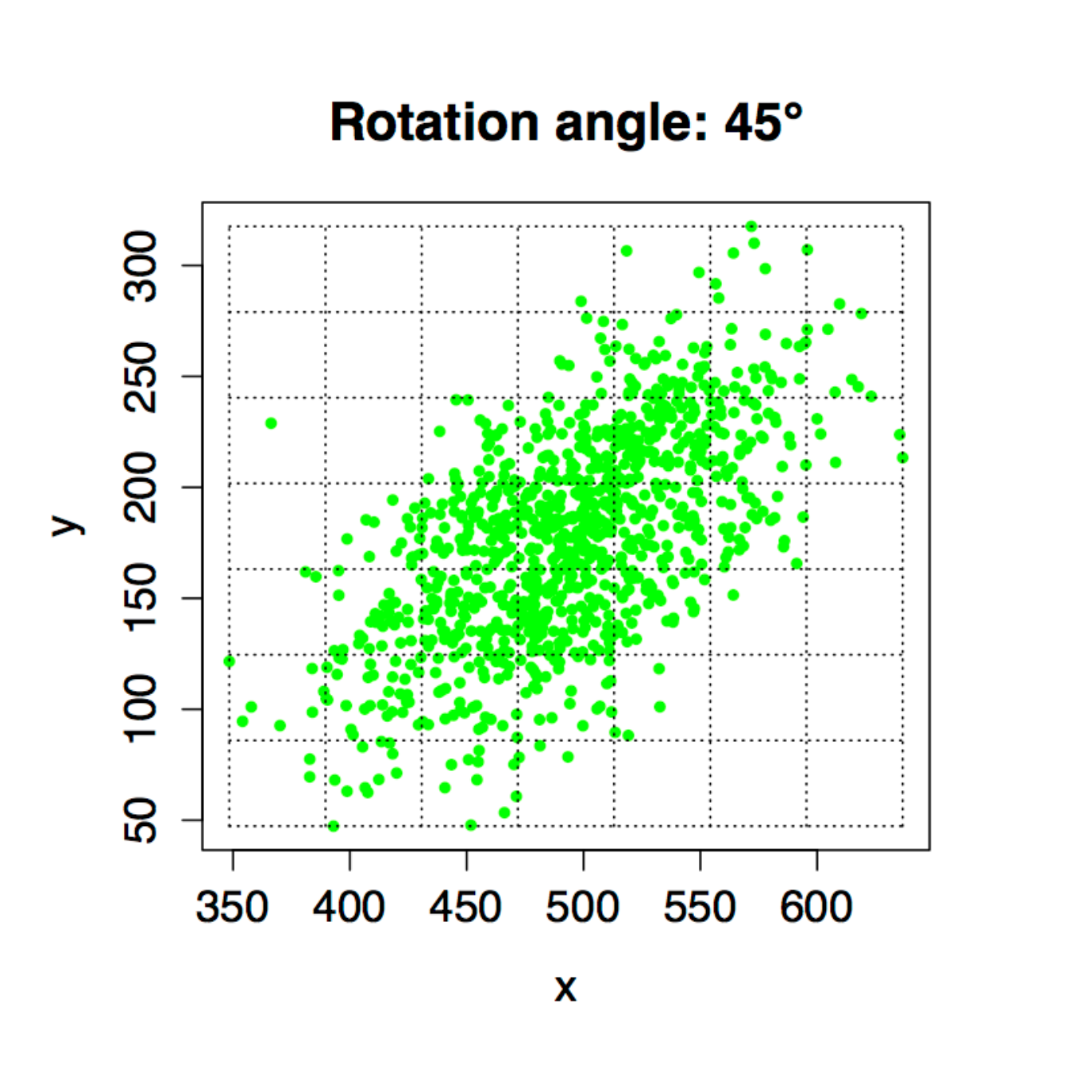} 
	\includegraphics[scale=0.4]{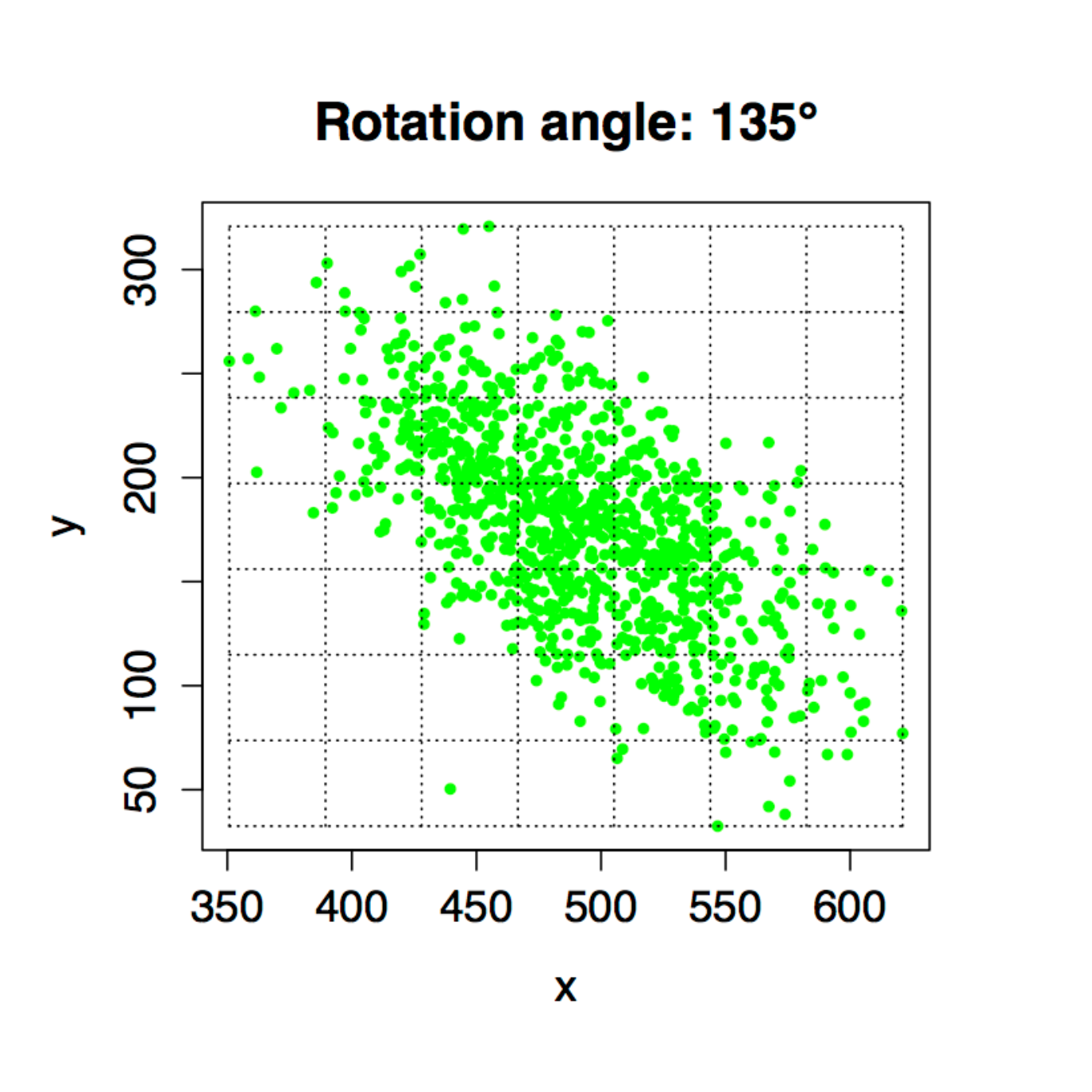}
	\caption{{The Knuth method's answer to anisotropic Gaussian clusters: plot of a dataset consisting of 1000 individuals clumped in one cluster with $\sigma_x=60 $ units and $\sigma_y=30$ units rotated of: $\ang{0}$, $\ang{90}$, $\ang{45}$ and $\ang{135}$ (top to bottom, left to right), arranged in the optimal grid returned by Knuth method. This latter well captures the anisotropic structure of clusters: firstly, it results sensitive to the inversion of variance's direction between $\ang{0}$-$\ang{90}$ and $\ang{45}$-$\ang{135}$ cases, which leads to the inversion of the optimal bin sizes; secondly, it responds well to the anisotropic structure, since in the $\ang{0}$ and $\ang{90}$ cases Knuth method sees the greater variance along $x$ with respect to the one along $y-$axis, while in the $\ang{45}$ and $\ang{135}$ cases, it captures the isotropy along the principal axes.}\label{fig:Knuth_anisotropy}}
\end{figure*}

We test the reliability of Knuth method on detecting the possible anisotropy of a point process. In particular, we apply the Knuth method on a dataset aggregated in a unique cluster distributed according to a bivariate Gaussian with standard deviation along the $x-$axis twice than along the $y-$axis. We then rotate the data points of different angles around the \emph{parent}'s coordinates and we apply the Knuth method to the new dataset. In \figurename~\ref{fig:Knuth_anisotropy} the generated datasets rotated of $\ang{0}$, $\ang{90}$, $\ang{45}$ and $\ang{135}$ are represented. The data are also arranged in the optimal grid returned by Knuth method. Knuth grid well captures the anisotropic structure of clusters: it gives $47\times30$ units in the first case, $30\times47$ units in the second, $41\times39 $ units in the third and $39\times41$ units in the last one.\\
Moreover it responds well to the anisotropic structure, since in the $\ang{0}$ and $\ang{90}$ cases Knuth method sees the greater variance along $x$ with respect to the one along $y-$axis, in the $\ang{45}$ and $\ang{135}$ cases, it captures the isotropy along the principal axes.

\section{Anisotropic uniform clusters}
\begin{figure*}
	\centering
	\subfloat[{}\label{fig:Knuth_horizontal}]
	{\includegraphics[width=70mm]{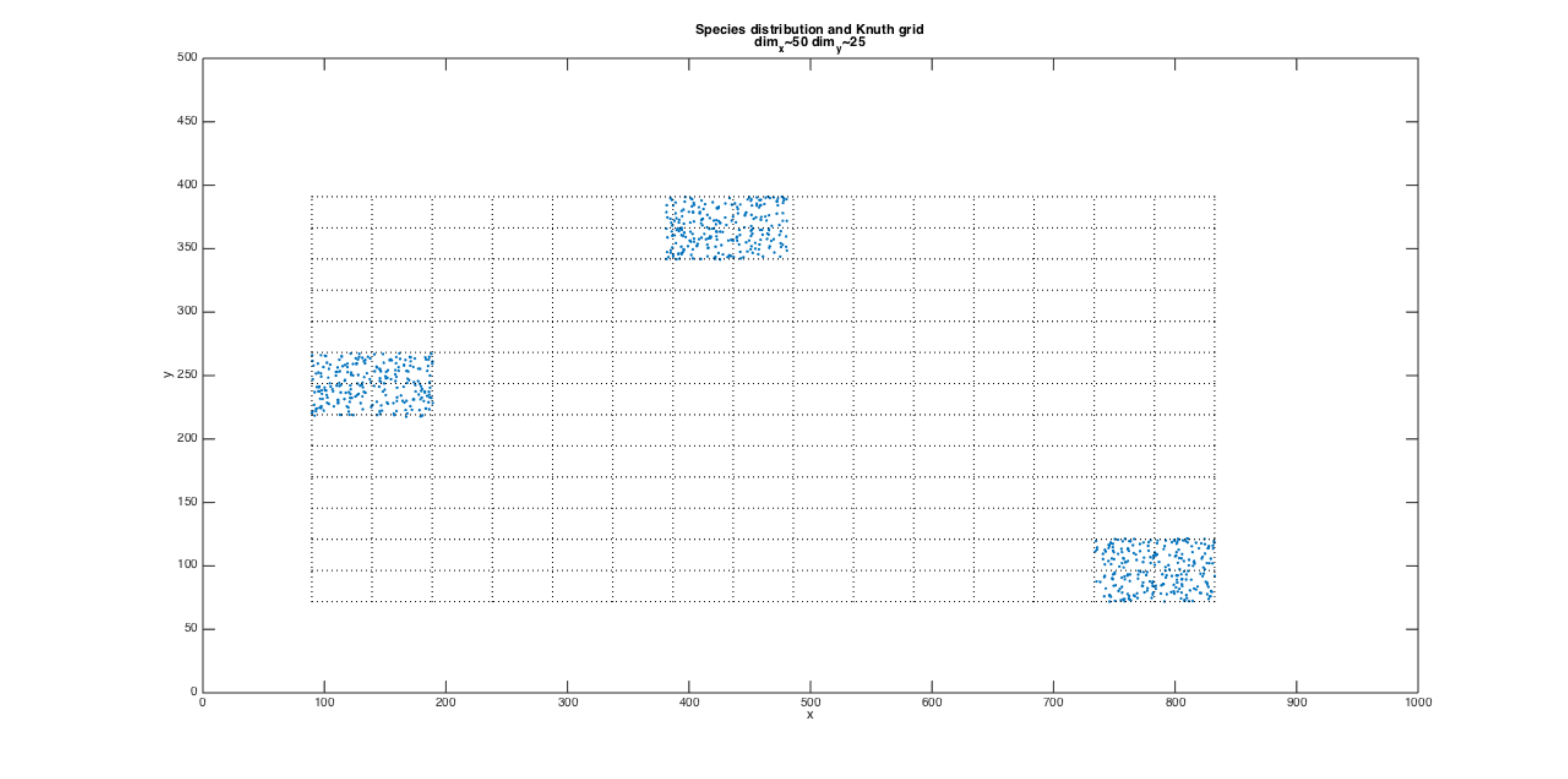}}
	\subfloat[{}\label{fig:pcp_horizontal}]
	{\includegraphics[width=70mm]{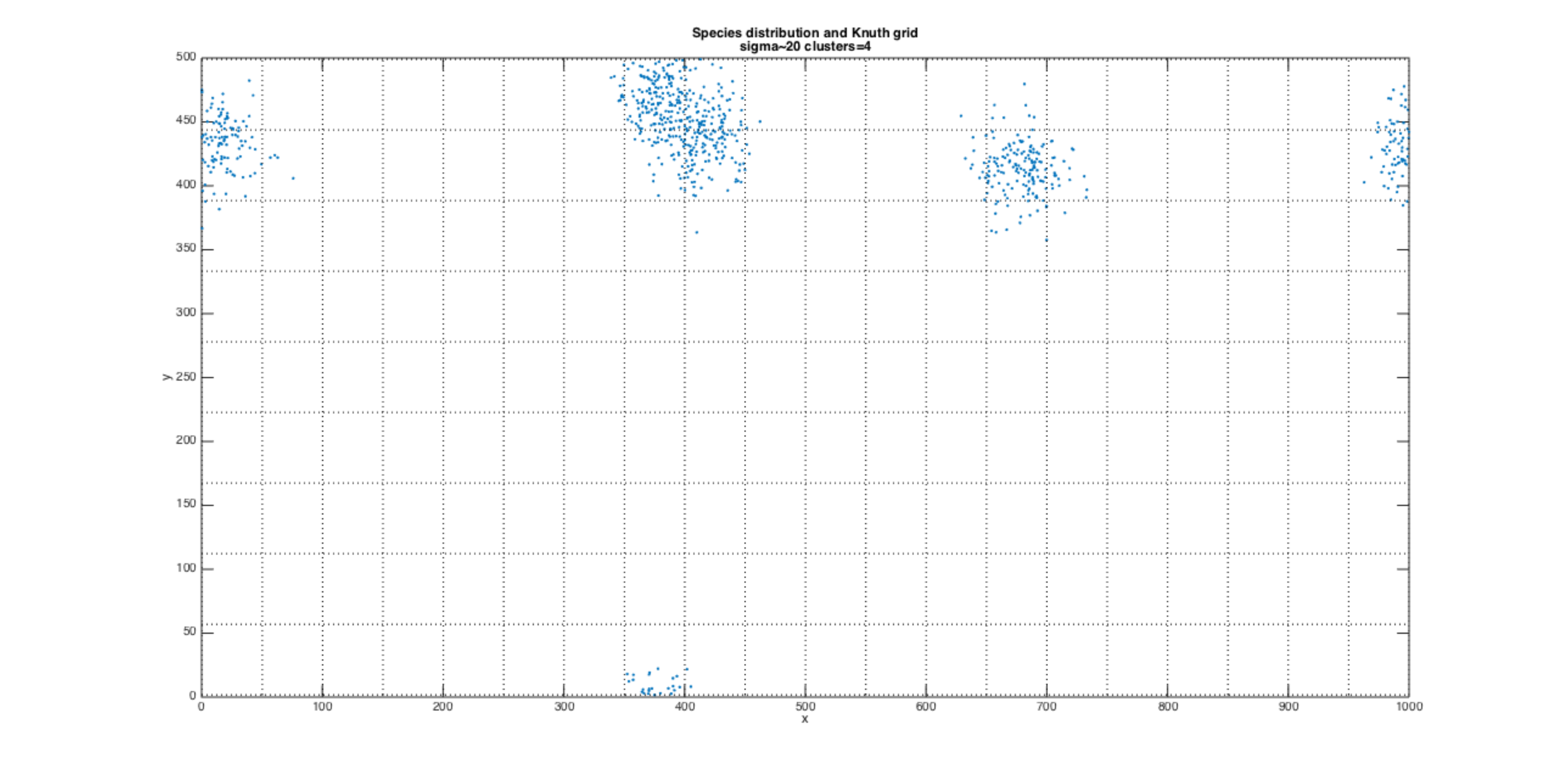}}\\
	\vspace*{-0.3cm} 
	\subfloat[{}\label{fig:Knuth_vertical}]
	{\includegraphics[width=70mm]{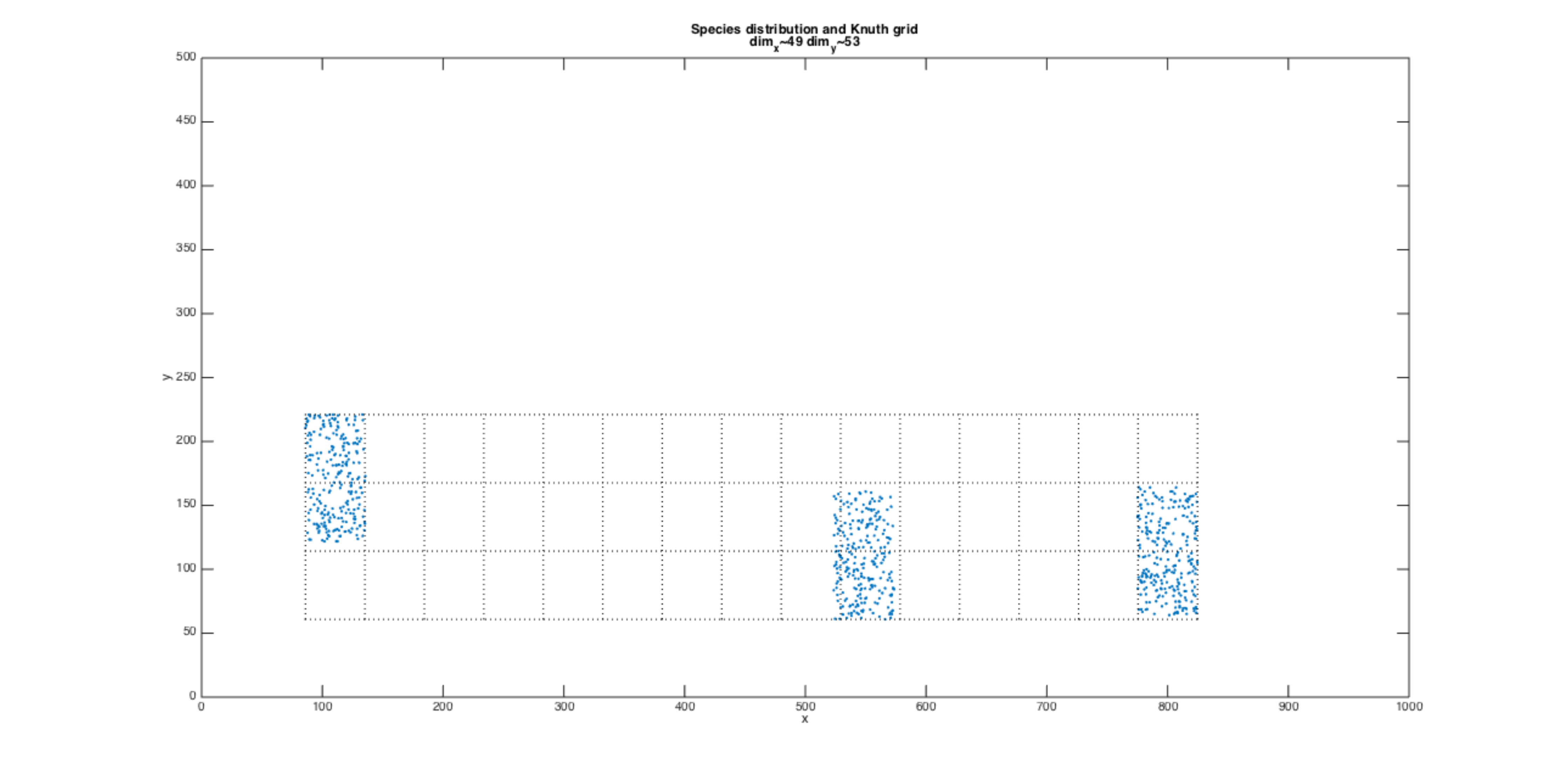}}
	\subfloat[{}\label{fig:pcp_vertical}]
	{\includegraphics[width=70mm]{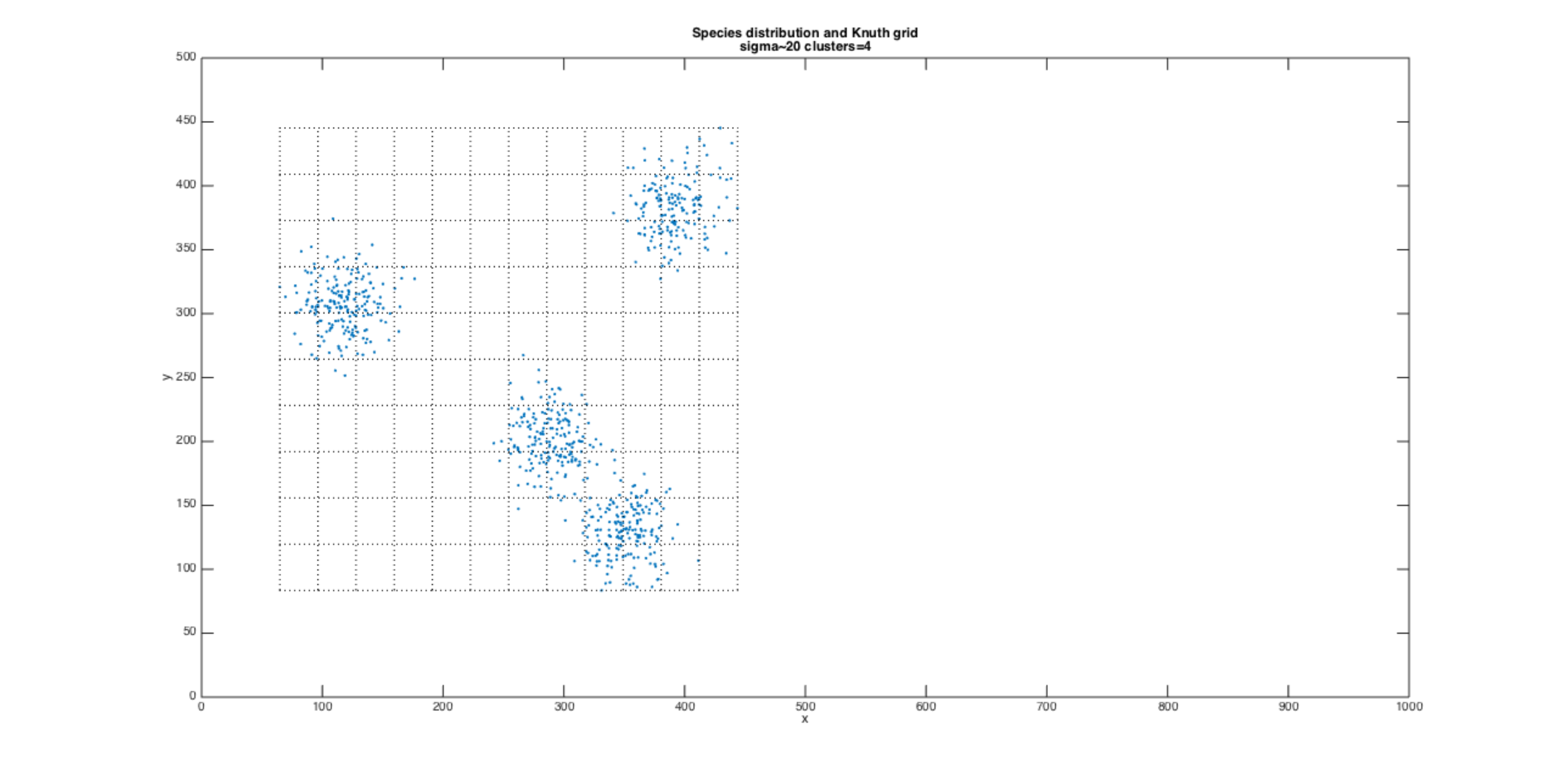}}\\
	\vspace*{-0.3cm} 
	\subfloat[{}\label{fig:Knuth_square}]
	{\includegraphics[width=70mm]{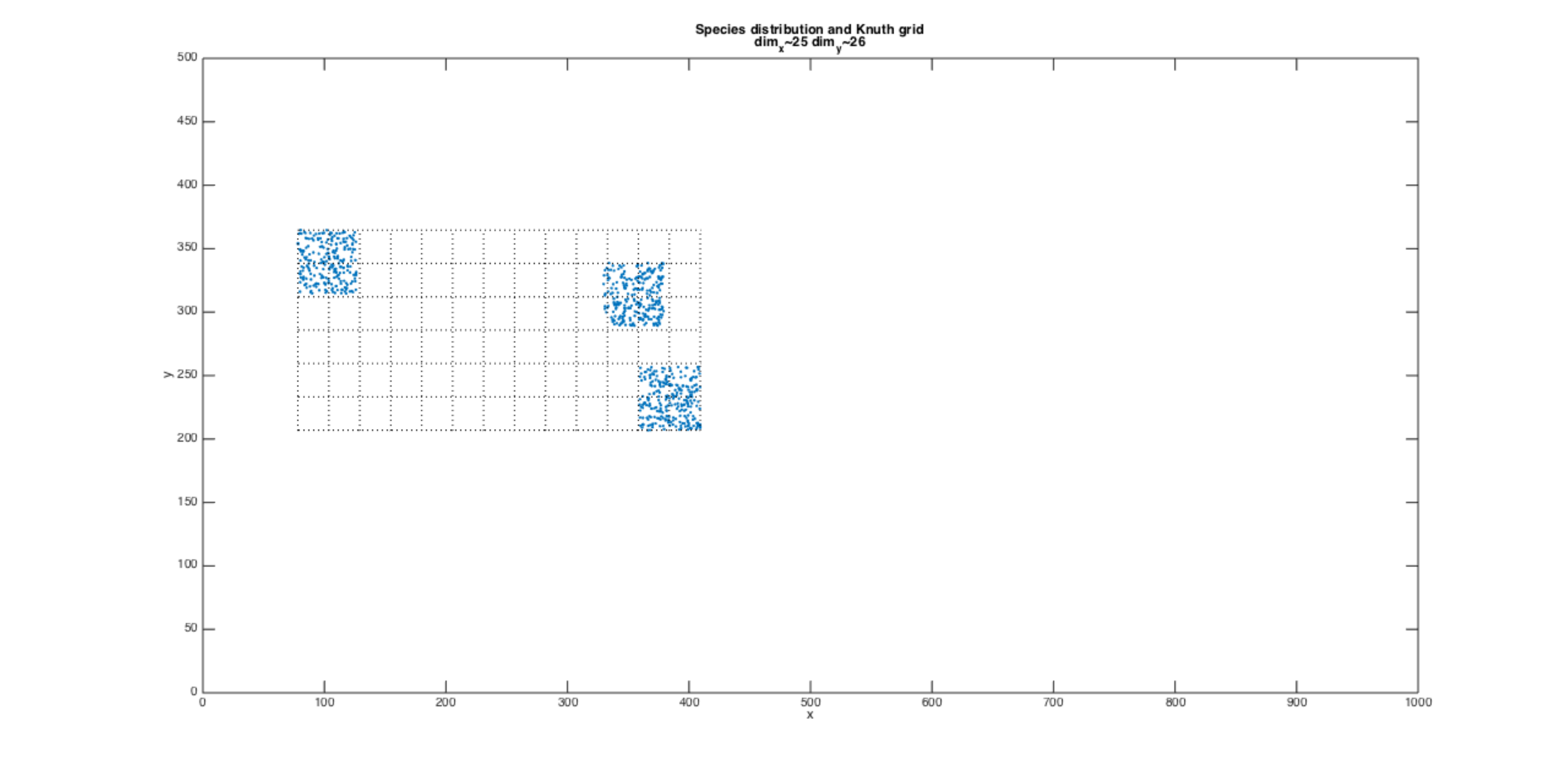}}
	\subfloat[{}\label{fig:pcp_square}]
	{\includegraphics[width=70mm]{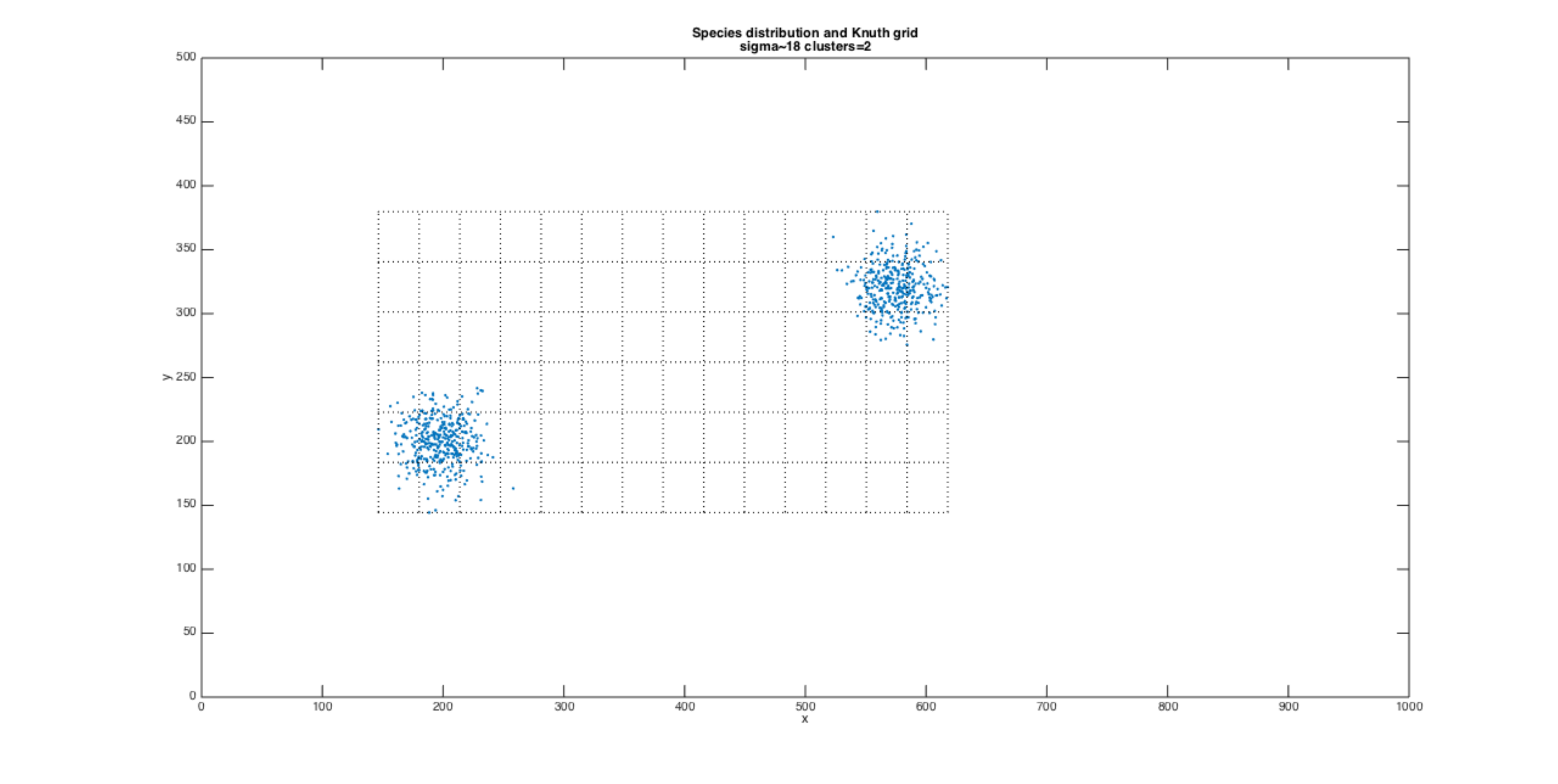}}
	\vspace*{0.3cm} 
	\caption{{Knuth and \textit{mTp} answer to anisotropic uniform clusters. On the left column we plot three datasets consisting of 750 individuals clumped in three clusters of different dimensions: (a) $100\times50$ m, (c) $50\times100$ m and (e) $50\times50$ m. On the right column we plot the datasets according to a modified Thomas process: parameters $\mu, \rho$ and $\sigma$ are chosen to best fit the hypothetical $K$-Ripley function for the process compared with the empirical one. As we can see, Knuth grid optimally captures both uniform and anisotropic clusters, while \textit{mTp} clearly not, since it does assume Gaussian isotropic clusters.}\label{fig:Knuth_vs_Pcp}}
\end{figure*}

We test now on data generated form a Poisson cluster process where the \emph{offspring} is distributed around its \emph{parent} according to a uniform distribution on clusters of dimension $100\times50$ (\figurename~\ref{fig:Knuth_vs_Pcp}\subref{fig:Knuth_horizontal}), $\ 50\times100$ (\figurename~\ref{fig:Knuth_vs_Pcp}\subref{fig:Knuth_vertical}) and $50\times50$ (\figurename~\ref{fig:Knuth_vs_Pcp}\subref{fig:Knuth_square}).\\ In \figurename~\ref{fig:Knuth_vs_Pcp} we inserted the point distribution in the whole observation window (left column).\\
As we can see, once again Knuth well captures the uniform structure of the data: it returns bin sizes proportional to the length of the chosen dimensions of the clusters: $50\times25$ for the first case, $49\times53$ for the second and $25\times26$ for the last one.\\
On the right column we plot the data generated according to an $mTp$ with the parameters that we get by fitting the original data.
We get four clusters with $\sigma=4$ for the first two cases, which therefore are considered equal in the \emph{mTp} fitting and two clusters with $\sigma=18$ for the last.
	
\section{Application of Knuth to BCI}
\begin{figure*}
	\centering
	\subfloat[{}\label{fig:hist_clumpsize}]
	{\includegraphics[width=70mm]{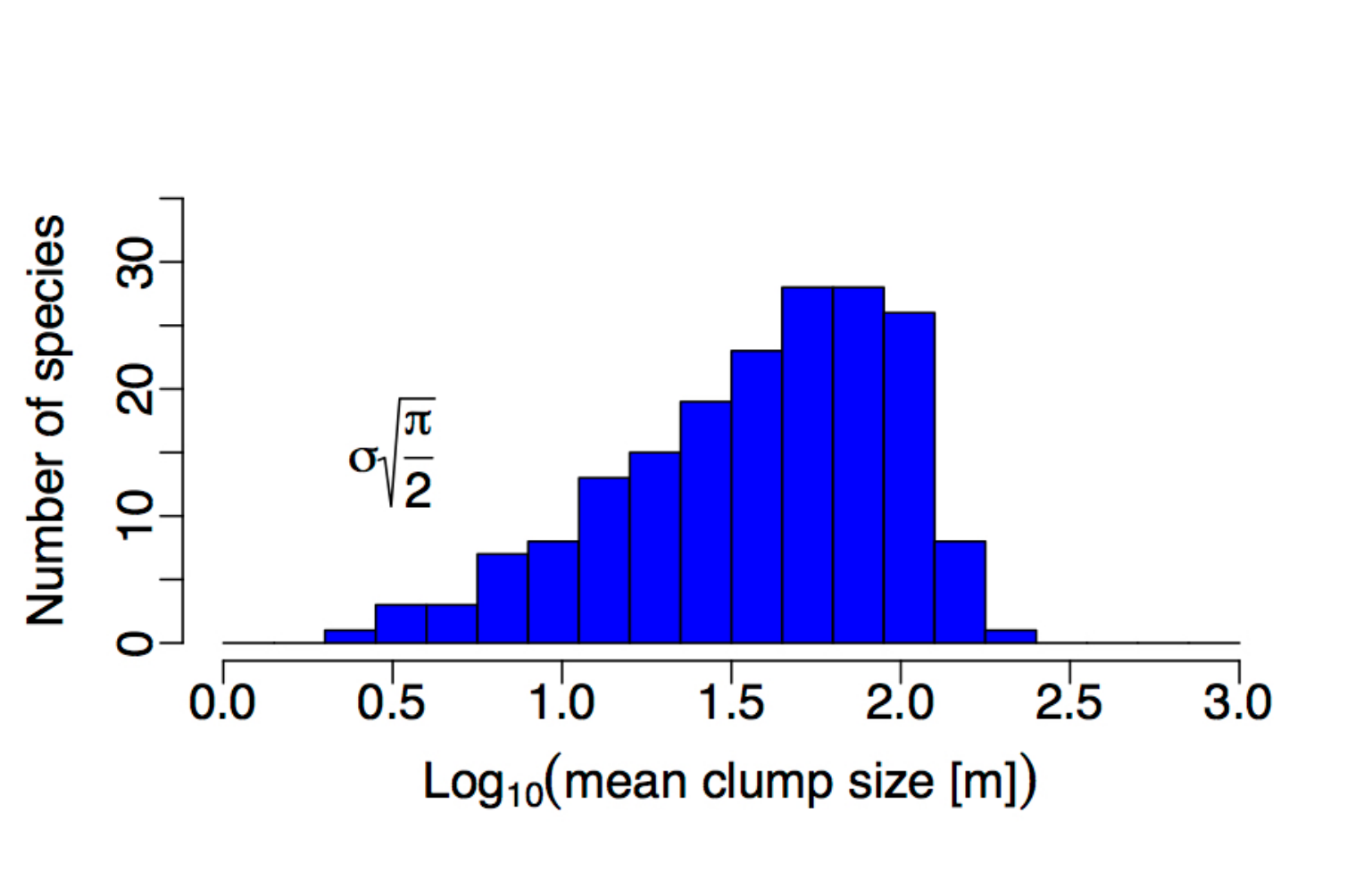}}
	\subfloat[{}\label{fig:hist_rho}]
	{\includegraphics[width=70mm]{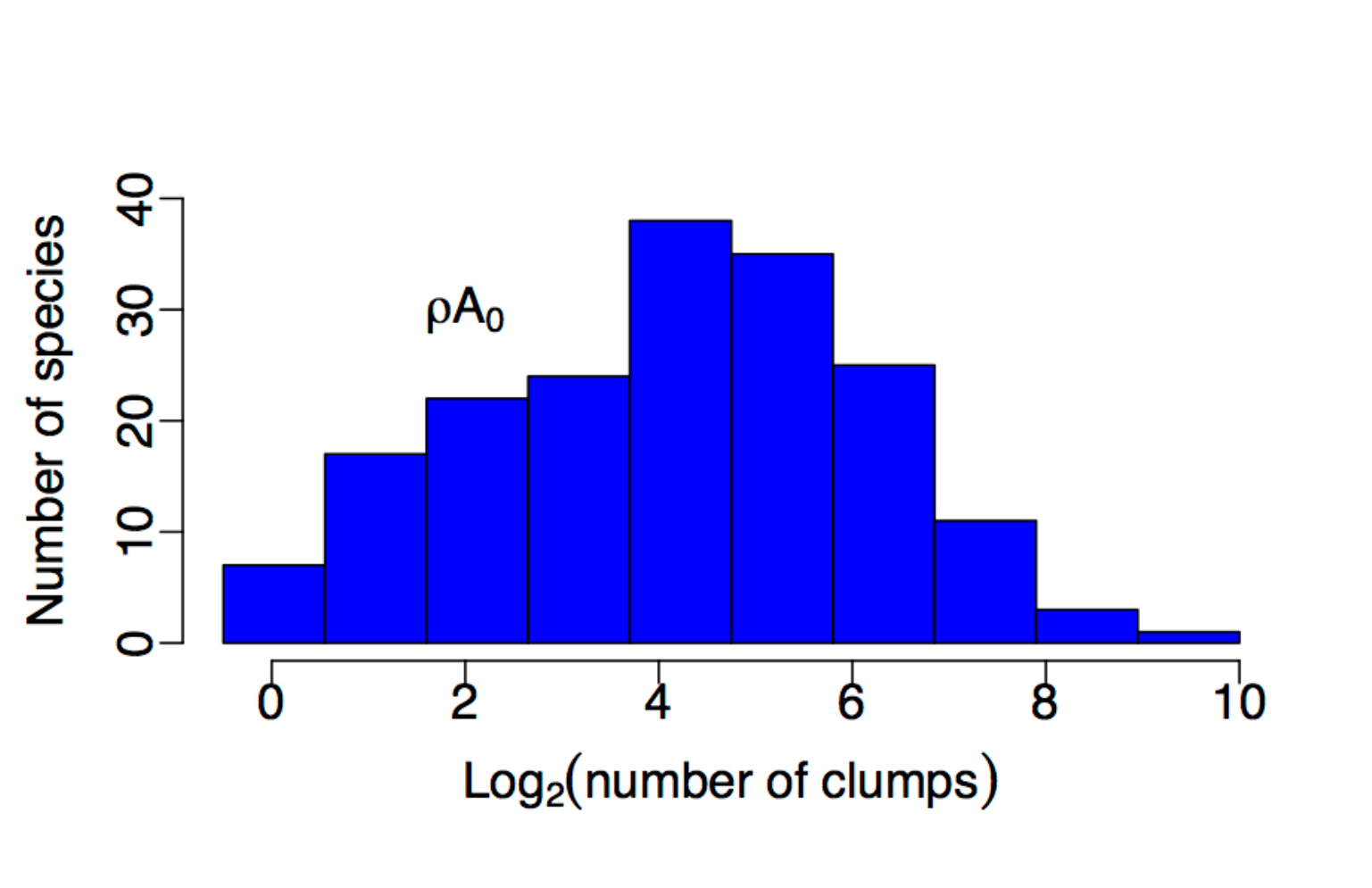}}\\
	\vspace*{-0.4cm} 
	\subfloat[{}\label{fig:hist_mu}]
	{\includegraphics[width=70mm]{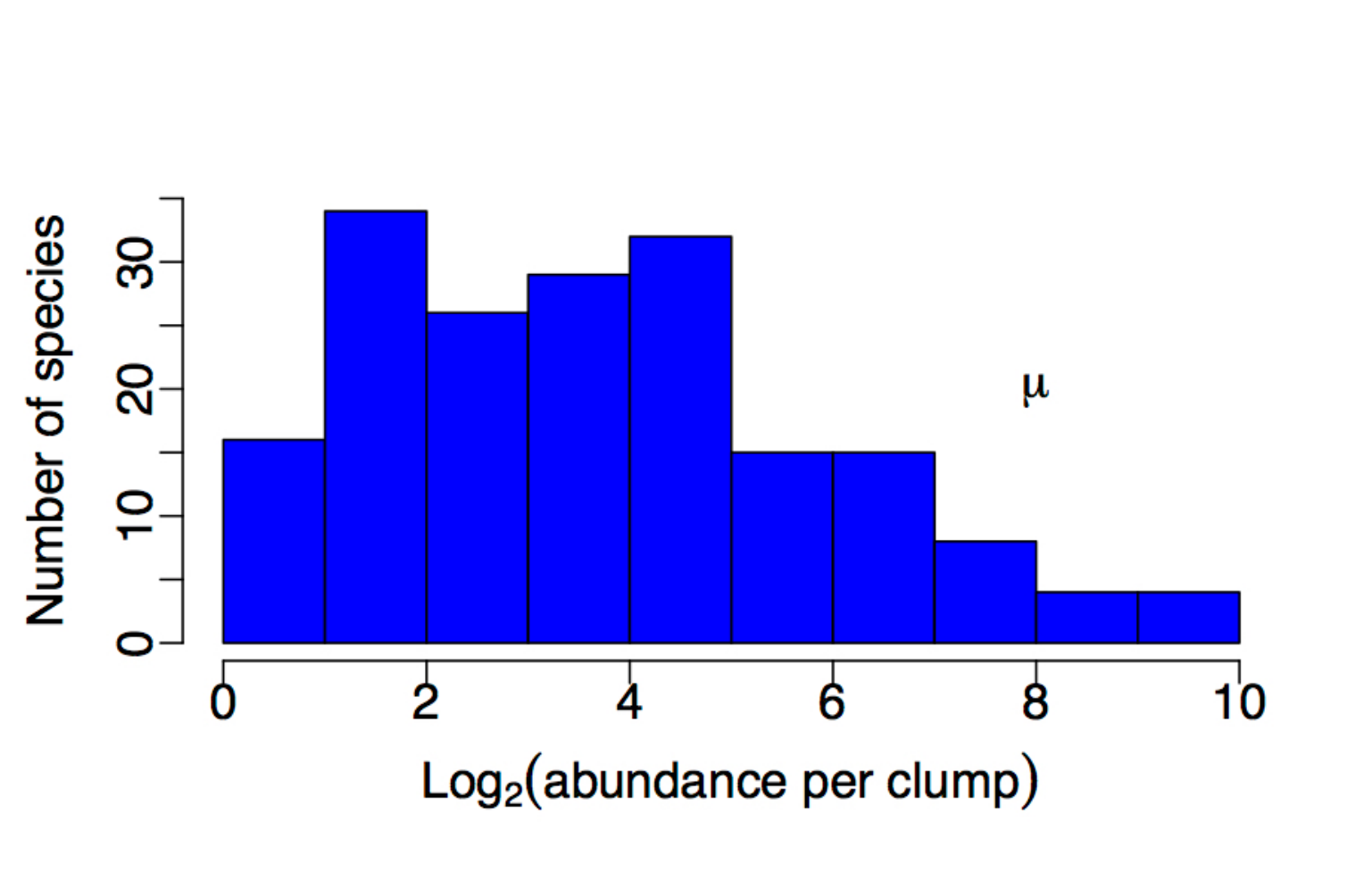}}
	\subfloat[{}\label{fig:hist_areabin}]
	{\includegraphics[width=70mm]{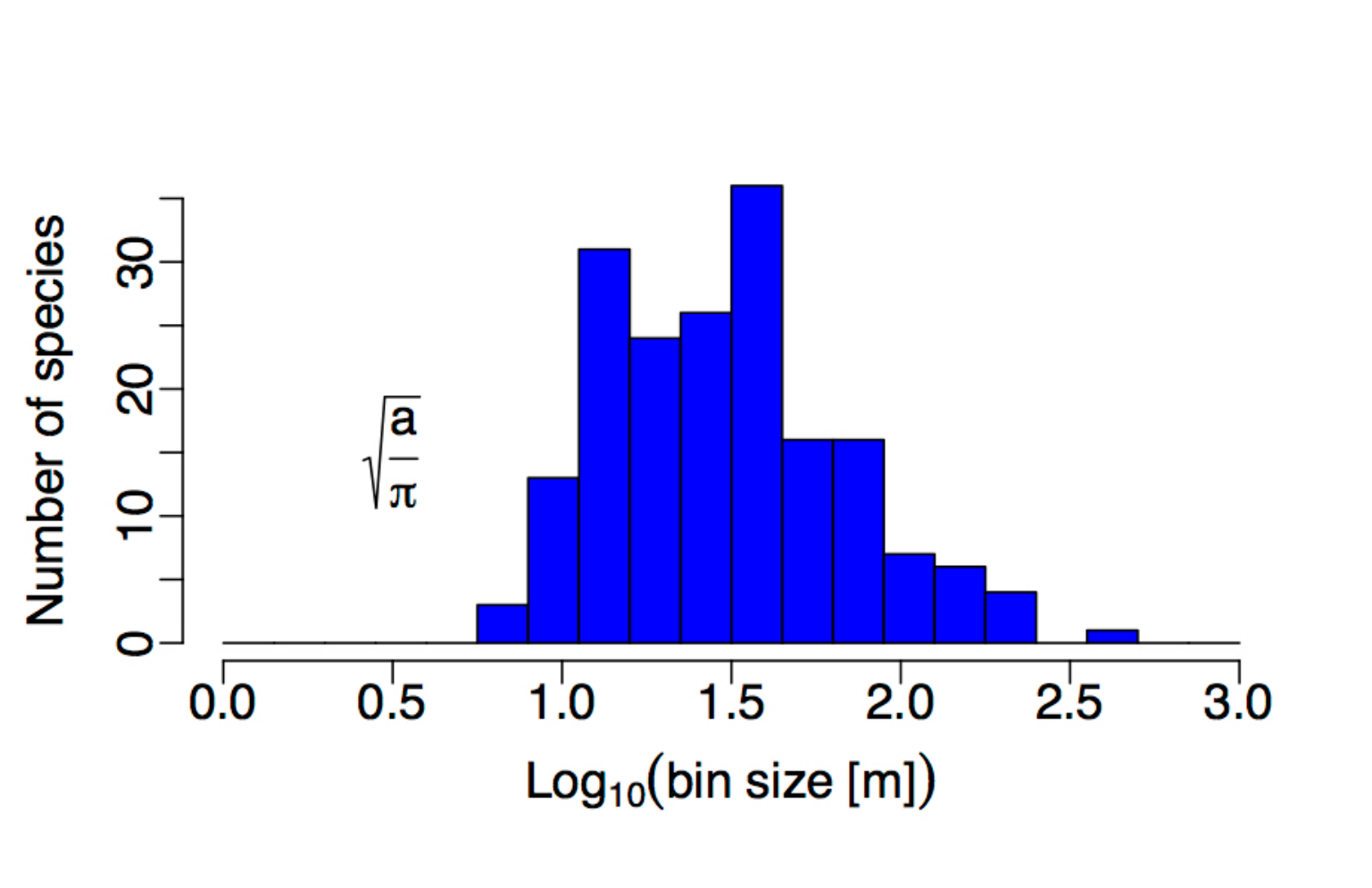}}\\
	\vspace*{0.3cm} 
	\caption{{Frequency histograms of the \textit{mTp} parameters and the Knuth one.}\label{fig:hist_parameters}}
\end{figure*}

In (a)-(c) of \figurename~\ref{fig:hist_parameters} we show the frequency histograms of the \textit{mTp} parameters for BCI selected species (see the main article).\\
We have computed the optimal bin number with Knuth method for each species to investigate how similar or different are the species' distributions of the rainforest within the 50~ha plot and to compare the results with the previous ones.\\
In \figurename~\ref{fig:hist_parameters}\subref{fig:hist_areabin}, we see the frequency histogram of $\sqrt{a/\pi}$, where $a$ is the optimal bin area. This corresponds to the radius of the circle equivalent to the rectangular bin we get by Knuth and which therefore gives us  measure of the size of the underlying minimal structure of the dataset.\\ 
As we can see from the histogram, there is no preferred choice for the common number of bins of all rainforest's species, since the values of the optimal bin area are quite different from each other spanning from $100$ to $ 5\cdot10^5$ circa. This is not surprising, since each species has its own distribution due to myriad of factors such as seed dispersal, gap recruitment or adaptation to the surrounding soil [\citealp{Augspurger};\citealp{Plotkin}].\\
In \ref{fig:num_parameters} we insert the plot the $mTp$ parameters and Knuth ones against the abundance of species to see if they are linearly correlated. In agreement with the literature \citep{Morlon}, we find a slightly positive correlation for the mean clump radius, number of clumps and abundance per clump and a negative one for the relative neighbourhood density. Knuth optimal clump radius and index of anisotropy resulted completely uncorrelated with the abundance.

%%----------------------------------------------------------------------------------------
%%	REFERENCE LIST
%%----------------------------------------------------------------------------------------
%%\phantomsection
%%
%%\printbibliography
%\bibliographystyle{plainnat}
%\bibliography{biblio}
%----------------------------------------------------------------------------------------

%----------------------------------------------------------------------------------------
%	FIGURES
%----------------------------------------------------------------------------------------
%\newpage
%\newpage
%\newpage
%\newpage
%\clearpage
%\newpage
%\newpage
\begin{figure*}
	\centering
	\begin{tabular}{cc}
	\includegraphics[width=70mm]{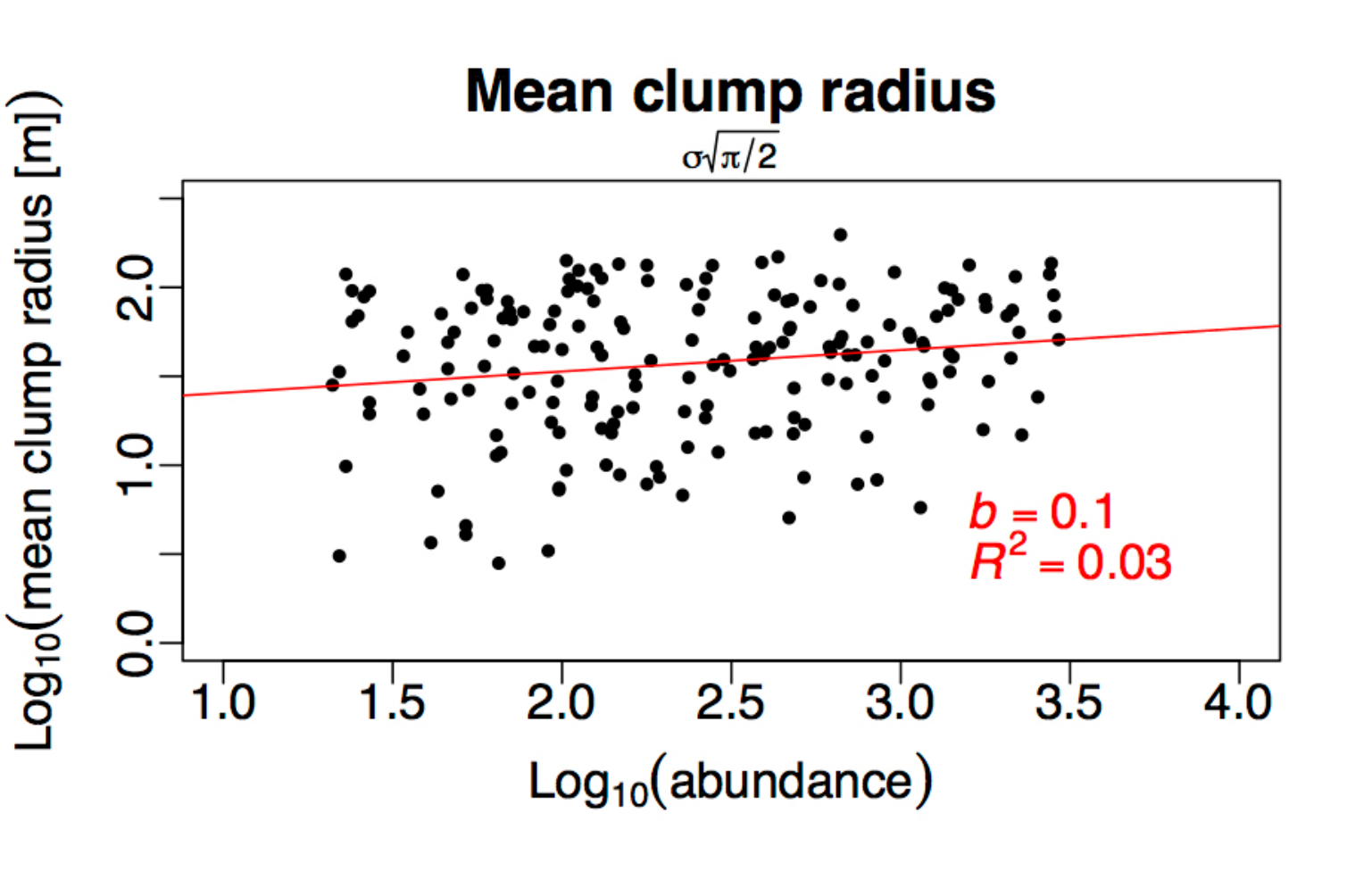}&
	\includegraphics[width=70mm]{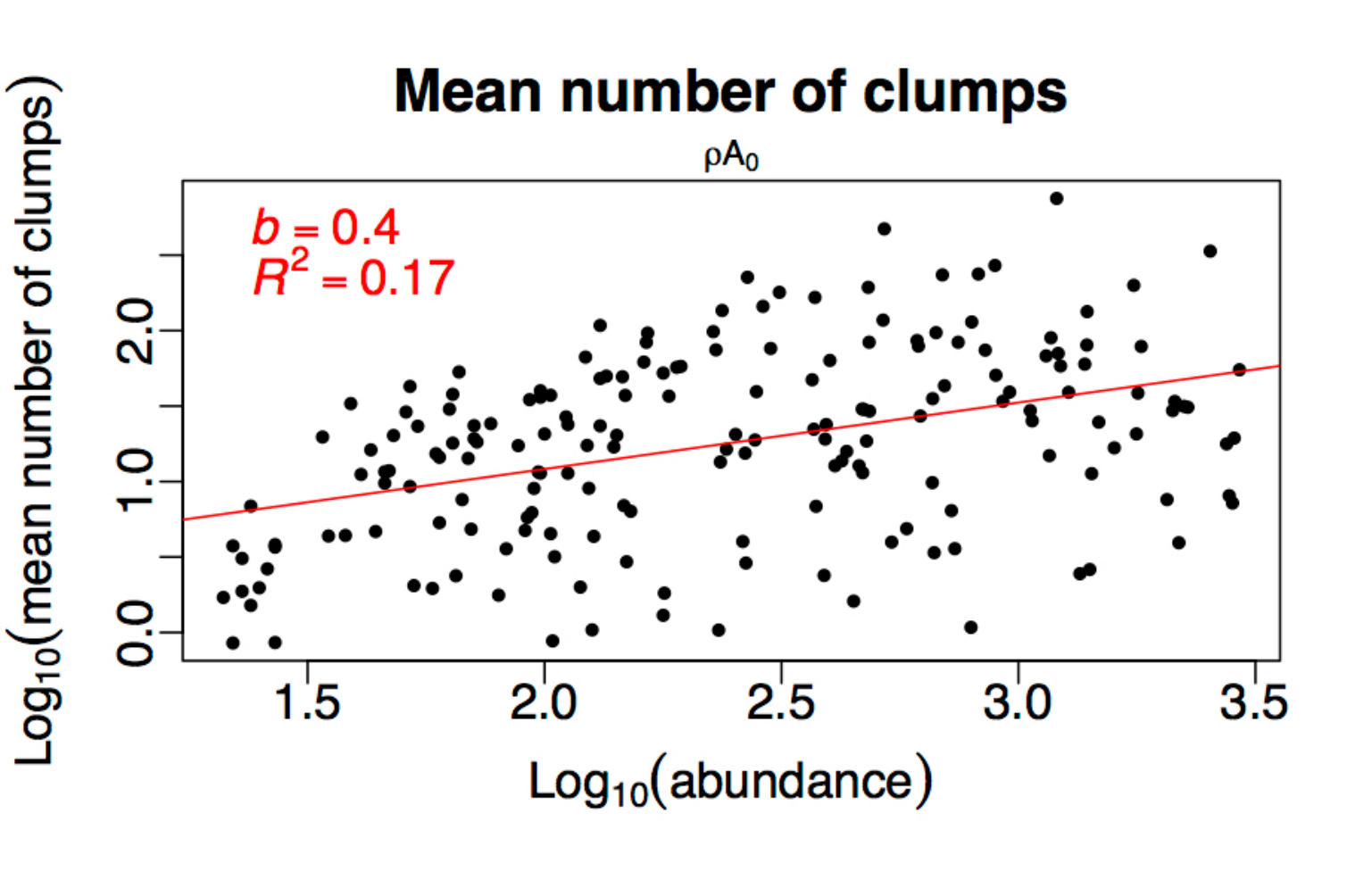}\\
	\includegraphics[width=70mm]{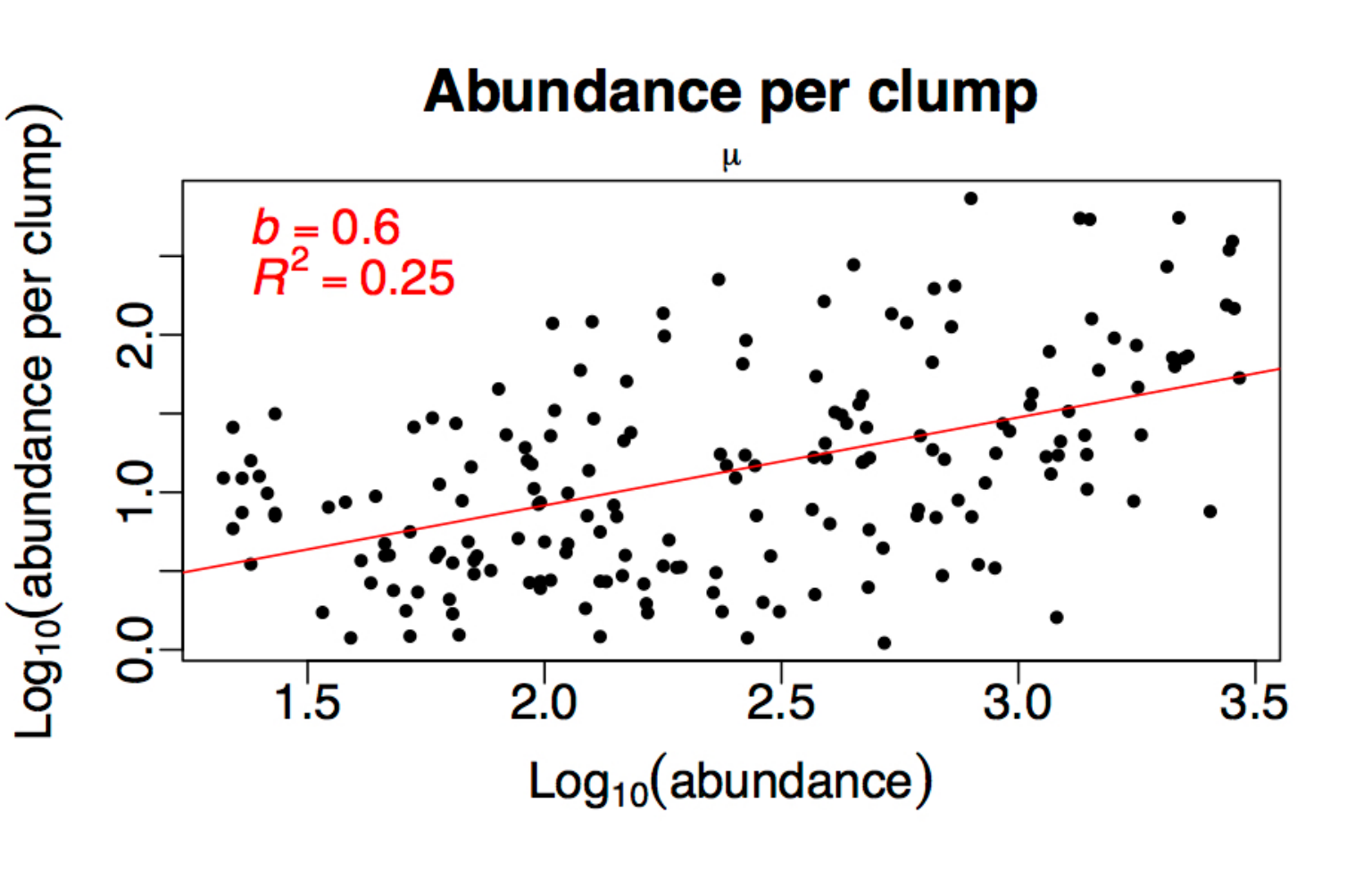}&
	\includegraphics[width=70mm]{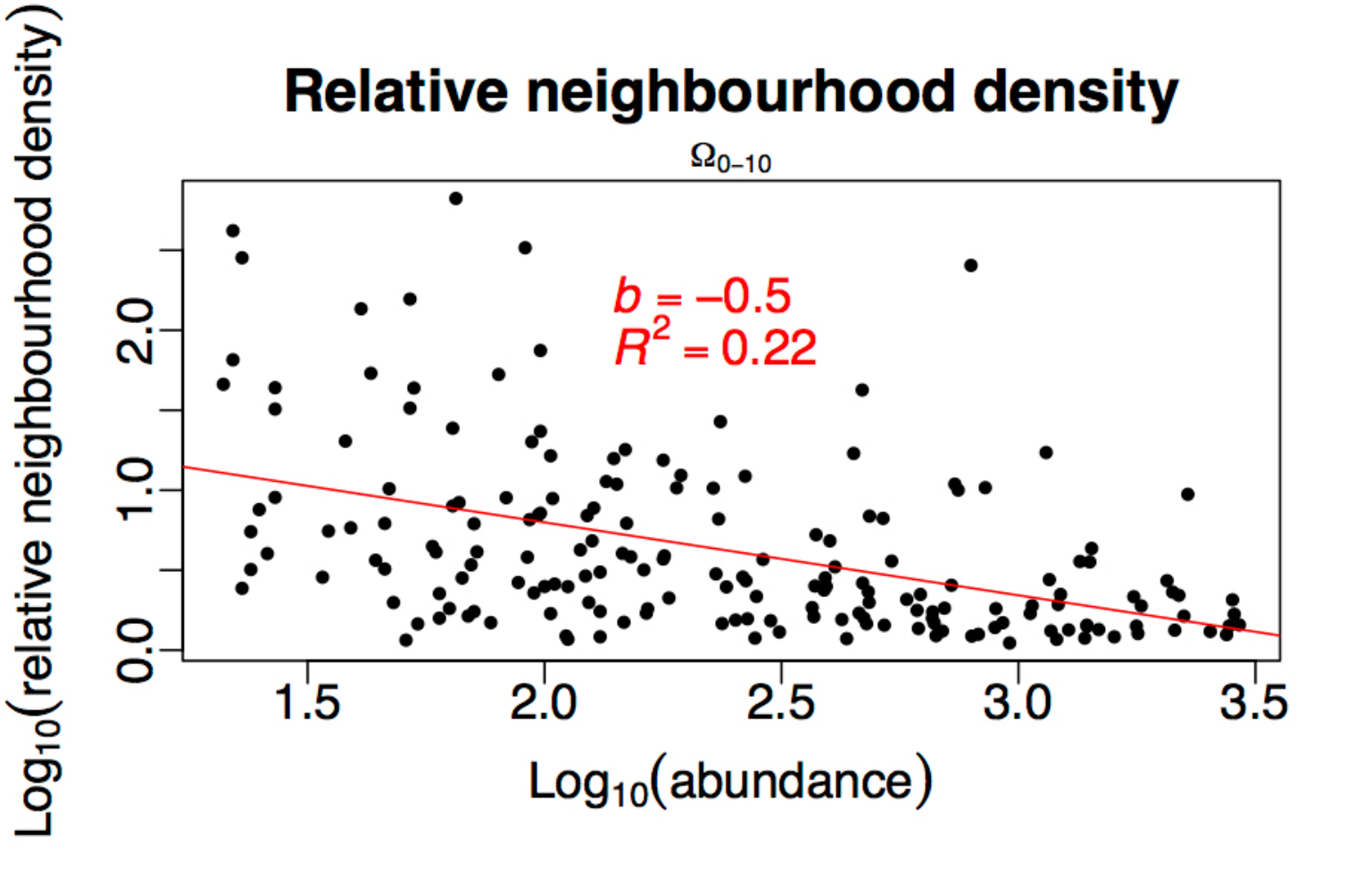}\\
	\includegraphics[width=70mm]{num_areabin}&
	\includegraphics[width=70mm]{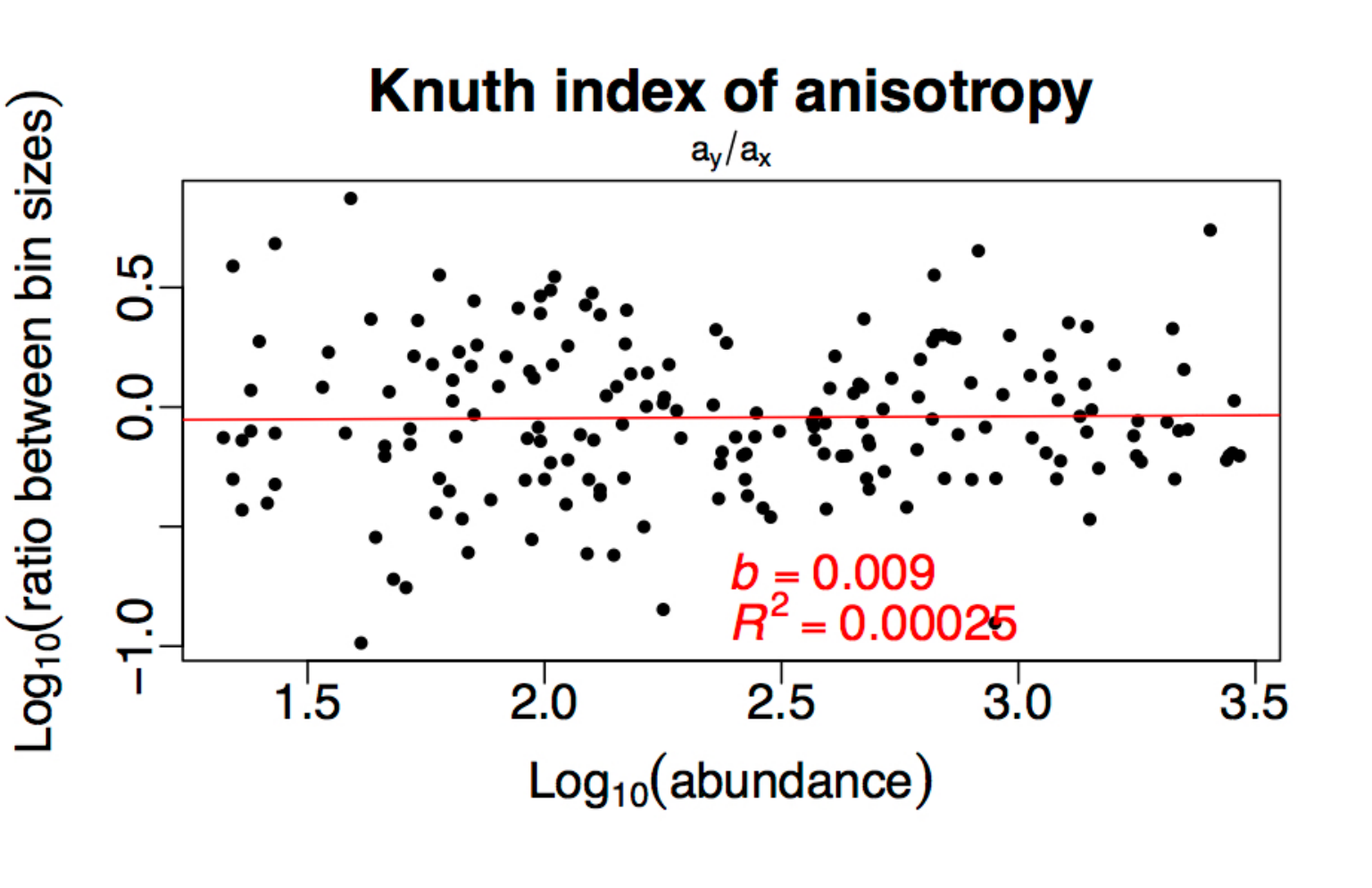}
	\end{tabular}
	\caption{Correlation between \textit{mTp} and Knuth parameters and the abundance of a species.}\label{fig:num_parameters}
\end{figure*}

%\part*{References}
%\bibliographystyle{plainnat}
%\bibliography{biblio}

\end{document}